\documentclass[reprint,longbibliography,footinbib,amsmath,amssymb,amsfont]{revtex4-1} 
\usepackage[colorlinks, allcolors=blue]{hyperref}
\usepackage{graphicx} 
\usepackage{enumitem}
\usepackage{eurosym}
\usepackage[cal=dutchcal]{mathalfa}
\usepackage{csquotes}

\newcommand{\myquote}[1]{ \textit{\textquote{#1}} }
\newcommand*\diff{\mathop{}\!\mathrm{d}}
\newcommand{\iseq}[1]{\underset{#1}{\simeq}}
\newcommand{\thlim}{\underset{\infty}{\simeq}}
\newcommand{\mylist}[6]{%
	\begin{itemize}[leftmargin=*]
		\item \textit{#1}\,: #2
		\begin{enumerate}[label={\alph*)}, leftmargin=*]
			\item #3
			\item #4
			\ifx&#5&%
			\else
			\item #5
			\fi
			\ifx&#6&%
			\else
			\item #6
			\fi
		\end{enumerate}
	\end{itemize}%
}
\newcommand{\myparadox}[5]{%
	\begin{itemize}[leftmargin=*]
		\item \textit{#1}\,: #2
		\begin{enumerate}[label=\Alph*),  leftmargin=*]
			\item #3
			\item #4
			\ifx&#5&%
			\else
			\item #5
			\fi
		\end{enumerate}
	\end{itemize}%
}

\newcommand{\law}[2]{
	\vspace{6pt}
	\noindent
	\textbf{#1}\,: 
	\textit{#2}
	\vspace{6pt}
}

\newcommand{\statement}[1]{
	\vspace{6pt}
	\noindent
	\textbullet~\textit{#1}
	\vspace{6pt}
}

\usepackage{nomencl}
\makenomenclature

\begin{document}
\author{D. Lairez}
\email{didier.lairez@polytechnique.edu}
\affiliation{Laboratoire des solides irradi\'es, \'Ecole polytechnique, \\ CEA, CNRS, IPP,
91128 Palaiseau, France}
\title{What entropy really is\,:\\ the contribution of information theory}
\date{\today}
	
\begin{abstract}
	Even today, the concept of entropy is perceived by many as quite obscure.
The main difficulty is analyzed as being fundamentally due to the subjectivity and anthropocentrism of the concept that prevent us to have a sufficient distance to embrace it. 
However, it is pointed out that the lack of coherence of certain presentations or certain preconceived ideas do not help. They are of three kinds\,:
1) axiomatic thermodynamics; 2) inconsistent solutions of certain paradoxes; 3) reluctance of physicists to the simplification provided by information theory.
The purpose of this paper is to examine these points in a didactic way by paying attention to the structure of the theory, what are the foundations, how ideas articulate, with a peculiar focus on their consistency and economy.
It is shown how entropy can be introduced in a more consistent and economical manner with the help of information theory, which from the start takes into account its subjective nature, finally allowing a more intuitive understanding.
\end{abstract}

\maketitle

\section*{Introduction}

According to the legend, J.~von Neumann (the father of the extension of statistical entropy to the field of quantum physics) would have said in 1945 to C.~Shannon (the father of information theory)\,: \myquote{no one knows what entropy really is,...}\,\footnote{This sentence is quoted in 
	\href{https://www.jstor.org/stable/10.2307/24923125}{M. Tribus and E. C. McIrvine
		\textit{Energy and information}, In: \textit{Scientific American}, 225 (1971),
		pp. 179-190}.
	About the fact that it is a legend read
	\href{https://ethw.org/Oral-History:Claude_E._Shannon}{Claude E. Shannon, an oral history conducted in 1982 by Robert Price. IEEE History Center, Piscataway, NJ, USA.}}. 
If so, it would be quite annoying for a notion that is at the heart of thermodynamics and statistical physics, two pillars of physics, but also a key concept to understand \myquote{what is life}\,\cite{Schrodinger_1944}.
Fortunately, the end of the sentence is much more optimistic\,:
\myquote{... so in a debate you will always
	have the advantage}, which (unless it was ironic) suggests
that von Neumann was confident that Shannon had the answer and that the information theory would certainly clarify the situation. However, seventy-seven years later, it is not certain that the haze has completely dissipated.

Everyone knows that entropy has something to do with energy dissipation. 
Also, it is often associated with the arrow of time,
rightly or wrongly (e.g. \cite{Prigogine_1998}, for a contradictory point of view see~\cite{Ben-Naim_2016}). 
However, while energy, time, and entropy are equally fundamentally puzzling, they clearly do not have the same status on the pragmatic level of everyday life for scientists.
If a child asks a scientist what energy or time is, he will probably get a quick answer, even if it is simplistic.
For entropy the answer is likely to be delayed and more confused.

What does it mean to understand\,?
\myquote{The world is comprehensible because the body has long been exposed (from the beginning) to its regularities} (P. Bourdieu\,\cite{Bourdieu_1997}).
A remarkable example is that of gravity, which is the most puzzling fundamental interaction, but also perfectly understood by children from an early age. The effectiveness of this comprehension is so great that they can excellently predict the direction taken by a falling object.
\myquote{Gravitation no longer disturbs anybody\,: it has become a common unintelligibility} (E. Mach\,\cite{Mach_1911}).
Getting used to something is part of an \textquote{intuitive understanding}, the importance of which should not be overlooked. 
In science, we also need a similar level of understanding,
which is necessary for the emergence of new concepts. \myquote{It is by logic that we prove, it is by intuition that we invent}
(H.~Poincar\'e\,\cite{Poincare_1920}).

Energy and time are understood in this intuitive manner.
Concepts like equilibrium and irreversibility, which are at the heart of the one of entropy, are also intuitive. 
Why does not this hold for entropy\,?

To understand is also to establish connections, to link and unify things to make a whole\,\cite{Deniau2008}. So that
the first difficulty for entropy is that the concept is protean\,\cite{Balian2004}. 
Entropy belongs to thermodynamics, statistical mechanics and information theory.
However, this difficulty to make things coherent is reinforced by a special feature of entropy.
Scientists are comfortable when the observer (the subject) marginally affects the object under study.
But this is not the case with entropy.
Entropy cannot be measured without being transformed entirely into something else.
So that the subject annihilates the object.
This subject-object relationship cannot be separated from the concept of entropy.
Hence, the difficulty to embrace it, we do not have a sufficient distance, we participate in the object of our study.

Entropy is a subjective concept, it is anthropocentric from the origin in thermodynamics with the ideas of energy grades and useful work. Thermodynamics is a theory of phenomena, perceived by our senses at a macroscopic scale, that is only defined as being the human scale.
Seventy years before Shannon, J.C.~Maxwell wrote\,:
\myquote{The idea of dissipation of energy depends on the extent of our knowledge}\,\cite{Maxwell_1878}.
We must recognize, however, that Maxwell was a pioneer and far ahead of his time. In most cases, this subjective point of view was much less explicit. 
This is how it gradually gave way to a presentation of thermodynamics that moved away from phenomenology in favor of an axiomatic approach.
By axiomatic, I mean a presentation of thermodynamics in the line of that of H. Callen\,\cite{Callen_1985} that starts by defining entropy from the mathematical properties it is supposed to have. For instance, extensivity, concavity, temperature defined as a partial derivative, etc.
Beyond the fact that this is probably not the most didactic and seems to forget that \myquote{all knowledge about reality begins with experience and terminates in it} (A. Einstein\,\cite{Einstein_1934}), this approach poses many problems of consistency which participate to the trouble.
But above all, it misses the major aspect linked to the subjective side of entropy.

Then comes statistical mechanics that aspires to derive everything with a bottom-up approach, from the microscopic to the macroscopic scale,
from the Newton's mechanics of collisions and their statistics.
What could be farther from subjectivity than atoms and molecules, than the mathematics of statistics\,?
The subjectivity is simply hidden behind the fact the statistical frequencies in question (mathematical \textit{a posteriori} expectations) are in most cases probabilities (human \textit{a priori} expectations). This ambiguity also participates to the trouble.
In particular it gives rise to certain paradoxes related to the irreversibility but also to the mixing of gases.
The latter known as Gibbs paradoxes are partially solved in the large majority of textbooks in terms of quantum physics by invoking particles indistinguishability.
It is true that we find the subject-object relationship in quantum mechanics, at the scale of particles. Schr\"odinger's cat is both dead and alive until we measure its state. Particles have no identity until they reach a detector.
But with entropy it is different. It is an emergent concept introduced to account for macroscopic observations. Also, statistical mechanics has been founded from the start in a classical framework, so the intrusion of quantum physics poses a problem of consistency and leads equally to confusion. 

In fact, the notion of information also displays a similar subject-object relation\,:
you cannot get the same information twice (because the second time it is no longer an information).
But it has the advantage of being more flexible and of allowing more consistent reasoning about the scale or the resolution at which phenomena are observed.
Information theory, born in 1948 with C. Shannon\,\cite{Shannon_1948}, permits after E.T. Jaynes\,\cite{Jaynes_1957} a great simplification of foundations of statistical mechanics, a major gain of consistency and provides the most economical solution of all current paradoxes.
However, we only have to note that in most cases it is either ignored, or presented as anecdotic, or as an interesting analogy.
It is not really exploited in most textbook (a noticeable exception is the book of A. Ben-Naim and D. Casadei\,\cite{Ben-Naim_2016}), and no more in \textquote{recent} (given the age of the problem) literature about Gibbs paradoxes\,\cite{Gibbsparadox2018}. Physicists seem embarrassed with information theory.

In my opinion this situation has two origins. The first, that was outlined by Jaynes\,\cite{Jaynes1992}, is the misunderstanding of the word \textquote{subjective}. It is not synonymous of lack of rigor or irrationality.
It is as far from that as the use of confidence intervals for measurements can be.
The second is likely that people just stop at the analogy between the formulas for statistical entropy of Boltzmann and Gibbs and that of Shannon, but do not realize the great advantage provided by the maximum-entropy principle, which is basically nothing but a rational criterion for extracting the maximum amount of information, but no more, from our knowledge. 
A criterion with which scientists accustomed to analyzing their data should feel quite comfortable.

The purpose of this essentially didactic paper is to attempt to shed light on these different aspects. The road map of the article is the following\,:
\begin{enumerate}
	\item The first part is a presentation of what we are speaking about. That is to say the phenomenology of thermodynamics. It is shown how introducing a state-quantity, named Clausius entropy, is needed to account for phenomena.
	The main idea that will be  finally exposed is that the state of a system is only a representation of its being (the tip of the iceberg) and is inherently subjective. It will be shown how this subjectivity, which is difficult to apprehend, leads to a famous paradox, namely Gibbs' paradox \#1.
	\item The second part presents statistical entropy as it was initially by J.W. Gibbs by identification of statistical quantities with thermodynamics quantities.
	The aim is not to detail some technical features, but to dissect the structure of the theory, what are the foundations, how ideas articulate with a peculiar accent on their consistency. The bottom-up approach of statistical mechanics (I mean the explanation of the macroscopic from the microscopic scale), leads us to forgot the subjectivity part that cannot disappear because of the link with thermodynamics. The main goal is to show how this subjectivity was moved to the root, at a cost of consistency and a paradoxical conflict with thermodynamics, namely the Gibbs paradox \#2.
	\item The third part is concerned with the Shannon entropy and the maximum-entropy principle. The goal is to show how together with thermodynamics and statistical mechanics, the three form the most economical and coherent framework. In brief, the economy comes from that, with Shannon, the famous formula $S=-\sum p_i\ln p_i$ becomes totally free from thermodynamics. Also, with information theory the subjectivity is accepted from the start providing a great gain of consistency and the solution of both Gibbs paradoxes.
	\item The last part is a discussion of some very popular alternative presentations of entropy with a particular focus on their compared consistency and on the economy of thought they permit. Two main points will be exposed\,: 1) In \textquote{pure} thermodynamics (i.e. without probabilities), the idea of entropy as being maximum at equilibrium obliges it to be extensive and concave at the cost of a conflict with statistical mechanics; 2)~All solutions of Gibbs paradoxes, alternative to the one from information theory, use in one way or another the Stirling approximation
	$\ln N! - N\ln N=0$, that is wrong in the thermodynamic limit of very large $N$. Information theory directly gives the correct result.
\end{enumerate}

Today, it is good form to highlight the novelty of an article. With such an old subject, which has caused so much ink to flow, the novelty is hard to reach. Everything on this subject is probably written here or elsewhere, so maybe the main novelty of this paper lies in the all-in-one it offers. The reader may also find novelty in the way Gibbs' two paradoxes are resolved and in the discussion of the last section which confronts the present approach to usual presentations of the subject.

\tableofcontents

\section{Thermodynamic entropy}\label{thermo}

The concept of entropy was first introduced in thermodynamics\,\cite{Clausius_1879}, a theory built in the first half of the 19th century to account for energy transformations.
Thermodynamics is a typical case of phenomenology in science. That is to say an approach that starts from experiments and particular observations, then proceeds by induction and infers general laws, often improperly called \textquote{principles}, improperly because these principles are not foundations. These laws introduce some concepts such as internal energy and entropy and intend to provide a \textquote{condensate} of possible phenomena allowing an economy of thought\,\cite{Mach_1911}. 
\myquote{From these, by pure logical reasoning, a large number of new physical and chemical laws are deduced, which are capable of extensive application, and have hitherto stood the test without exception.} (M. Planck\,\cite{Planck_1903}).
M. Planck, who widely contributed to initiate statistical mechanics, is also himself the ardent defender of this phenomenological thermodynamics.
The last point to understand the spirit of thermodynamics is that at this time the atomist theory was still far from being generally accepted\,\cite{Brock_Knight1965a}. 
So that thermodynamics deliberately keeps a distance from assumptions about the \textquote{ultimate} nature of things but  starts direct from a few very general empirical facts\,\cite{Planck_1903}.

\subsection{Beginnings}

Initially, thermodynamics dealt with gas, pressure, pumps, motors and with work and heat.
Temperature is omnipresent and its definition is simple\,: temperature is what thermometers measure\,\cite{Maxwell_1872}. Heat is treated as a fluid, the caloric, that can flow and spread.
An important concept is that of thermal equilibrium of a body which is understood as a steady state, but also a state with a uniform temperature, just like the level of a fluid in communicating vessels.
A body that is not in equilibrium tends to be.
The key observation is that it is from the reestablishment of equilibrium that work can be produced.
So some statements come naturally\,\cite{Carnot_1872}:
\begin{enumerate}[label={\arabic*)}, nosep]
	\item Temperatures of two bodies in contact tend to equalize (tendency to reach equilibrium).
	\item Heat cannot  spontaneously pass from a cold body to a hot body (otherwise it would be contradictory to 1).
	\item Producing work needs at least two reservoirs at different temperatures (by forcing the system to be out-of-equilibrium).
	\item Work can be completely converted into heat, but not heat into work. Some dissipation occurs.
	\item Consequence of 4\,: perpetual motion is impossible.
\end{enumerate}
Formalizing these statements mathematically is the challenge of thermodynamics which, to do this, introduced the notion of entropy.

\subsection{Zeroth law and equilibrium}\label{equilibrium0}

Thermodynamics is concerned with \textquote{systems} and \textquote{processes}. A system is just a set of  things, the simplest being a given volume of an homogeneous substance, for instance a simple gas or liquid. More complex systems are composite and made of multiple simple systems, for instance a liquid with its vapor, or a container with two compartments etc\,\footnote{This definition of what is \textquote{simple} and \textquote{complex} is a lighter version of the one given by Callen\,\cite{Callen_1985} p.9 the later being formulated in a more rigorous way.}. In this article, the examples provided will be limited to simple systems, but of course thermodynamics does not.

Thermodynamics starts by defining what it is talking about, that is to say the nature of equilibrium. 
A simple system is at equilibrium when it can be entirely characterized with some physical quantities\,: pressure, temperature, volume, amount of matter... called state-quantities, which take certain values that do not change over time. Thus equilibrium is a steady state.
But the definition of equilibrium needs a supplementary  preliminary statement that is now often called \textquote{zeroth law} of thermodynamics, as if something had been forgotten in its foundations. This is indicative of the shift from the historical inductive phenomenology to the modern axiomatic approach.
Usually, the zeroth law is concerned with thermal equilibrium and stated like this\,: \myquote{if A is in thermal equilibrium with B, and B is in thermal equilibrium with C, then C will be in thermal equilibrium with A}\,\cite{Atkins_2010}. This transitive relation applies in particular to a thermometer, allowing to measure and define temperature. This leads finally to the statement\,:\myquote{if the temperatures of two systems are the same, then they will be in thermal equilibrium}\,\cite{Atkins_2010}.
So that if A, B and C are subparts of a simple system, the thermal equilibrium of the system is the state where the temperature is uniform.
If we deal with \textquote{equilibrium}, including thermal and mechanical equilibrium, then the uniform distribution criterion for a simple system must also apply to other state-quantities such as pressure and density. These quantities make it possible to define a homogeneous state as being that where the values taken by these quantities are the same for the whole and for its subparts. These state-quantities (temperature, pressure, density) are said to be intensive. 
Also, an important implicit point in this definition of the equilibrium is that it is supposed to exist. So that in this paper, we propose to reformulate the zeroth law as\,:

\law{Zeroth law}{There exists a state of equilibrium such as
	\begin{enumerate}[label={\alph*)}]
		\item The equilibrium is stable.
		\item At the equilibrium, intensive state quantities of a simple system are spatially uniform.
	\end{enumerate}
}
\noindent Note that in a phenomenological approach, the existence of the equilibrium means that it can be observed at one time or another. So that implicitly, the equilibrium is accessible.

A composite system is at equilibrium if its simple subsystems are themselves at equilibrium.
A~process is an action or an event that modifies the state of a system.

The usual presentation of the zeroth law as the expression of a transitive property of thermal equilibrium, is often viewed as a definition of temperature. But it can be viewed also as a definition of equilibrium. Here, this later alternative is preferred for two reasons. Firstly, in the spirit of phenomenological thermodynamics, it is not necessary to define the temperature as being different from the quantity measured by the thermometers. Secondly, the notion of equilibrium is central in thermodynamics, it will also be the starting point of statistical mechanics (\S\ref{statmech}) and is also central in the contribution of information theory (\S\ref{Shannon}). Actually, classical thermodynamics is only concerned with equilibrium. In his famous book Callen writes\,: \myquote{a system is in an equilibrium state if its properties are consistently described by thermodynamic theory\,!}\cite{Callen_1985} p.15. To which it immediately follows that\,: a system out-of-equilibrium is not described by thermodynamics.
This is a paradox of thermodynamics. On one side, thermodynamics deals with state variables, which by definition define a state (so they are static and are not expressed as a function of the time variable), but on the other side deals with motors and motion (so that intends to approach dynamics).
When a system is out-of-equilibrium by definition it undergoes a process between an initial equilibrium and a final equilibrium. Classical thermodynamics describes these two extremities, but not the intermediate stage. When it is possible, this issue is solved by the quasi-static approximation (see the next section \S\ref{rev_irrev}), i.e. by approximating the process with a succession of equilibrium states\,\cite{Planck_1903}, and the time variable is introduced in the manner of A.~Einstein\,\cite{Einstein_1906}. This open the door to a new thermodynamics of out-of-equilibrium systems that includes the physics of transport of matter and energy. This field is out of the scope of the present paper.

\subsection{Reversible versus irreversible}\label{rev_irrev}

\begin{figure}[!hbtp]
	\begin{center}
		\includegraphics[width=1\linewidth]{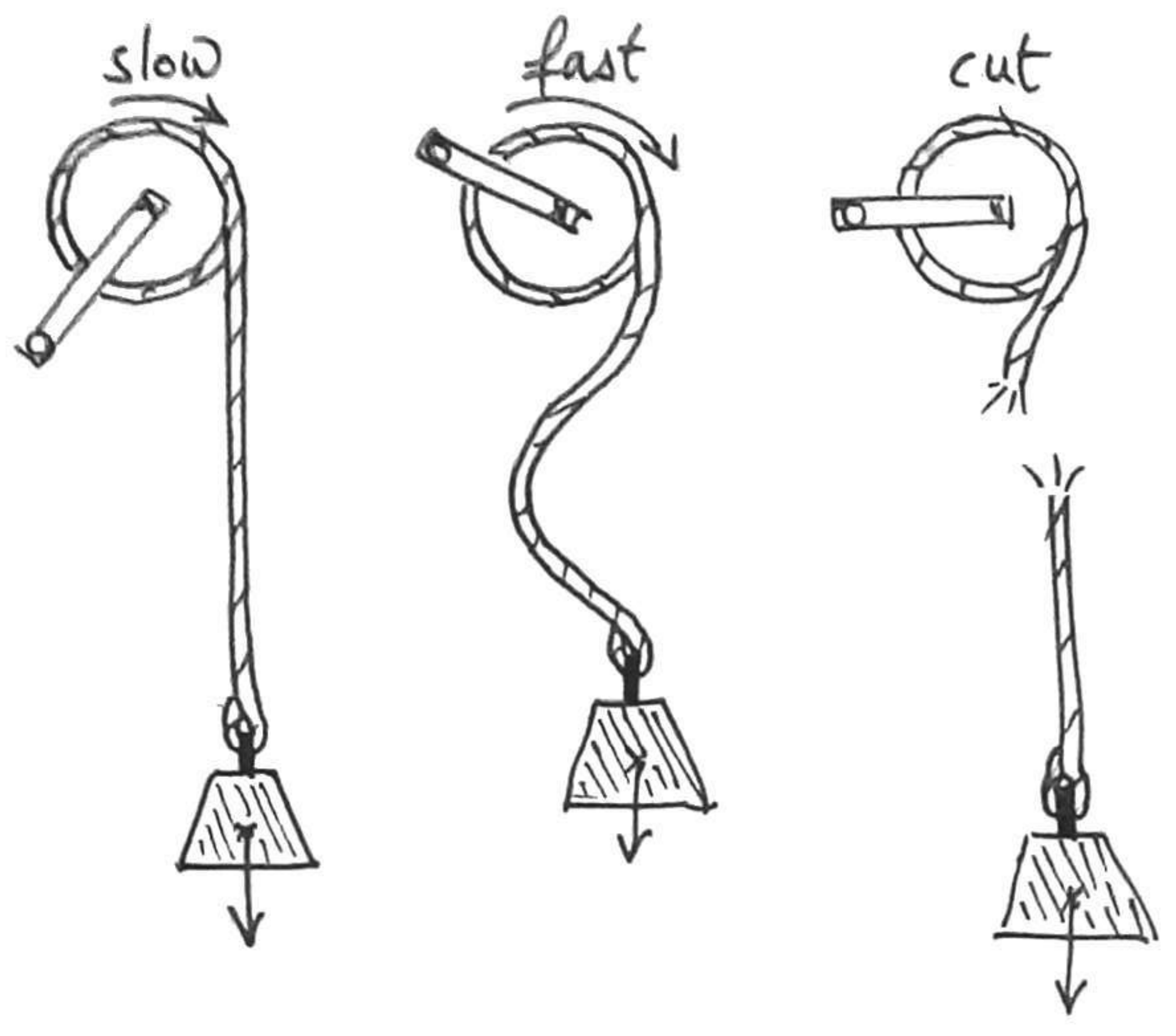}
		\caption{Mechanical analogy for reversible (left), irreversible quasistatic (middle) and irreversible non-quasistatic (right) processes.}\label{analogy_mech}
	\end{center}
\end{figure}

What emerges from experiments is that an isolated system, that is to say a subsystem in its environment taken as a whole, tends towards equilibrium in an irreversible manner.
In the sense that if it is filmed, assuming that all the components of the system are visible and their temperature too (using an infrared camera), the film played backwards is implausible.
This irreversibility is at the heart of thermodynamics so that some vocabulary points must first be clarified.
Imagine a weight at the end of a string winded up around a pulley (see Figure~\ref{analogy_mech}) and film it.
\begin{enumerate}[label={\arabic*)}]
	\item Unwind slowly the string.
	The film is made of a succession of images on which the cause (the string unwinds a little) and the effect (the weight goes down a little) are always in phase, it is made of a succession of quasi-equilibrium as the weight and the reaction of the string almost equilibrate each other every time. \textquote{Quasi-equilibrium} and  \textquote{almost equilibrate} because the weight goes down anyway.
	The process is reversible. 
	\item Unwind the string faster than the weight can go down due to its inertia. The delay makes the string becoming slack and not fully extended. Weight and string-tension do not equilibrate.
	The process is irreversible. 
	
	However, even if you stop and restart unwinding, the elasticity of the string allows its tension to always have a derivative.
	As in the first case, if we only look at the string behavior, it appears always as a succession of "quasi-equilibrium". Both cases are said quasistatic.
	
	\item Cut the string at a given point in the descent. The forces applied to the weight suddenly change. The tension of the string does not have a derivative. The process is no longer quasistatic. It is also irreversible, but this is not due to a delay, since cause and effect are concomitant, it is rather due to this non-quasistatic feature.
\end{enumerate}

Thermodynamic processes obey to the same classification.
In our mechanical analogy, the subsystem can be thought of as the string, the surroundings the rest. So that, a reversible process is a succession of quasi-equilibrium of the whole, and a quasistatic process a succession of quasi-equilibrium of the subsystem. 

The non-differentiability provides a mathematical criterion of the irreversibility of the latter kind.
But formalizing the difference between the first two cases and identifying a unique criterion for the irreversibility of the last two is more difficult.
This is precisely the role of the notion of entropy.

\subsection{First law}

The theory of caloric was abandoned (but the metaphor of heat spreading should be kept in mind) mainly because it was observed that heat is not a property of a system that is conserved, but rather a quantity related to a given process that the system undergoes.
It has been observed that the heat $Q$ and the work $W$ that a system receives from its environment are two forms of the same physical quantity\,\cite{Joule_1850} which both contribute to increasing what is called the internal energy $U$ of the system, which in turn is conserved and can be measured later by an inverse transformation.
For a system undergoing a quasistatic process one can write\,:
\begin{equation}\label{conservation}
	\diff U = \diff Q + \diff W
\end{equation}
\nomenclature{$U$}{internal energy}%
\nomenclature{$Q$}{heat}%
\nomenclature{$W$}{work}%
$U$ is a characteristic of the state of the system, it is a state-quantity. $Q$ and $W$ are not, they are exchanged during the process and both converted into internal energy. 
$Q$ and $W$ depend on the process,  whereas $U$ depends only on the result (different processes can have the same result). Imagine a cyclic process which after many exchanges of heat and work leaves the system to its original state. The variation of internal energy over the cycle is zero. One writes:
\begin{equation}\label{statefun}
	\oint \diff U= 0
\end{equation}
$Q$ and $W$ not being state-quantities but path-quantities, Eq.\ref{conservation} does not mean that $U$ is a function of $Q$ and $W$. The actual variables for $U$ remain to be determined.

An important point is that $U$, as the quantities $Q$ and $W$ received by the system, are additive.
For a given system, quantities are additive when their values are the sum of those of its subparts.
Volume, amount of matter and internal energy are additive. 
This leads to the statement of the first law of thermodynamics\,:

\law{First law}{There exists an additive state-quantity, named internal energy, which variations is the sum of heat and work exchanged by the system.}

This first law is now understood as a special case of the law of conservation of energy, that is more general in physics.

\subsection{Perfect gas}

The favorite system of thermodynamicists is a given amount of an \textquote{ideal} gas.
Many relationships between pressure $P$, volume $V$ and temperature $T$ were known since the 17th century and will later lead to the ideal gas law\,:
\begin{equation}\label{ideal}
	PV = N T
\end{equation}
\nomenclature{$P$}{pressure}%
\nomenclature{$V$}{volume}%
\nomenclature{$N$}{amount of matter, number of particles}%
\nomenclature{$T$}{temperature in unit of energy}%
with $N$ the amount of matter expressed in number of molecules
and $T$ the temperature in Joule obtained by multiplying the absolute temperature by the Boltzmann constant $k$\,\footnote{Actually, placed here in the text, Eq.\ref{ideal} contains two anachronisms: 1)~in a non-atomistic world $N$ should be rather expressed in mole and $k$ replaced by the perfect gas constant $R$; 2)~it is precisely the work of Clausius and the introduction of the notion of entropy that lead to definitely adopt the absolute temperature scale (\cite{Maxwell_1872} p.155 and following). But for reasons of didactics and of coherence with the following, the modern expression of the ideal gas law is preferable. Incorporating the Boltzmann constant $k$ into temperature $T$ provides the double advantage of being more concise and underlying the physical meaning of temperature that should be more conveniently called thermal energy. This will permit also an equality between statistical and Shannon entropies without this dimensional prefactor.}.
An \textquote{ideal gas} is a gas that obeys to the \textquote{ideal gas law}.
In this paper for the sake of simplicity, we will consider a  \textquote{perfect gas} that is usually understood as being a gas with no interaction (an ideal gas too, but here it is explicit). No interaction means in particular no hydrodynamic interaction, no friction, no viscosity and thus no time delay 
between two equilibrium states.
A perfect gas reaches equilibrium instantaneously, or in other words, a perfect gas is always at equilibrium.
A perfect gas does not exists. It is an exercise of thought that allows us to attribute any delay responsible for irreversibility (see \S\ref{rev_irrev}) to the container that plays the role of a \textquote{black box} containing all the features I do not want to discuss in this paper.

It was found that
the amount of heat $\diff Q$ needed to increase the temperature of a given volume of gas by $ \diff T$ is proportional to its amount of matter\,: $\diff Q=c_v N \diff T$, where $c_v>0$ is a constant for a given gas species and is named isochoric specific heat capacity. By receiving $\diff Q$, the gas increases its internal energy by $\diff U$. By integration and assuming a zero integration constant for simplicity, we have\,:
\begin{equation}\label{ideal2}
	U = c_v N T
\end{equation}
\nomenclature{$c_v$}{isochoric specific heat capacity}%
This is after Eq.\ref{ideal} the second fundamental equation for the perfect gas.

Initially, thermodynamics considered only closed systems with a given amount of matter. The work in Eq.\ref{conservation} was only mechanical and the product of a force to its displacement, thus for a gas is only linked to variations of volume. This was extended by J.W.~Gibbs\,\cite{Gibbs1874} to enlarge the scope of thermodynamics to processes involving composite systems that exchange matter.
Thus the work traditionally encompasses these two contributions and writes for a quasistatic process\,: 
\begin{equation}\label{work}
	\diff W = -P\diff V + \mu \diff N
\end{equation}
\nomenclature{$\mu$}{chemical potential}%
where $\mu$ is an intensive quantity named chemical potential. The negative sign in front of $P$ means that by increasing its volume ($\diff V>0$), the gas produces work  ($\diff W<0$). This energy lost tends to decrease the internal energy (Eq.\ref{conservation}), so that it must be heated ($\diff Q>0$) for $U$ to be kept constant. Variation of the amount of matter can be understood in the same manner.
At constant volume ($\diff V=0$), by increasing the amount of matter ($\diff N>0$), the system needs more heat ($\diff Q>0$) to maintain $U$ constant, thus  $\mu$ must be negative.

At this stage, from Eq.\ref{conservation} and \ref{work} one can write
\begin{equation}
	\diff U = \diff Q -P\diff V + \mu \diff N
\end{equation}
The work $W$ has been replaced by two additive state-variables, $Q$ not yet.

\subsection{Isotherms and adiabats}\label{rever}

\begin{figure}[!hbtp]
	\begin{center}
		\includegraphics[width=1\linewidth]{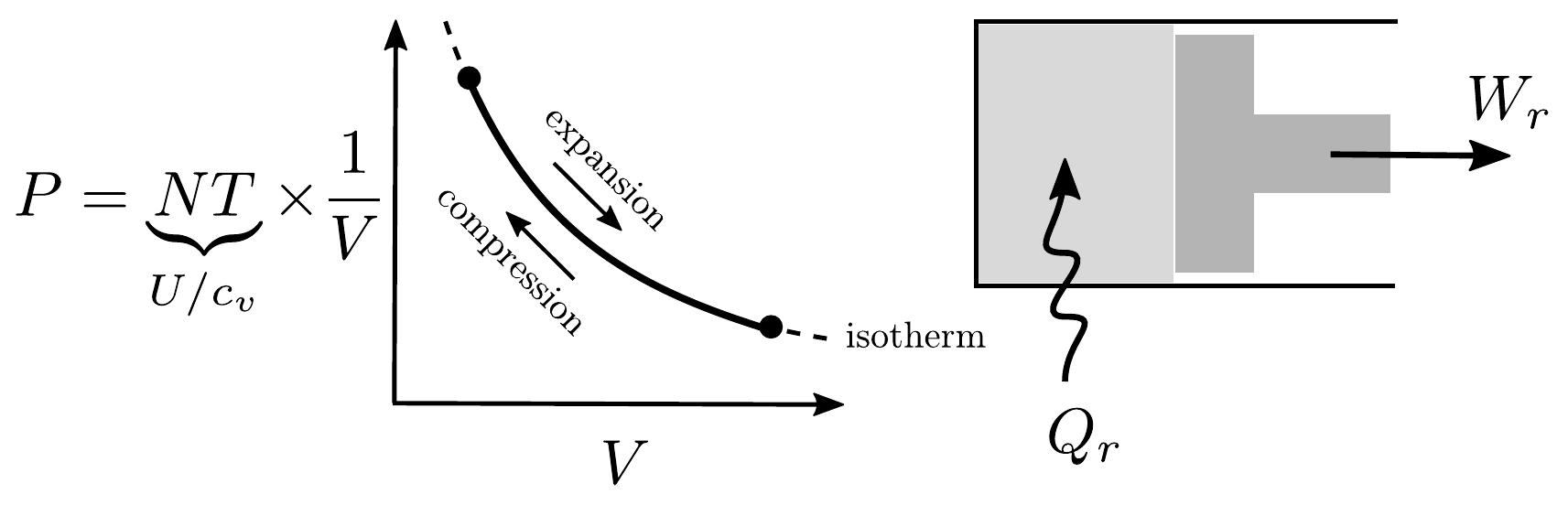}
		\caption{Isothermal expansion-compression of a perfect gas in an ideal diathermal container. The process is reversible.}\label{isoth1}
	\end{center}
\end{figure}

\begin{figure}[!htbp]
	\begin{center}
		\includegraphics[width=1\linewidth]{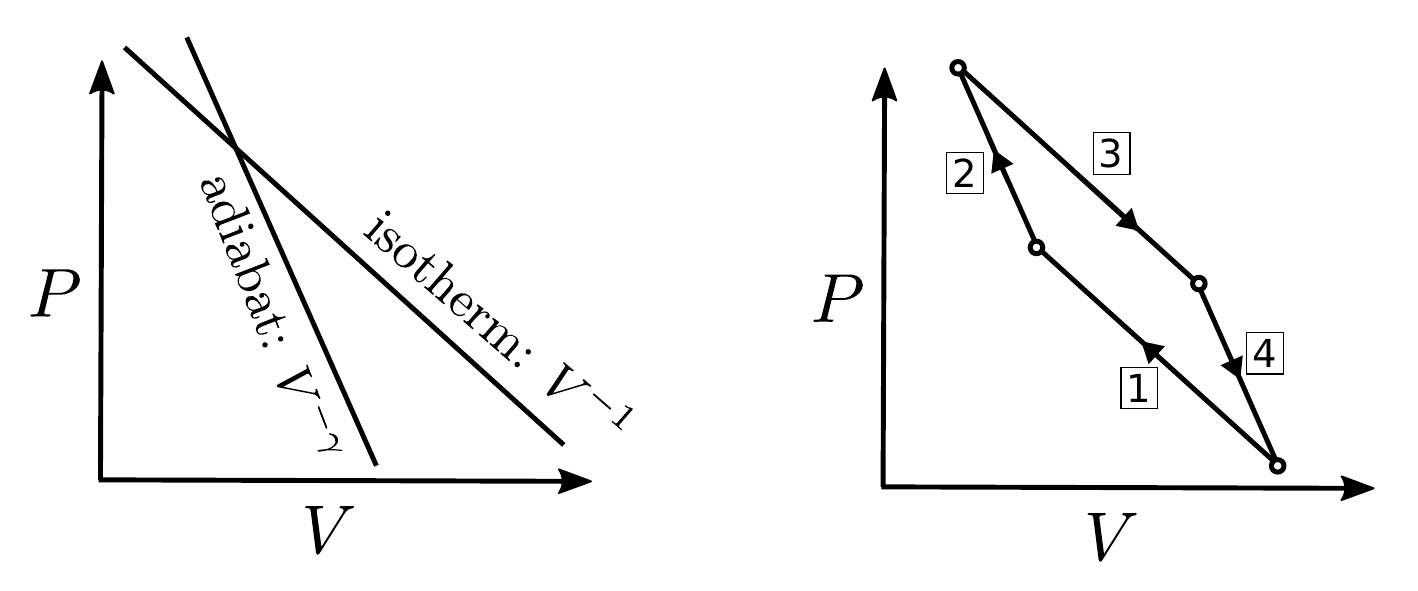}
		\caption{Pressure-volume diagram (in log-log scale)\,: Left - isotherm and adiabat;
			Right - clockwise Carnot cycle (power cycle) that produces work (the area of the loop) and consumes heat.}\label{pv}
	\end{center}
\end{figure}

For a given amount of perfect gas, processes can be visualized in a pressure-volume diagram (Figure~\ref{isoth1} and \ref{pv}), where isotherms are lines obeying to Eq.\ref{ideal}. To force processes to stay on such a line, the gas is put in an ideal diathermal container, i.e. made in a material allowing instantaneous heat exchanges. So that, if the gas is already at equilibrium itself, it is also in thermal equilibrium with the surroundings. If the latter is sufficiently large to be considered at constant temperature, the process is isothermal and reversible.

An isothermal expansion from $V_1$ to $V_2$ can be achieved by pulling a piston. The work is $W_r= -\int_{V_1}^{V_2}P\diff V=-NT\int_{V_1}^{V_2}\frac{\diff V}{V}=-NT(\ln{V_2}-\ln{V_1})$, where the second equality is obtained by using Eq.\ref{ideal}. Here the subscript \textquote{$r$} stands for \textquote{reversible}.
Since temperature is constant $\Delta U=Q_r+W_r=0$, so that the energy balance is:
\begin{equation}\label{isoexp}
	\textrm{isotherm }\left\{{\begin{array}{l}
			Q_r= NT \Delta(\ln V)\\  
			W_r=- NT \Delta(\ln V)\\  
			\Delta U=0
	\end{array}}\right.
\end{equation}
\nomenclature{$Q_r$}{heat for a reversible process}%
\nomenclature{$W_r$}{work for a reversible process}%
Which means that the gas produces work ($-W_r>0$ is the area under the curve $P$ versus $V$) exactly compensated by a gain of heat pumped from the surroundings. A reversible compression from $V_2$ to $V_1$ do exactly the reverse. So that net heat and net work are both zero for a complete cycle (see Figure~\ref{isoth1}).

An adiabatic container prevents all heat exchanges between its contents and the surroundings.
The latter can thus be ignored. 
In adiabatic container, if a perfect gas is itself in equilibrium, the process it undergoes is thus reversible.
Compared to isotherm, an adiabatic expansion with a piston does not permit the gas to pump heat from the surroundings\,: $Q_r=0$ and $\Delta U = W_r$. From Eq.\ref{ideal2}, $\Delta U=c_vN \Delta T$,
so that the energy balance is\,:
\begin{equation}
	\textrm{adiabat }	\left\{{\begin{array}{l}
			Q_r=0\\ 
			W_r=c_vN\Delta T\\
			\Delta U=c_vN\Delta T
	\end{array}}\right.
\end{equation}
The gas produces work to the detriment of its internal energy. Conversely, an adiabatic compression increases the internal energy by the amount of work received. Again, net heat and net work are both zero for a complete cycle.

An important fact is that adiabats are more inclined than isotherms. As $\diff U=c_vN\diff T=-P\diff V$, one has $c_v \diff T/T = -\diff V/V$. By integration one gets $T^{c_v}\propto V^{-1}$, so that using again Eq.\ref{ideal2}, instead of $P\propto 1/V$ for isotherms, one has for adiabats\,:
\begin{equation}
	P\propto 1/V^\gamma
\end{equation}
with
\begin{equation}\label{CV1}
	\gamma=1+\frac{1}{c_v} 
\end{equation}
A consequence is that any two points of the pressure-volume diagram can be connected by a path made only of isotherms and adiabats. 
By generalizing, it is possible to approach any path as closely as desired by a series of infinitely small segments of isotherms and adiabats.

\subsection{Cycles}

Transformations of energy must be based on a cycle in order to be repeated.
However as seen just before, a non-zero net transformation 
requires different outward and return paths and thus at least two different temperatures.

What is the most direct way to increase temperature by $\Delta T$\,? 
To illustrate the ideas, imagine we intend to run a cycle clockwise (see Figure~\,\ref{pv}) and we have to increase temperature from isotherm (1) to (3). 
This can be done, for instance, by using the adiabat~2. 
Another option to reach the same point of the diagram should be: 1)~compress the gas a little more; 2)~then heat the gas at constant volume. But during the first step, the gas has been cooled in order to stay on the isotherm. So that this way is less direct in raising the temperature.
Actually, the most direct cycle, known as Carnot cycle, is composed of four reversible stages\,: an isothermal compression (1), followed by an adiabatic compression (2) and the same for the expansion (3 and 4) to close the loop (Figure~\ref{pv}). 
The energy balance of a Carnot cycle is: $W_{1}=-Q_{1}$; $W_{2}=c_vN\Delta T$ and $Q_{2}=0$; $W_{3}=-Q_{3}$; $W_{4}=c_vN\Delta T$ and $Q_{4}=0$. 
So that the net work is $W=-(Q_1+Q_3)$. 
For a motor, the efficiency $\epsilon$ is the ratio of the net work delivered, $-W$, to the heat $Q_3$ supplied by the hot reservoir that is usually the energy provided by the fuel: $\epsilon=-W/Q_3=1 + Q_1/Q_3$.
The key observation of R.~Clausius\,\cite{Clausius_1865} is that ${Q_1}/{Q_3} = -T_1/T_3$ or
\begin{equation}\label{Clausius1}
	\frac{Q_1}{T_1} + \frac{Q_3}{T_3}=0
\end{equation}
By generalizing to any cycle approached by a series of infinitely small segments of adiabats and isotherms, 
we can write:
\begin{equation}\label{Clausius2}
	\oint \frac{\diff Q_r}{T} = 0
\end{equation}
where the subscript \textquote{$r$} stands for \textquote{reversible} and is used to keep in mind that this equation is only valid for reversible processes. 

\subsection{Clausius entropy}

Equation \ref{Clausius2} simply allows us to turn the path-quantity $Q_r$ into a state-quantity $S$ called entropy and defined by the differential:
\begin{equation}\label{diffS}
	\diff S = \frac{\diff Q_r}{T}
\end{equation}
\nomenclature{$S$}{entropy}%
This is the definition of the Clausius entropy\,\cite{Clausius_1865}.
Directly due to this definition and the additivity of $Q_r$ itself, $S$ is additive over the subparts of the system.
Thus, from Eq.\ref{conservation}, \ref{work} and \ref{diffS}, it is now possible to express the differential $\diff U$ only in term of other additive state-variables:
\begin{equation}\label{diffU2}
	\diff U = T\diff S - P \diff V+\mu \diff N
\end{equation}
The entropy $S$ is therefore the third state-variable on which the internal energy depends and that we were looking for.
Once all variables elucidated, $U$ is simply the value taken by a multi-variable mathematical function of the additive variables $(S, V, N)$. The intensive quantities ($T,-P,\mu)$ are its partial derivatives:
\begin{equation}
	U = \mathcal{U}(S, V, N)
\end{equation}

However, this paper deals with entropy. As $\partial U/\partial S = T$ is always positive, $\mathcal U$ is a strictly increasing function of $S$. Thus, it is possible to take the inverse and express $S$ as a function of $U$. Eq.\ref{diffU2} gives:
\begin{equation}\label{diffS2}
	\diff S = \frac{1}{T}\diff U + \frac{P}{T} \diff V-\frac{\mu}{T} \diff N
\end{equation}
which shows that $S$ is the value taken by a three-variable function of $(U, V, N)$:
\begin{equation}
	S = \mathcal{S}(U, V, N)
\end{equation}
\nomenclature{$\mathcal{S}$}{entropy as a three-variable function}%
which gradient is 
\begin{equation}\label{gradient}
	\nabla\mathcal{S}(U,V,N) 
	=\left(\begin{array}{c} 1/T \\  P/T \\ -\mu/T \end{array}\right) 
	= \left(\begin{array}{c} c_vNU^{-1} \\  NV^{-1}\\ \mathcal S_{N}\end{array}\right) 
\end{equation}
\nomenclature{$\mathcal{S}_N$}{first partial derivative of $\mathcal{S}$ with respect to $N$}%
\nomenclature{$\mathcal{S}_{NN}$}{second partial derivative of $\mathcal{S}$ with respect to $N$}%
the last equality being obtained by using Eq.\ref{ideal} and \ref{ideal2}. The partial derivative $\partial  \mathcal S/\partial N=\mathcal S_{N}$ cannot be calculated because an expression for $-\mu/T$ as a function of $N$ is lacking.
Actually, in classical thermodynamics $N$ is not properly speaking a variable. It was introduced here for reason of consistency with the following. Clausius entropy is only defined for closed systems with constant amount of matter.

\subsection{Irreversibility}\label{irrev}

\begin{figure}[!hbtp]
	\begin{center}
		\includegraphics[width=1\linewidth]{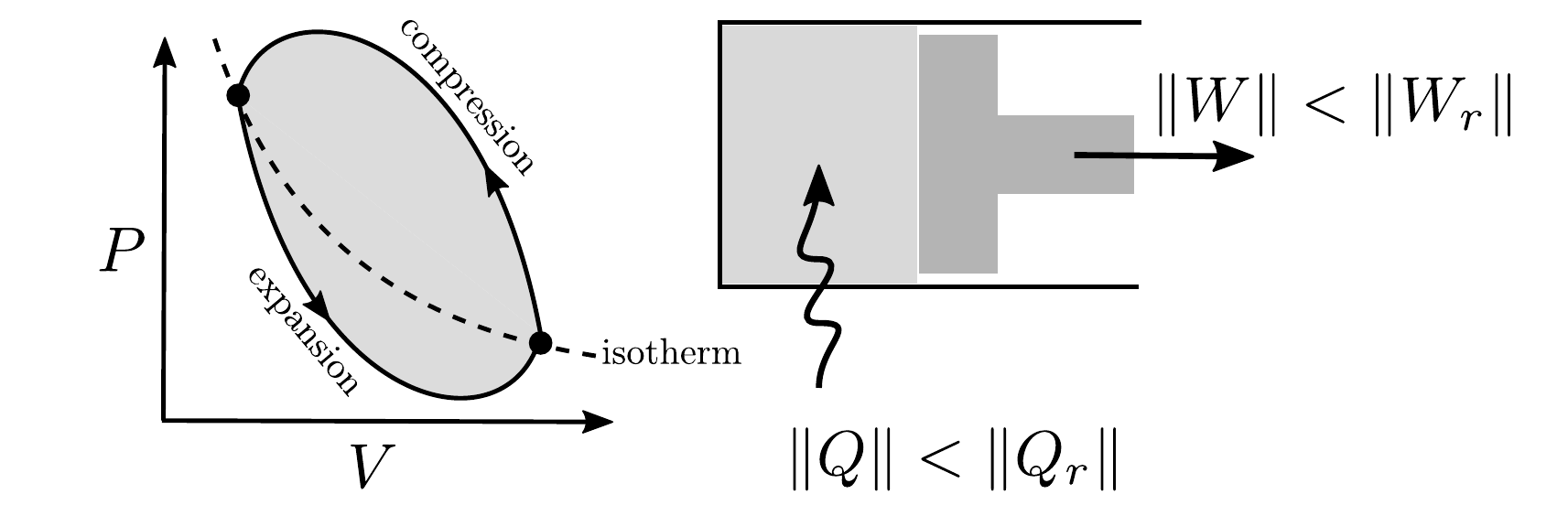}
		\caption{Monotherm irreversible expansion-compression of a perfect gas in a real container: thermalization delay causes a net heat transfer to the surroundings (the area of the loop). To be compared to Figure~\ref{isoth1}}\label{realgas1}
	\end{center}
\end{figure}

\begin{figure}[!htbp]
	\begin{center}
		\includegraphics[width=1\linewidth]{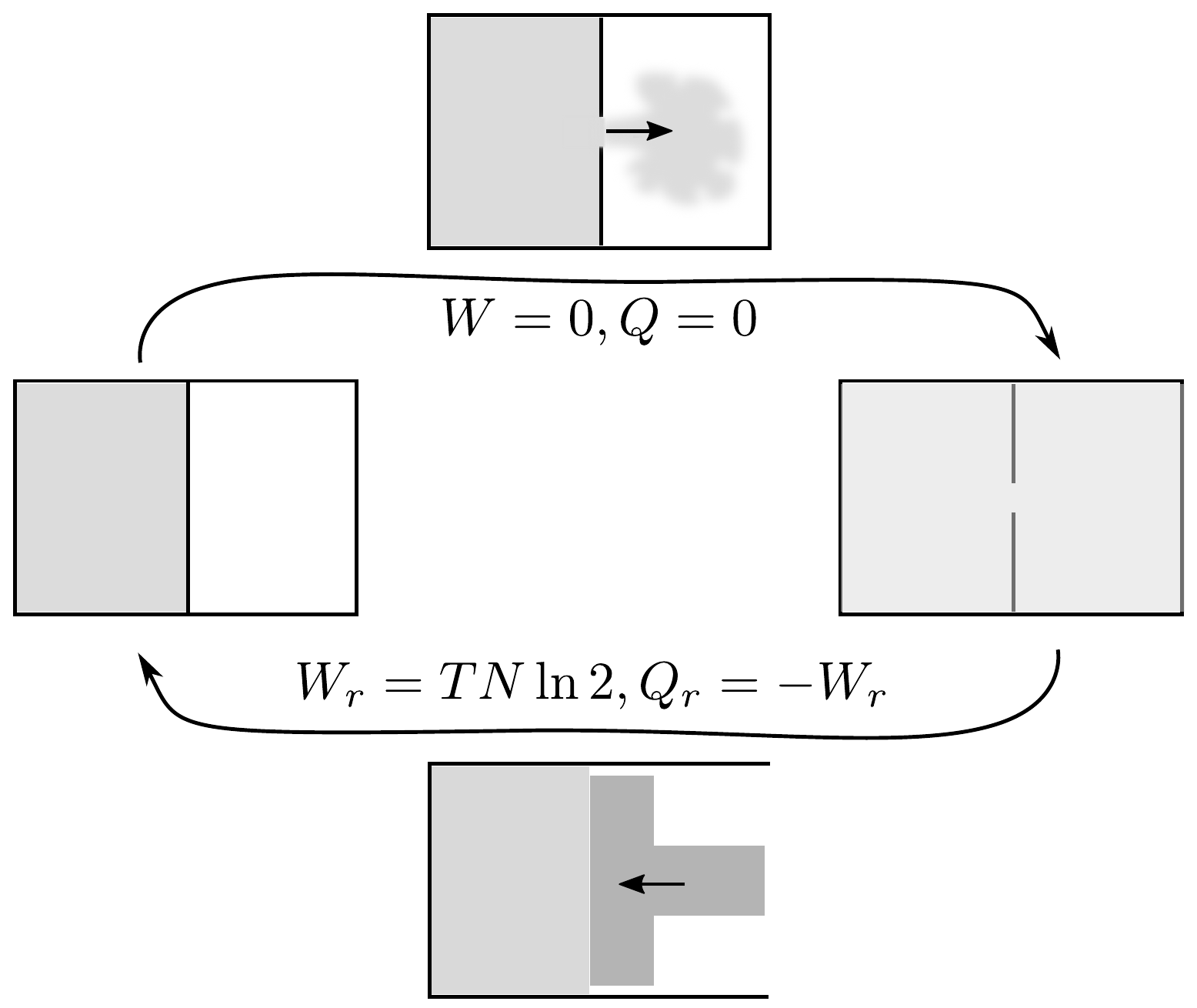}
		\caption{Adiabatic free expansion (left to right, top). The cycle is completed with a reversible isotherm compression (right to left, bottom).}
		\label{freeexp}
	\end{center}
\end{figure}

Let us examine the irreversible counterparts of the two reversible processes of  \S\ref{rever}\,:
a composite system consisting in a temperature reservoir  (the surroundings) that embodies a subsystem made of a perfect gas in a container. 
\begin{enumerate}[label={\arabic*)}]	
	\item Monothermal compression-expansion with a piston (Figure~\ref{freeexp} left)\,: the container is diathermal but not perfect and heat needs time to diffuse through, so that the equilibrium with the surroundings is delayed.
	The process is thus quasistatic but irreversible.
	The cycle deviates from the isotherm and opens in a loop comparable to that of a hysteresis (Figure~\ref{realgas1}). 
	During expansion, the gas takes less heat from the surroundings than if the process were reversible, and during compression it supplies more.
	The cycle spontaneously runs counterclockwise, running clockwise is implausible as we would see heat passing from a cold to a hot body. 
	The complete cycle requires a net work to be provided (the area of the loop), transferred and dispersed as heat to the surroundings. 
	\item Adiabatic free expansion (Figure~\ref{freeexp} right)\,:  a hole suddenly opens between a compartment containing the gas and another under vacuum.
	The gas expands throughout the available volume. 
	During the process the pressure is not defined and not differentiable and cannot be visualized in a pressure-volume diagram The process is not quasistatic and thus irreversible.
	There is neither heat exchange ($Q=0$), nor force applied against a piston ($W=0$). The internal energy is thus constant ($\Delta U=Q+W=0$) and so the temperature (as there is no temperature change, note that there is no need for an adiabatic container).
	But something clearly happened and the return to the initial situation would require an energy expense.
	It can be done in a reversible manner by an isothermal compression with a piston. So that here again, the complete cycle requires a net work that is transferred and dispersed as heat to the surroundings. 
\end{enumerate}


Let us consider only the expansion stages from $V_i$ to $V_f$.
The initial and final states are the same in the above two cases, and also the same as if the expansion were isothermal and reversible. 
Only the three paths are different. Using subscript 1 to refer to the expansion stage, in this three cases one can write\,:
$$
\begin{array}{c}
	\Delta U_1-W_1=Q_1\\
	T\Delta S_1  =\Delta U_1-{W_r} = Q_r\\
\end{array}
$$
To restore subsystems to their original state, the process (subscript 2) that requires the least work is an isothermal compression such as\,: $Q_2=-Q_r$ and $W_2=-W_r$. It gives the subsystems more work than that produced during expansion\,:  $W_2\ge(-W_1)$, where the equality holds only for the reversible expansion.
So that in all cases one can write\,: 
\begin{equation}\label{Ssub0}
	\begin{array}{ccc}
		\underbrace{T\Delta S_1=\Delta U_1-W_r }
		_{\begin{array}{c} \textrm{cost in energy} \\ \textrm{to go back}\end{array}}
		& \ge & 
		\underbrace
		{\Delta U_1-W_1}
		_{\begin{array}{c} \textrm{energy yielded by} \\ \textrm{the expansion process}\end{array}}
	\end{array}
\end{equation}
Here, $\Delta U_1=0$ and from Eq.\ref{isoexp}, $Q_r=NT \ln(V_f/V_i)$.

Inequality \ref{Ssub0} is the criterion for irreversibility we were looking for in \S\ref{rev_irrev}. It can be rewritten as follows\,:
\begin{equation}\label{Ssub}
	\Delta S_1 \ge \left(\frac{Q_1}{T} =\frac{ \Delta U_1- W_1}{T}\right)
\end{equation}
Or alternatively as:
\begin{equation}\label{Ssub22}
	\Delta S_1= \frac{Q_1 }{T}+\frac{ \mathcal Q}{T} \quad \textrm{with}\quad  \frac{\mathcal Q}{T} \ge 0
\end{equation}
${\mathcal Q}/{T}$ is the amount of entropy produced by the process.
The variation of entropy $\Delta S_2$ due to the reversible restoring compression is  $\Delta S_2=Q_2/T=-Q_r/T$. Thus, for the complete cycle (1- irreversible expansion; 2- reversible restoring compression) the net variation of entropy is\,: $\Delta S=\Delta S_1+\Delta S_2=(Q_1+\mathcal Q)/T+Q_2/T=0$. Thus\,:
\begin{equation}
	\displaystyle\left({ \sum_{i=1}^{2}\frac{Q_i}{T} }\right)_\textrm{cycle}\displaystyle =\frac{-\mathcal Q}{T} \le 0
\end{equation}
That has to be compared to Eq.\ref{Clausius2}.
The negative sign of this sum, means that the net energy ($-\mathcal Q$) received by the subsystem is not stored by the subsystem (as $\Delta U=0$) but spread outside and transferred as heat throughout the large volume of the environment.
This spreading, dispersal or dissipation allows us to identify the process as being irreversible.

For adiabatic free expansion $Q_1=0$. It is a case of thermally insulated system that do not have a larger environment with which to exchange heat (note that \textit{a fortiori} totally isolated systems enter in this category).
In this case Eq.\ref{Ssub} becomes:
\begin{equation}\label{ineq2nd}
	\Delta S_1\ge 0
\end{equation} 
What about the spreading metaphor in these cases\,? 
$T\Delta S_1$ is the cost in energy to restore the system in its original state. Doing so, the system is no longer thermally insulated and $T\Delta S_1$  is effectively dissipated in the environment.
While waiting for this restoring process, the expansion step amounts to spread heat as much as possible throughout the available space. So that $T\Delta S_1$  would be the cost in energy in case we attempt to \textquote{unspread} it.

\subsection{Evolution, equilibrium and stability}\label{equilib}

The inequality \ref{Ssub} obtained for two examples is confirmed by all experimental results, for all systems without exception.
In other words the inequality \ref{Ssub} is observed each time a thermodynamic process occurs.
Thus it has been raised to the rank of a law of evolution of the state of a system and becomes the condition to which any process must obey in order to take place\,: a process that violates Eq.\ref{Ssub} is impossible. This law of evolution can be formulated as\,:

\statement{Any process increases entropy by an amount equal to the minimum net amount of energy (in unit of thermal energy) that must be supplied to the system to restore it to its original state.}

\noindent or more shortly\,:

\statement{The entropy of the state of a system cannot decrease at no cost in energy.}

\noindent These statements explicitly refer to the only quantity that can be directly measured\,: the cost in energy to restore the system to its original state.
A slight deviation from this requirement leads equivalently to\,:

\statement{The entropy of the state of a system cannot spontaneously decrease.}

\noindent where \textquote{spontaneous} means \textquote{without cost in energy}.

A change in meaning very often encountered is to regard the subsystem plus its environment as a whole and Eq.\ref{ineq2nd} instead of Eq.\ref{Ssub}, which leads to the statement\,: \textquote{The entropy of a totally isolated system cannot decrease}.
The word \textquote{spontaneous} disappears as an isolated system only behaves spontaneously. Deviation from experiment is clear as no measurement can be done on isolated systems (a system allowing measurements is no longer isolated). In addition, Eq.\ref{ineq2nd} concerns thermally insulated systems, so that this last statement is at the same time more speculative than the previous ones and weaker because isolated systems form a subset of insulated ones. 

Once the law of evolution has been stated, it is often added a mention of the type\,: \textquote{Entropy is maximum at equilibrium}.
The idea is to ensure the stability of the equilibrium. 
But actually this mention is not necessary as it is already in the zeroth law.
In fact, behind the notion of stability is hidden the one of fluctuations that is originally absent of thermodynamics\,: small fluctuations are likely to cause the system to move away from equilibrium, so something is needed to restore it. If\,: 1) Entropy is maximum at equilibrium; 2) Entropy cannot spontaneously decrease; these two statements act as a restoring force and the equilibrium is stable.
Reciprocally\,: 1) the definition of the equilibrium as being a stable accessible state; 2) plus the statement that entropy cannot spontaneously decrease, automatically imply that it is maximum at equilibrium (otherwise the equilibrium is not accessible). An assertion like \textquote{entropy is maximum at equilibrium} is not directly inferred from experiments, contrary to \textquote{the equilibrium is stable}. So that in the framework of a phenomenological approach, the latter is preferred.

The idea that \textquote{the equilibrium is the state that \textquote{maximizes} or \textquote{minimizes} something with respect to something else}
is deeply rooted in our scientific culture because of our everyday-life experience of mechanics. Nevertheless it does not belongs to phenomenological thermodynamics. 
Actually this idea was encouraged since the origin by R.~Clausius himself who, in his paper introducing entropy for the first time, based on the fact that nothing is more isolated than universe concludes\,:
\myquote{1)~The energy of the universe is constant. 2)~The entropy of the universe tends to a maximum}\,\cite{Clausius_1865}. However, it is very likely that these statements were only metaphorical and not really serious. A clue to this is that a few years later in his book\,\cite{Clausius_1879}, Clausius makes no such mention.
In any event, these statements were later taken literally.

\subsection{Second law}\label{sndlaw}

Finally the second law of thermodynamics can be stated.
It is basically twofold.

\law{Second law}{
	\begin{enumerate}[label={\alph*)}]
		\item There exists a state-quantity named entropy which variation for a reversible transformation is the heat exchanged expressed in temperature unit.
		\item The entropy of the state of a system cannot decrease without cost in energy.
	\end{enumerate}
}
\noindent
With the zeroth law this can form an axiomatic system for thermodynamics (the first law of conservation of energy is implicit as it is not specific to thermodynamics) that is elliptically\,:

\law{Principles of thermodynamics}{
	\begin{enumerate}[label={\alph*)}]
		\item There exists a stable state of equilibrium.
		\item At equilibrium intensive state-quantities are spatially uniform.
		\item $\displaystyle \diff S =\frac{\diff Q_r}{T}$
		\item $\displaystyle\sum_{i\in \textrm{cycle}} \frac{Q_i}{T} \le 0$
	\end{enumerate}
}
\noindent
Starting from these principles, it is possible to reconstruct or deduce everything and predict phenomena. 

\subsection{Free energy}

Let us apply the second law to a very common class of processes occurring at constant temperature. From Eq.\ref{Ssub} one has $\Delta U - T\Delta S\le W$, as $T$ is constant this writes:
\begin{equation}
	\Delta (U-TS) \le W
\end{equation}
The variation of the quantity $U-TS$ is at best equal to the useful work $W$. 
This introduces a new physical quantity\,\cite{Planck_1903} named free energy (or Helmholtz free energy):
\begin{equation}\label{freeEnergy1}
	F=U-TS
\end{equation}
\nomenclature{$F$}{(Helmholtz) free energy}%
which differential is $\diff F=\diff U-\diff(TS)$ or
\begin{equation}
	\diff F = - S\diff T - P \diff V+\mu \diff N
\end{equation}
Let us express the relation between free energy and internal energy that can be derived from Eq.\ref{freeEnergy1}\,: $F/T=U/T-S$, the partial derivative with respect to $T$ is $\partial (F/T)/\partial T = -U/T^2 + T^{-1} \partial U/\partial T -\partial S/\partial T$, yet $\partial U/\partial T=c_vN$ and $\partial S/\partial T=c_vN/T$, thus
\begin{equation}\label{Gibbs-Helmholtz}
	U=-T^2\frac{\partial (F/T)}{\partial T} = \frac{\partial (F/T)}{\partial (1/T)}
\end{equation}
This equation is one of the Gibbs-Helmholtz relations.

Free energy (Eq.\ref{freeEnergy1} and \ref{Gibbs-Helmholtz}) is the link we will need to bridge the gap between thermodynamics and statistical mechanics in \S\ref{GibbsEntropy}.

\subsection{Subjectivity of dissipation and mixing}\label{mixing}

Let us come back to the adiabatic free expansion of a gas.
Imagine two transparent compartments, one containing a transparent and colorless gas, the other being empty.
\begin{enumerate}
	\item Imagine we have no information about the exact contents of the two compartments, except that they are apparently identical (same color and transparency). So that we can open the hole between the two compartments without any observable consequence (no heat, no work exchanged). We can close it back and the observable quantities are exactly as before. Based on our experience and senses, the process is reversible.
	\item Imagine that we know that one compartment is empty and the other is not. Now, the same process become irreversible. It would become reversible if we used a piston, but the idea of using a piston only comes if we know that there exist a difference of pressure.
\end{enumerate}
\noindent The sentence of Maxwell (\myquote{The idea of dissipation of energy depends on the extent of our knowledge}\,\cite{Maxwell_1878}) must be understood in this way.

\begin{figure}[!htbp]
	\begin{center}
		\includegraphics[width=1\linewidth]{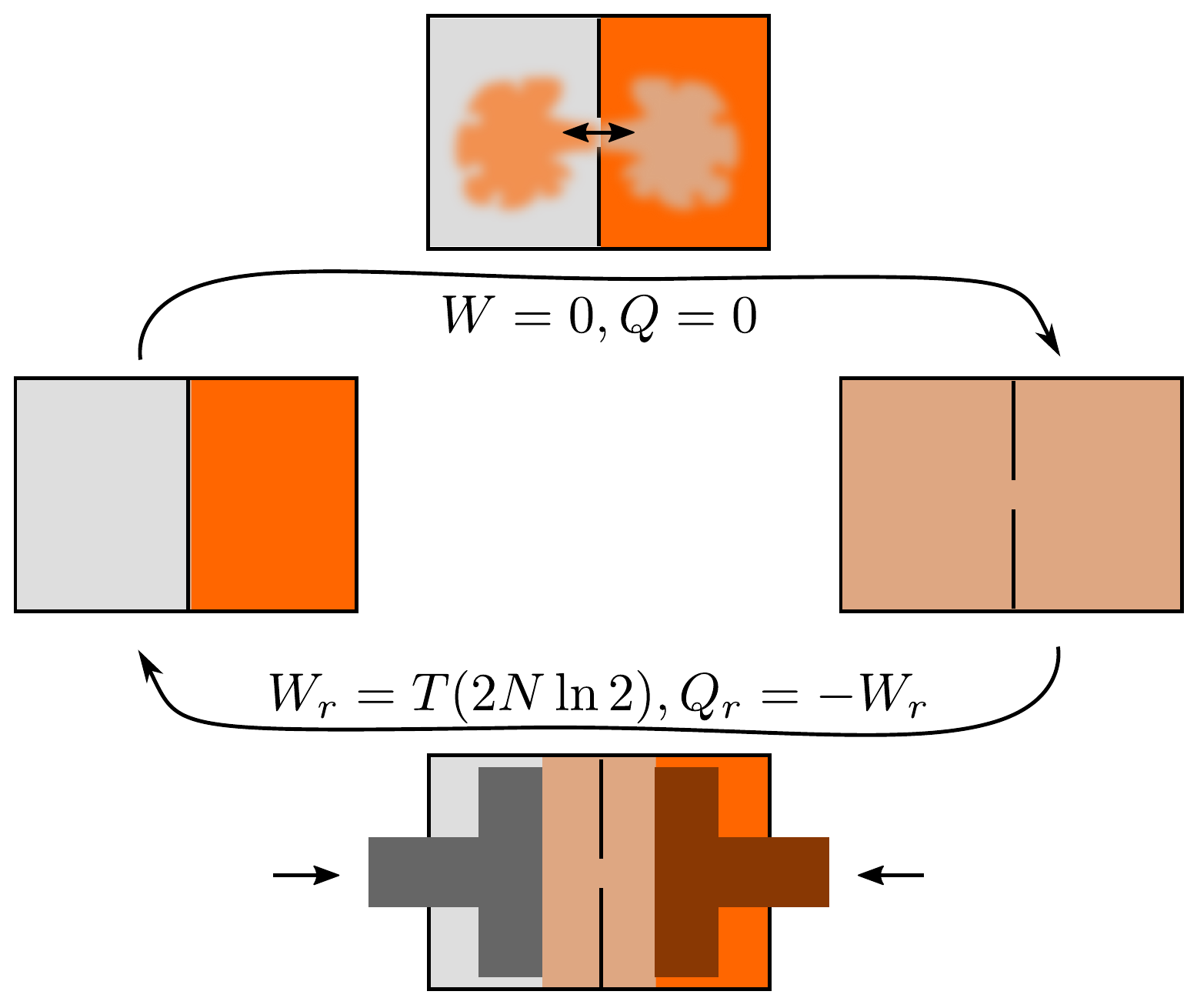}
		\caption{Mixing-unmixing cycle\,: the mixing (left to right, top) is irreversible and corresponds to the free expansion of each gas (see Figure~\ref{freeexp}); the unmixing is reversible (right to left, bottom) and completed with two isothermal compressions with semi-permeable pistons.}
		\label{figMix1}
	\end{center}
\end{figure}

The mixing process is another famous illustration of this subjectivity of entropy.
Once again consider a composite system made of two separated compartments $A$ and $B$ of same volume filled by the same amount of gas at the same temperature (and so at the same pressure):
\begin{equation}\label{eqmix}
	V_A=V_B=V, \quad N_A=N_B=N, \quad P_A=P_B=P
\end{equation}
Suppose the separation is removed at time $t_0$ and consider three cases\,:
\begin{enumerate}[label={\arabic*)}]
	\item The gas is a pure compound. At equilibrium
	uniformity of pressures implies\,:
	\begin{equation}
		N_{A}=N_{B}
	\end{equation}
	As the system already fulfills this condition before $t_0$, nothing special happens after. 
	Putting the partition back in place restores the system to the initial situation at no cost in energy.
	
	\item The gas is a  mixture of two chemical species in equal amounts, say "black" ($\bullet$) and "white" ($\circ$), such as\,:
	\begin{equation}\label{cond1}
		N_\bullet = N_\circ ,\quad P_\bullet = P_\circ
	\end{equation}
	At low pressure it behaves as a perfect gas. The total pressure is the sum of the two partial pressures (Dalton law, $P=P_\bullet +P_\circ$).
	After $t_0$ and at equilibrium, the two species are uniformly distributed between the two compartments\,: 
	\begin{equation}\label{eqcondmix}
		N_{A\bullet}=N_{B\bullet},\quad 
		N_{A\circ}=N_{B\circ}
	\end{equation}
	So that, if it is already the case before $t_0$, removing the separation does not produce anything special.
	
	\item Suppose the initial conditions \ref{cond1} are the same except that compartment $A$ contains only black species and $B$ only white\,: 
	$$
	\begin{array}{cc}
		N_A=N_{A\bullet}=N_{\bullet},& N_{B\bullet}=0 \\
		N_{A\circ}=0 ,& N_B=N_{B\circ}=N_{\circ}
	\end{array}
	$$
	The equilibrium condition \ref{eqcondmix} is unchanged. 
	At the opening, because temperatures and total pressures are equal there is neither heat transfer nor mechanical work.
	However, something happens as at $t_0$ the equilibrium condition Eq.\ref{eqcondmix} is not fulfilled.
	The two gases mix and the process resembles their free expansion.
	It would have a cost in energy to return the system to its previous state. The mixing is irreversible and goes with an increase in entropy, namely the entropy of mixing.
	
	The entropy of mixing can only be assessed when restoring the original state by a reversible process.
	For free expansion it is an isothermal compression against a piston.
	Here, it can be done by two isothermal compressions against two pistons equipped with different semi-permeable membranes\,\cite{Planck_1903} (see Figure~\ref{figMix1}).
	Semi-permeable membranes are those being permeable to one chemical species and impermeable to another. 
	The first piston is only able to compress black species and the second only white species.
	The first piston achieves a work equal to $TN_A\ln({(V_A+V_B)}/{V_A})$ whereas the second achieves one equal to $TN_B\ln({(V_A+V_B)}/{V_B})$. So that the total work needed to separate the two gases is $W=T\times 2N\ln 2=Q_r=\Delta S_\textrm{mix} $. The mixing entropy is thus
	\begin{equation}
		\Delta S_\textrm{mix} = 2N\ln 2
	\end{equation}
	\nomenclature{$\Delta S_\textrm{mix}$}{mixing entropy}%
\end{enumerate}

Finally, whether the contents of the two compartments are initially the same or not, neither mechanical work nor heat is produced by their mixing.
The entropy of mixing is not a directly measurable thermodynamic quantity.
Only unmixing is measurable.
In other words, mixing two gases produces no observable effect except those related to what we know about their difference.

In thermodynamics the concept of mixing only makes sense in the framework of such a "mixing-unmixing" cycle.
If for some reasons, the unmixing is not achieved because impossible, irrelevant, or judged unnecessary, the entropy of mixing is undetermined and this will have absolutely no consequence on the calculation of the thermodynamics of the problem under study.

\subsection{Useful energy}

Consider a perfect gas undergoing an isothermal process ($\Delta U=0$). It is a particular case where $Q_r =-W_r=T\Delta S$. It follows that the variation of entropy, between given initial and final states, is the maximum \textquote{useful} work that can be obtained from any transformation between these two states. The term \textquote{useful} is quite anthropocentric. It is likely that a bacterium would have a different opinion on the matter. So that, the classification of \textquote{energies} in term of grade is now judged obsolete, all forms of energy transform into each other.
However, the anthropocentric character of entropy remains unavoidable.

Let us come back to the mixing-unmixing of two different gases species as in the previous section (Figure~\ref{figMix1}). The membranes that are used to unmix the two gases could be equally used to perform the exact symmetric expansion in a reversible manner. So that a maximum amount of work equal to $T\Delta S_\textrm{mix}$ could be obtained from the mixing and an equal amount of heat pumped from the surroundings. This time however, the work that can be extracted, depends on whether or not we have the appropriate membranes at our disposal, depends on whether or not we know how to build them. Jaynes\,\cite{Jaynes1992} mentions the case of a gas which we do not know is composed of two different species, for instance two different isotopes before their discovery. How to conceive the idea to extract useful work from their mixing\,? There is no clue for that, such as an exchange of heat that would occur when mixing. 
Heat and work only come into play if these membranes are used.
Clearly, they depend on our knowledge.

\subsection{Gibbs paradox \#1}\label{GBthermo}

The question of mixing is far from anecdotal and has occupied scientists for 150 years.
In 1875, Gibbs wrote on this question:

\myquote{Now we may without violence to the general laws of gases which are embodied in our equations suppose other gases to exist than such as actually do exist, and there does not appear to be any limit to the resemblance which there might be between two such kinds of gas. But, the increase of entropy due to the mixing of given volumes of the gases at a given temperature and pressure would be independent of the degree of similarity or dissimilarity between them... In such respects, entropy stands strongly contrasted with energy.} (J.W. Gibbs\,\cite{Gibbs1874} p.228).

Although Gibbs himself did not view it as a paradox, this point was later called the \textquote{original} version of Gibbs paradox. In fact, it was the first of a series of others that are still debated today (see e.g. \cite{Gibbsparadox2018}). Let us reformulate it in a different way allowing to stress the paradox\,:
\myparadox{Gibbs paradox \#1}{}{The internal energy of a simple system is expected to vary continuously with the variables $(S, V, N)$  on which it depends and therefore these variables too.}{If we define a continuous variable $D$ for the degree of dissimilarity of two species, the Clausius entropy of mixing is the discontinuous step function: 0 if $D=0$, $2N\ln 2$ if $D\ne 0$.}{A and B seem to be contradictory.}

\noindent The logic of paradoxes following W.V.~Quine\cite{Quine1976} will be presented in \S\ref{GBdiscuss}. In our case, statements A and B follows from a traditional presentation of entropy in thermodynamics, as it was done in this section. Here, it is neither a question of analyzing or judging their validity, but only of emphasizing their contradiction. 

The relevance of this paradox is often overlooked with the argument that the degree of dissimilarity of two atoms is never continuous. 
But this argument is much less clear if the gas is not monatomic but replaced by a solution of colloids or macromolecules, where each individual object may consist of millions of atoms.
Then, their dissimilarity can vary, not continuously because of the discrete nature of matter, but at least gradually with their size.
Thermodynamics alone cannot resolve this paradox.
We will come back to this question later.

\subsection{What thermodynamic entropy is}\label{whatis1}

Thermodynamics is phenomenological. In this spirit, it is less concerned with what things are and more with how they behave, that is to say how they manifest to an observer.
To understand energy transformations, entropy was introduced as a state-quantity of a system. 
So, the first point is that entropy is not a property of the system but a property of its state.
This sounds like a truism, but it is actually more profound as a state is only defined by how it is perceived by an observer. A clue for that is the choice of the description scale, i.e. the human or macroscopic scale at which temperature and pressure are only defined. A definitive proof is the study of mixing.

State-quantities such as volume, quantity of matter or temperature can be measured directly and the measurement in classical physics is supposed to marginally affect these quantities.
Entropy cannot be measured directly.
Entropy can only be apprehended if the state changes in a way that allows entropy to be entirely transformed into something else, i.e. by using a reversible path.
So for now, from thermodynamics, to the question \textquote{What is entropy\,?} the only rigorous answer based on experimental facts is:

\statement{Increase in entropy is the minimum net cost in energy (in units of thermal energy) to restore the system to its original state as it was known to us.}

Entropy being a state-function, it is however legitimate to ask what characteristics or features of this state, entropy is a function of.
The answer is less rigorous and more metaphoric than the previous one, but from this section it can be proposed that\,:

\statement{Entropy is the degree of energy spreading of a system.}

\noindent With this metaphor, a system can evolve spontaneously only by spreading and flattening its energy repartition.
Unspreading requires a cost in energy.

\section{Statistical entropy}\label{statmech}

During the second half of the 19th century the evidence of atoms and molecules prevailed, the microscopic too. With molecules, what is outstanding is their large number and their incessant movements. Questions arise: 
How to go from microscopic to macroscopic scale\,? 
How can the microscopic profusion be reduced to a small number of variables\,? 
How, from movement and dynamics, to account for equilibrium, statics and thermodynamics\,?
Statistical mechanics aims to answer these questions by the use of probability theory and Newton's mechanics with the minimum additional assumptions. 

The chronology of this approach starts with J.C. Maxwell and the kinetics theory of gases; then L. Boltzmann with his H-function that intends to account for a non-equilibrium evolution; then M. Planck, the actual author of the {Boltzmann's } entropy; and finally J.W. Gibbs who first writes the statistical entropy. However, as the latter is concerned with equilibrium and is directly linked to thermodynamics we will use the reverse chronology instead.

The basic idea of statistical mechanics is that a macroscopic system can adopt many different microscopic configurations without any difference to an observer.
These configurations are said compatible with the observed macroscopic state. Next, the probability for a macroscopic system to be observed in a given state is proportional to the number of compatible microscopic configurations. Then, the equilibrium state is the most probable and an irreversible process is the passage from an unlikely to a very likely state. This is the overall story.

Let us first emphasize how this approach is completely different from the phenomenology of thermodynamics, quite simply because the microscopic configurations cannot in any way be measured or counted in a laboratory.
To derive everything from probabilities, four premises are first required\,: the definition of equilibrium, the hypothesis of ergodicity, the one of independence of probabilities and the so-called fundamental postulate.
The rest comes from deductions.

The axiomatic approach, the deductive reasoning and the rigor of mathematics, lead us to think statistical mechanics as being concerned with the objective properties of matter, rather than their subjective perception which the phenomenology of thermodynamics deals with. Since the origin the ambition of statistical mechanics is to provide \myquote{the rational foundations of thermodynamics} (J.W. Gibbs, second title of his book \cite{Gibbs_1902}).
In reality, statistical mechanics is not totally emancipated from all subjectivity, even if it wishes to be. Subjectivity has simply been moved to the root.

Classically, a probability is understood\,\cite{Keynes_1921, Dubs_1942}\,: 1)~either as \textit{a priori} human expectation, that is a prediction about an event that we just know could happen (this is the primary meaning of probabilities); 2)~or as \textit{a posteriori} mathematical expectation, that is to say the average fraction of the number of occurrences of an event to a total number of observations  (this is usually the meaning of statistics).
In the first acceptation, a probability is subjective and depends on our knowledge. It is an expression of the uncertainty linked to our perception of phenomena. In the second acceptation, a probability is an objective property of the matter under study. Despite many efforts to avoid ambiguities between these two meanings and to tend towards objectivity, statistical mechanics remains full of subjectivity, as we will see (for a review on this feature see the paper of J.~Uffink\,\cite{Uffink_2006}).

\subsection{Definition of equilibrium}\label{equilibrium2}

The definition of what is exactly the state of equilibrium as viewed in statistical mechanics if often eluded. It is thought obvious, just a definition and not really a fundamental postulate or hypothesis. In reality it is, exactly as it was in thermodynamics (\S\ref{equilibrium0}).

In thermodynamics, the equilibrium is a steady state. It is also the condition required by Gibbs the founder of statistical mechanics (\cite{Gibbs_1902} p. 18).
The equilibrium is therefore by definition the state in which the system spends all its time.
Thus, for consistency, if probabilities are understood as frequencies of occurrence, the equilibrium must be defined as the most probable state.
The justification of this definition, as we will see, is subjected to the next hypothesis of ergodicity, but most importantly it is subjected to the definition given in thermodynamics. Therefore to liberate from this dependency, in accordance with the ambition of a bottom-up approach, this definition must be conceived as a postulate.

\subsection{Hypothesis of ergodicity}\label{ergo}

Let us take the example of the random variable which would be the color of the cars parked in the street.
One uses a camera to determine the random variable distribution and its average value. There are two ways to do this\,: 
\begin{enumerate}
	\item One can place the camera in a fixed position in the street, point the lens in a given car-place and take photos at regular interval during the renewal of cars. Like this one can compute the time average of the color. 
	\item One can move the camera along the street and photograph the cars that are parked. This would lead to an ensemble-average color.
\end{enumerate}
However in the first case, the camera field (the space resolution) can be larger that one car place so that one given photo is an ensemble-average that is used to compute a time-average. In the second case, during the time needed to take one photo (the time-resolution) a car-renewal may occur. Thus, one given photo is a time-average that is used to compute an ensemble-average.
Here, due to the finite resolution in space and in time, the measurement results always from a combination of an ensemble-average with a time-average. The easiest way to take this into account is to say it does not matter. This is the ergodic hypothesis.
A system is ergodic when both averaging procedure give the same result, or in other words when the probability of an event and its relative frequency of occurrence are equal. 
In brief, the property of ergodicity\,\cite{Moore_2015} supposes that\,:
\begin{enumerate}[label={\arabic*)}]
	\item The system is dynamical, that is to say it is continuously changing.
	\item  The system is metrically transitive, that is to say it can always come back to any already observed outcome and accomplishes all possibilities at one time or another.
\end{enumerate}

Ergodicity is not a general property and not all systems are ergodic, e.g. those which are frozen or blocked in a state or in a direction of evolution without any possibility to come back are not ergodic.
So that ergodicity is not a general principle, but only the framework in which statistical mechanics is placed, at least in its first developments.

The first condition for a system to be ergodic, i.e. be dynamical,  poses no problem.
The second property of metric transitivity is much less obvious for systems with which statistical mechanics deals.
Quite simply because the number of possibilities offered by an assembly of atoms is so enormous that it would take geological time to explore them all.
In this context, the total number of possibilities is unmeasurable, unknown or arbitrary, and the frequencies of occurrence are not defined for lack of the normalization value.
The hypothesis of ergodicity therefore seems incompatible with the vision of probabilities as objective properties of a system and cannot possibly escape the subjective point of view.

\subsection{Hypothesis of independence}\label{indep}

In Hamiltonian mechanics, a system made of $N$ particles of a perfect gas is entirely characterized by the $3N$ coordinates of the position $\mathbf r$ and those of momentum $\mathbf p$ that define a \textquote{microstate}, that is to say one possibility offered to the system. The time-evolution of the system, which results from collisions of molecules, is a continuous trajectory in a close volume, namely the phase-space, which belongs to $\mathbb{R}^{6N}$. 
Suppose the system entirely known at a given time $t_0$ with a measurement uncertainty $\mathfrak{h}^3$ that has the dimension $\textrm{L}^3\times (\textrm{M}\textrm{L}\textrm{T}^{-1})^3 =(\textrm{M}\textrm{L}^2\textrm{T}^{-1})^3$ and writes:
\begin{equation}\label{volunit}
	\mathfrak{h}^3=\sigma_\mathbf{r} \sigma_\mathbf{p}
\end{equation}
\nomenclature{$\mathbf r$}{particle position}%
\nomenclature{$\mathbf p$}{particle momentum}%
\nomenclature{$\sigma_\mathbf{r}$}{resolution on position}%
\nomenclature{$\sigma_\mathbf{p}$}{resolution on momentum}%
\nomenclature{$\mathfrak{h}$}{resolution on action}%
\nomenclature{$h$}{Planck constant (quantum of action)}%
\nomenclature{$e$}{Euler number}%
To measure the velocity (or momentum) of an object is to measure something over a spatial extent. The larger this extent, the better the accuracy of the measurement, but also the more indeterminate the place to which it corresponds.
It follows that the uncertainty in Eq.\ref{volunit} is necessarily finite\,\footnote{
	Eq.\,\ref{volunit}, with $\mathfrak h$ equal to the Planck constant, is the Heisenberg uncertainty principle. As W. Heisenberg is the father of quantum mechanics, it is often viewed as coming from quantum physics ideas.
	Actually, it is between two worlds, classical and quantum. It comes from the wave-particle duality, from its probability interpretation and from the fact that the probability densities of $\mathbf r$ and $\mathbf p$ of a given particle are Fourier transforms of each other. Then, from the mathematical property of Fourier transform, increasing the resolution in the domain of one variable, degrades that of its conjugate.}. 
If we focus on one of the two molecules involved in the first collision that occurs, although the microstate of the system can entirely be calculated after this collision, the uncertainty on the position of the molecule grows with time as $t(\sigma_\mathbf{p}/m)$. So that, there exists a given time after which it is impossible to predict with which molecule (or where) the next collision will occur. Beyond this time, correlations are lost and the parameters of the next collision appear as independent random variables, the microstate of the system too, as well as its energy. Statistical mechanics assumes that macroscopic observations are always made in this situation. 

Although Newton's mechanics is fully deterministic, the resulting microstates can only be treated as independent random variables because we are not able to perceive correlations. As this perception depends on the observer \textit{via} the resolution, subjectivity intrudes here again into the theory.

\subsection{Fundamental postulate}\label{fp}

Imagine a large isolated system, with total energy $E$, divided into identical subparts with number $i=1\cdots N$ and $N\gg 1$.
The finite resolution of Eq.\ref{volunit} and the existence of a upper boundary for the energy, allow us to discretize the possible energy levels into a set of $z$ values\,:
$$
E_i \in \{F_0, F_1, \cdots F_{z-1}\} \quad \textrm{with} \quad F_k=k\epsilon
$$
where $\epsilon$ is the smallest observable energy exchange.
\noindent For the sake of simplicity, let us define a microstate by the multiplet $$\textrm{microstate} = (E_1, E_2,\cdots,E_N),$$ so that the system can adopt a finite number of different microstates.
Also, it is supposed to be dynamical and at every time an elementary transition can occur from the microstate $(E_1,\cdots, E_i,\cdots, E_j,\cdots,E_N)$ to the neighboring microstate $(E_1,\cdots, E_i-\epsilon,\cdots,E_j+\epsilon,\cdots,E_N)$. The microstate in which an observer can found the system is thus a random variable.

The fundamental postulate of statistical mechanics is a variation of the Laplace's \textquote{principle of insufficient reason} (renamed \textquote{principle of indifference} by J.M. Keynes\,\cite{Keynes_1921})\,: \myquote{When one does not know anything the answer is simple. One is satisfied with enumerating the possible events and assigning equal probabilities to them} (R. Balian \cite{Balian_1991} p.143). 
Here, it becomes\,:

\law{Fundamental postulate of statistical mechanics}{At the equilibrium all the microstates of an isolated system are equiprobable.}

Since the system is isolated, the total energy $E=\sum_i E_i$ is constant, however those $E_i$ of subparts fluctuate due to transitions.
The energy of a given subpart is also a random variable with a probability distribution $f$ that is the same for all subparts as they are assumed to be identical.
Consider two given subparts $(i,j)$ sufficiently far the one from the other so that they are uncoupled\,: $E_i$ and $E_j$ are independent random variables. This is possible since $N\gg 1$. Denote
$$
\mathcal E=E_i+E_j
$$
The probability for $\mathcal E$ to have a given value $x$ is\,:
\begin{equation}\label{sume}
\mathbb{P} (\mathcal E = x) = \sum_{k=0}^{z-1} f(F_k)f(x -F_k)
\end{equation}
Once given $(i,j)$ the rest of the $N-2$ subparts can adopt a finite number of different microstates that only depends on $\mathcal E$ and is independent of $k$. All of them are equiprobable. 
It follows that in Eq.\ref{sume}, the $z$ terms of the sum are identical\,:
\begin{equation}\label{sume2}
\mathbb{P} (\mathcal E=x) = z f(F_k)f(x -F_k),\quad \forall k\in\{0,\cdots z-1\}
\end{equation}
Whatever $a$ and $b$ in $\{0,\cdots z-1\}$, one can write
\begin{equation}\label{sume2}
\begin{array}{rcl}
	\mathbb{P} (\mathcal E=F_a+F_b) &=& z f(F_a)f(F_b)\\ \\
	&=&z f(0)f(F_a+F_b)
\end{array}
\end{equation}
Thus
$
f(F_a) f(F_b)= f(0) f(F_a+F_b)	
$ whatever $F_a$ and $F_b$,
which is a definition for the exponential function. If in addition the total energy is not diverging and proportional to the temperature (as the internal energy of a gas) one finally obtains the probability distribution of microstates in terms of their energy: 
\begin{equation}\label{boltzmann}
f(E_i)\propto e^{-E_i/T}	
\end{equation}
This exponential distribution comes from the three ingredients: 
\begin{enumerate}[label={\arabic*)}, nosep]
\item Additivity of energy.
\item Independence of probabilities.
\item Principle of insufficient reason.
\end{enumerate}
Therefore it is quite general.
One of the two subparts can be any thermalized system in contact with a very large temperature reservoir.
The so obtained exponential distribution is known as Boltzmann or canonical distribution.

A particular case would consists in considering subparts as small as individual molecules. The Boltzmann distribution (Eq.\ref{boltzmann}) holds for their energy distribution.
For a perfect gas, the energy of a molecule of mass $m$ reduces to its kinetic energy: $E=\mathbf{p}^2/2m=(\mathsf{p}_x^2+\mathsf{p}_y^2+\mathsf{p}_z^2)/2m$. A change of variable in Eq.\ref{boltzmann} gives the distribution of momentum vector $\mathbf{p}$\,:
\begin{equation}\label{distmoment}
f(\mathbf{p})=\displaystyle\left({\frac{1}{2\pi m T}}\right)^{3/2} e^{\frac{-\mathbf{p}^2}{2mT}} \end{equation}
\nomenclature{$m$}{particle mass}%
We recognize the gaussian distribution. 

Three probability distributions have been introduced.
The uniform distribution is nothing more than an expression of our ignorance. The exponential distribution comes from the fact that the random variable is allowed to fluctuate but remains positive and has an average value. The gaussian distribution is obtained by an additional change of random variable that we expect as being centered. Finally, the three distributions result from a minimum knowledge and from the principle of insufficient reason. The subjectivity of the observer, its knowledge or ignorance, are still in these fundaments of statistical mechanics.

\subsection{From micro- to macroscopic}\label{micmac}

From the previous distribution of momentum (Eq.\ref{distmoment}), the distribution of the modulus $\mathsf{p}$ is obtained by summing the probability over the spherical shell of volume $4\pi \mathsf{p}^2 \diff \mathsf{p}$, leading to the Maxwell-Boltzmann distribution (note in passing that this corresponds to the hypothesis that the system is isotropic, behind which hides the principle of insufficient reason\,\cite{Uffink_2006}).
Then pressure, temperature and internal energy are understood as average quantities related to the average kinetic energy of the molecules which emerges from this distribution.
In particular one obtains for the internal energy of $N$ molecules of a perfect monoatomic gas\,: $U= N\left<{ (\mathsf{p}_x^2+\mathsf{p}_y^2+\mathsf{p}_z^2)/2m }\right>$:
\begin{equation}\label{U}
U = \frac{3}{2}NT
\end{equation}
Here, number 2 comes from the gaussian distribution of momentum along one axis, whereas number 3 comes from the triple integral. Actually, it can be shown that this number is more properly replaced by the number $\mathfrak{f}$ of freedom of molecules: $\mathfrak{f}=3$ for monoatomic (only 3 translation axis), $\mathfrak{f}=3+2$ for diatomic (in order to account for rotations), etc. Interestingly, this gives a microscopic interpretation for the different heat capacities $c_v$ of gases\,:
\begin{equation}\label{CV2}
c_v =  \frac{\mathfrak{f}}{2}
\end{equation}
With Eq.\ref{CV1} it follows that
\begin{equation}\label{gamma}
\gamma = 1+\frac{2}{\mathfrak{f}}
\end{equation}
In this way statistics provide a microscopic basis for macroscopic quantities.

\subsection{Gibbs entropy}\label{GibbsEntropy}

A consequence of the finite accuracy with which microstates can be discriminated is the discretization of the phase-space. Since the latter has a finite volume, a continuous and infinite number of possibilities is therefore reduced to a finite number of discernable microstates, leading to a discrete number of energy levels and a discrete distribution of their probabilities. This allows us to rewrite Eq.\ref{boltzmann} as\,:
\begin{equation}\label{pi}
p_i=\frac{1}{Z}e^{-E_i/T} = e^{({-T\ln Z-E_i})/{T}}
\quad\textrm{with}\quad
Z=\sum_{i=1}^{\mathcal{W}} e^{-E_i/T}
\end{equation}
\nomenclature{$Z$}{partition function}%
\nomenclature{$\mathcal{W}$}{total number of discernable allowed microstates}%
\nomenclature{$p_i$}{probability of microstate $i$}%
\nomenclature{$E_i$}{energy of microstate $i$}%
where $\mathcal{W}$ is the total number of discernable allowed microstates. $Z$ is called the partition function.
It is the normalization factor for probabilities $p_i$, so that
its physical meaning depends on the one given to them.
In the position-momentum phase-space, each microstate occupies a given \textquote{volume} fraction $p_i$.
The Arrhenius-like form of probabilities $p_i$ shows that they can also be conceived as the relative lifetime of a given microstate.
So that consistently with ergodicity, the probability $p_i$ for one given microstate is also the ratio of the time spent in this volume to the total time needed to explore the whole. The partition function is the latter.

From the discrete probability distribution of microstates, some additive state-quantities used in thermodynamics can be derived. The internal energy is the mathematical expectation of the energy of microstates\,:
\begin{equation}\label{intU}
U = \sum_i p_i E_i
\end{equation}
The derivative of $\ln(Z)$ is:
$$
\begin{array}{rl}
\displaystyle\frac{\diff(\ln(Z))}{\diff (1/T)} 
& =\displaystyle \frac{1}{Z} \frac{\diff Z}{\diff(1/T)}\\
& =\displaystyle  \frac{1}{Z}  \sum_{i=1}^{\mathcal{W}}-E_i e^{-E_i/T}\\
& =\displaystyle - \sum_{i=1}^{\mathcal{W}}p_iE_i \\
&= -U
\end{array}
$$
Identification with the Gibbs-Helmholtz relation \ref{Gibbs-Helmholtz} gives the free energy:
\begin{equation}\label{Zfree}
F = - T \ln(Z)
\end{equation}
This allows us to rewrite Eq.\ref{pi} as
\begin{equation}\label{pi2}
p_i=e^{(F-E_i)/T}
\end{equation}
or $E_i= F -T\ln(p_i)$ and to rewrite Eq.\ref{intU} as:
$$
\begin{array}{rl}
U  &= \sum_i p_i \left[{F -T\ln(p_i)}\right]\\
&= F -T\sum_i p_i\ln(p_i) 
\end{array}
$$
Identification with Eq.\ref{freeEnergy1} provides an expression for the entropy in terms of the energy distribution of microstates\,:
\begin{equation}\label{S}
S= - \sum_{i=1}^{\mathcal{W}} p_i\ln(p_i)  
\end{equation}
which is called Gibbs entropy. 
Despite a different name, the way it has been derived by identification with a thermodynamic equality, shows that it is the same physical quantity as the Clausius entropy.
It is in this spirit that Gibbs, the first, got this expression (see \cite{Gibbs_1902} Eq. 108 to 116).

\subsection{Heat and work}\label{hear}

The differential of Gibbs entropy given by Eq.\ref{S} is\,: 
$\diff S =  - \sum_i p_i\diff \ln(p_i)    - \sum_i \ln(p_i) \diff p_i $.
With $\diff \ln(p_i)=p_i^{-1}\diff p_i$, one gets\,:
$	\diff S =  - \sum_i \diff p_i    - \sum_i \ln(p_i) \diff p_i $. Since $\sum_i  p_i =1$, so $\sum_i \diff p_i =0$ and 
\begin{equation}
\diff S = -  \sum_{i=1}^{\mathcal{W}} \ln(p_i) \diff p_i  
\end{equation}
With Eq.\ref{pi2} one obtains $	T\diff S =   \sum_i (E_i-F) \diff p_i =  \sum_i E_i \diff p_i  -F \sum_i \diff p_i $, and finally
\begin{equation}
T\diff S =  \sum_{i=1}^{\mathcal{W}}  E_i \diff p_i  	
\end{equation}
that is from Eq.\ref{diffS} the differential of the heat exchanged. Differentiating Eq.\ref{intU} gives $\diff U = \sum_i E_i \diff p_i +\sum_i p_i \diff E_i=\diff Q_r+\diff W_r$, so that we can write:
\begin{eqnarray}
\diff Q_r &= - \sum_{i=1}^{\mathcal{W}} E_i \diff p_i  	 \\ 
\diff W_r&=  - \sum_{i=1}^{\mathcal{W}} p_i \diff E_i  	
\end{eqnarray}
Exchanges of heat modify the population of energy levels, whereas exchanges of work modify their values.

\subsection{Boltzmann entropy}\label{eqwhole}

Let us write the probability $p_i$ as the ratio $w_i/\mathcal{W}$, with $\sum_iw_i=\mathcal{W}$, Eq.\ref{S} gives
\begin{equation}\label{eqbase}
\begin{array}{rl}
	\displaystyle - \sum_{i=1}^{\mathcal{W}} p_i\ln(p_i) 
	&\displaystyle =  - \sum_{i=1}^{\mathcal{W}} \frac{w_i}{\mathcal{W}}\ln\left({\frac{w_i}{\mathcal{W}}}\right)\\
	&\displaystyle = \ln \mathcal{W}- \sum_{i=1}^{\mathcal{W}} \frac{w_i}{\mathcal{W}}\ln(w_i)\\
\end{array}
\end{equation}
For a uniform probability distribution, all $w_i$-values are equal to 1, so the second term of this difference is zero and $S= \ln \mathcal{W}$.
Let us keep $\mathcal{W}$ unchanged but increase some $w_i$-values to the detriment of others put to zero. The second term increases and $S$ decreases, so that in any case:
\begin{equation}\label{inegS}
- \sum_{i=1}^{\mathcal{W}} p_i\ln(p_i)  \le \ln(\mathcal{W})
\end{equation}
Thus, the Gibbs entropy is maximum for the uniform probability distribution of microstates, which according to the fundamental postulate corresponds to the equilibrium of an isolated system. In this situation, one gets what is known as the \textquote{Boltzmann entropy}\,\cite{Boltzmann_Lectures}:
\begin{equation}\label{Sboltzmann}
S= \ln(\mathcal{W})
\end{equation}
where $\mathcal{W}$ is the number of discernable allowed microstates of the isolated system.
Note that actually Boltzmann did not write Eq.\ref{Sboltzmann} but something similar (\cite{Boltzmann_Lectures} p.57) in relation with his H-function (see \S\ref{irrev2}) that involves different probabilities than those of microstates. Eq.\ref{Sboltzmann}  is due to M. Planck (\cite{Planck_1914} p.141) but he attributed the idea to Boltzmann. 

In the following, we will refer to Gibbs entropy (Eq.\ref{S}) and Boltzmann entropy (Eq.\ref{Sboltzmann}) as statistical entropy,
the context making the distinction. The Gibbs entropy refers to a thermalized system with an exponential distribution of microstates (canonical),  whereas Boltzmann entropy refers to an isolated system uniformly distributed (microcanonical).
The nomenclature is due to Gibbs\,\cite{Gibbs_1902}.
The important point is that the statistical entropy of Eq.\ref{S} is actually a function of the probability distribution.

As Clausius entropy and thanks to the logarithm, statistical entropy is additive over subparts. Let $A$ and $B$ be two independent isolated systems, with entropy $S_A=\ln \mathcal{W}_A$ and  $S_B=\ln\mathcal{W}_B$, respectively. The mathematical union of these two sets $A\cup B$ can adopt  $\mathcal W=\mathcal{W}_A\mathcal{W}_B$ different microstates, so that $S_{A\cup B}=\ln(\mathcal{W}_A\mathcal{W}_B)=S_A+S_B$.

\subsection{Absolute entropy ?}\label{absEntr}

Equation \ref{Sboltzmann} suggests that statistical mechanics provides an absolute expression for the entropy, contrary to thermodynamics that only gives access to differences. Actually, in  Eq.\ref{eqbase} or \ref{Sboltzmann} the constant is hidden behind the number $\mathcal{W}$ 
that depends on the discretization of the phase-space and in reality is only known up to a factor.

Suppose a temperature reservoir in contact with one single monoatomic gas molecule which position $\mathbf{r}$ lies in a volume $V$ and which momentum $\mathbf{p}$ lies in the volume $(2\pi mT)^{3/2}$ (the normalization divisor in Eq.\ref{distmoment}). 
The partition function $Z_{(1)}$ for this molecule is given by the ratio of the \textquote{volume} $V (2\pi mT)^{3/2}$ to the elementary volume $\mathfrak{h}^3$:
\begin{equation}\label{Z1}
Z_{(1)}=\frac{V (2\pi mT)^{3/2}}{\mathfrak{h}^3}
\end{equation}
Ideally $\mathfrak{h}^3$ should correspond to the precision with which two microstates can be experimentally discerned (Eq.\ref{volunit}). But in classical statistical mechanics, for the motions of large molecules, polymers, colloids etc, we have no idea about this precision so that $\mathfrak{h}$ is arbitrary (with no consequence as we will see later). Eventually it can be taken as small as possible, that is $h^3$ with $h$ the Planck constant. Equations Eq.\ref{freeEnergy1} and \ref{Zfree} give $S=U/T+\ln(Z)$, so that with Eq.\ref{U} for $N=1$ the entropy of one single molecule is\,:
\begin{equation}
S_{(1)}=\frac{3}{2}+\ln Z_{(1)}
\end{equation}
By setting $S_{(1)}=\ln \mathcal{V}$, one obtains the number of microstates:
\begin{equation}\label{W1}
\mathcal{V}= \frac{V (2\pi mTe)^{3/2}}{\mathfrak{h}^3} 
\end{equation}
That can also be written as:
\begin{equation}\label{W1bis}
\mathcal{V} =\frac{V}{\lambda^3} \quad\textrm{with}\quad \lambda={\mathfrak{h}}{(2\pi mTe)^{-1/2}}
\end{equation}
\nomenclature{$\mathcal V$}{adimentional volume (normalized)}%
\nomenclature{$\lambda$}{length resolution}%
\nomenclature{$\lambda_T$}{thermal length of de Broglie}%
If $\mathfrak{h}$ is taken as equal to the Planck constant, then $\lambda_T=\lambda e^{1/2}$ is the thermal length of de Broglie.
Provided that  $\mathfrak{h}$ and $T$ are constant and identical particles are involved, $\lambda$ is constant. Thus $\lambda^3$ can be viewed as the unit of volume and $\mathcal{V}$ the volume expressed in this unit.
For $N$ identical molecules one has:
\begin{equation}\label{WN}
\mathcal{W}=\mathcal{V}^N
\end{equation}
So that Eq.\ref{Sboltzmann} gives\,:
\begin{equation}\label{SN}
S=N\ln(\mathcal{V})
\end{equation}
Let us expand this expression for a perfect monoatomic gas. From the previous equations, one obtains:
\begin{equation}\label{Sgaz}
\begin{array}{c}
	S =N\left({\ln(V) + \frac{3}{2} \ln(T) + c(m,\mathfrak{h})}\right)\\
	\textrm{with} \quad c(m, \mathfrak{h})=\frac{3}{2}\ln(2\pi m) - 3\ln(\mathfrak{h}) + \frac{3}{2}
\end{array}
\end{equation}
\nomenclature{$c(m, \mathfrak{h})$}{term in the entropy that is independent of $U$, $V$ and $N$}%
$c(m, \mathfrak{h})$ is independent of $N$, but is also unknown unless $\mathfrak{h}$ is proven to be equal to $h$. 
In order to be compared with thermodynamics, temperature can be replaced by internal energy using Eq.\ref{U}. Then we obtain an expression  for $\mathcal S(U,V,N)$\,:
\begin{equation}\label{sackur0}
\mathcal S(U,V,N)  = \frac{3}{2}N\ln(U) + N\ln(V)  
- \frac{3}{2}N\ln(N) + Nc(m,\mathfrak{h})
\end{equation}
Note that the gradient of $\mathcal S(U,V,N)$ calculated from this last equation agrees with the one given by Eq.\ref{gradient} with $c_v=3/2$. In addition, an expression for its third component $\mathcal S_N$ can now be proposed. 
Contrary to Clausius definition, the statistical entropy is now an explicit function of~$N$. 

\subsection{Maximum probability and entropy}\label{max1}

\begin{figure}[!hbtp]
\begin{center}
	\includegraphics[width=0.5\linewidth]{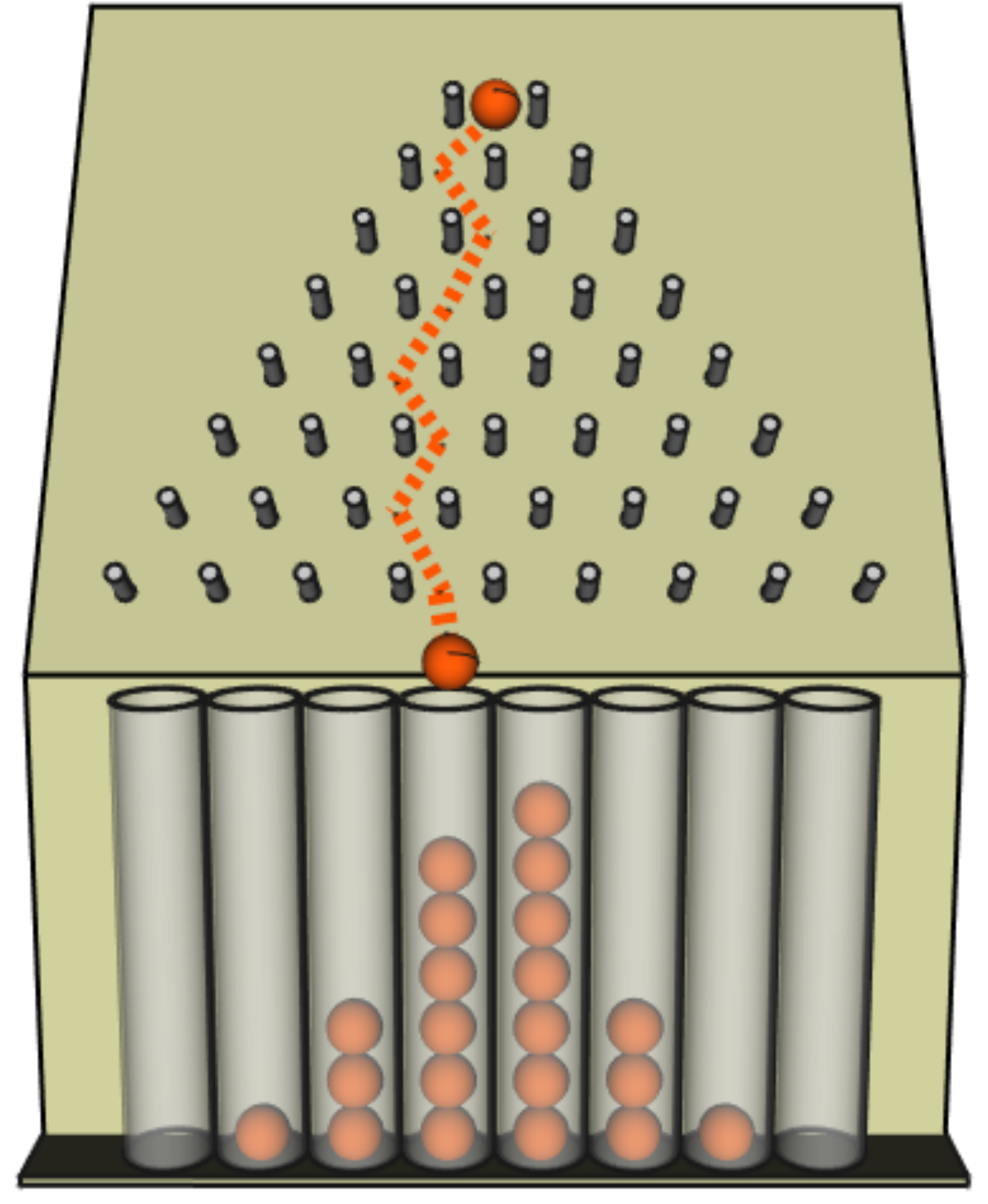} 
	\caption{Galton board: balls roll down a slope finding their way among obstacles. Each way (dashed red line) is equiprobable and corresponds to one microstate. The number of ways reaching a given output is its multiplicity. Each output corresponds to one macroscopic state.		
	}\label{galton}
\end{center}
\end{figure}

Due to its dynamics, the internal variables of a system continuously change and fluctuate. How these fluctuations can be consistent with the steady state of equilibrium\,?
Consider again a composite system made of a simple gas distributed between two compartments $A$ and $B$ with a permeable separation wall, with number of particles $N=N_A+N_B$ and volume $\mathcal{V}=\mathcal{V}_A+\mathcal{V}_B$. The quantities $\mathcal{V}, \mathcal{V}_A,\mathcal{V}_B$ and $N$ are constant, but $N_A$ and $N_B$ are internal random variables that can fluctuate. 
The total number of discernable allowed microstates is: 
\begin{equation}
\mathcal{W}_{\textrm{tot}}(N) \displaystyle = \mathcal{V}^{N}
\end{equation}
Among them, envisage a particular state with exactly $N_A$ molecules in compartment $A$ (and $N_B=N-N_A$ in $B$) and denote $\mathcal{W}(N,N_A)$ its multiplicity, i.e. the number of microstates that fulfill this requirement. 
From the fundamental postulate of equiprobability of microstates, the probability to find the system in this particular state can be written as:
\begin{equation}\label{pNa}
p(N,N_A) =\frac{\mathcal{W}(N,N_A)}{\mathcal{W}_{\textrm{tot}}(N)} 
\end{equation}
There are $\mathcal{V}_A^{N_A}$ different microstates for $N_A$ molecules deposited in compartment $A$, and $\mathcal{V}_B^{N_B}$ different microstates for $N_B$  molecules deposited in $B$. Also, there are ${N!}/({N_A! (N-N_A)!})$ different combinations to distribute $N$ molecules in two compartments with exactly $N_A$ in $A$\,\cite{Boltzmann_Lectures}. Thus one gets\,:
\begin{equation}\label{WNa}
\mathcal{W}(N,N_A) \displaystyle =\mathcal{V}_A^{N_A} \mathcal{V}_B^{N_B}\frac{N!}{N_A! N_B!}
\end{equation}
The corresponding probability is\,:
\begin{equation}\label{pNa1}
p(N,N_A)= \left({\frac{\mathcal{V}_A}{\mathcal{V}}}\right)^{N_A}\left({\frac{\mathcal{V}_B}{\mathcal{V}}}\right)^{N_B}\frac{N!}{N_A! N_B!}
\end{equation}
$p(N,N_A)$ is the binomial distribution for $N$ Bernouilli trials of probabilities $p=\mathcal{V}_A/\mathcal{V}$ and $q=1-p=\mathcal{V}_B/\mathcal{V}$. In figure \ref{galton}, it is illustrated for $p=q=1/2$. The mode and the mean of this distribution are 
\begin{equation}\label{barNa}
\bar N_A=\frac{\mathcal{V}_A}{\mathcal{V}} N
\end{equation} 
and its standard deviation 
\begin{equation}
\sigma_{N_A}=\left({\frac{\mathcal{V}_A}{\mathcal{V}}\frac{\mathcal{V}_B}{\mathcal{V}}N}\right)^{1/2}
\end{equation}
So that the relative characteristic amplitude of fluctuations is\,:
\begin{equation}
\frac{\sigma_{N_A}}{\bar N_A} = \left({\frac{\mathcal{V}_B}{\mathcal{V}_A}}\right)^{1/2}\times \frac{1}{N^{1/2}}
\end{equation} 
This is negligible being given that $N$ is of the order of the Avogadro number $6\times 10^{23}$. This is an expression of the law of large numbers.

The most probable state is therefore also the state in which we are sure to find the system. It is the equilibrium state.
In this state, from Eq.\ref{barNa} one can see that the average densities in both compartments are equal to the density of the whole:
\begin{equation}\label{eqPhase}
\frac{\bar N_A}{\mathcal{V}_A}=\frac{\bar N_B}{\mathcal{V}_B}=\frac{\bar N}{\mathcal{V}}
\end{equation}
This situation corresponds to the most uniform state, in agreement with thermodynamics. 

The Botzmann entropy of the system is\,: 
\begin{equation}\label{Stot}
\mathcal S(N)=\ln(\mathcal{W}_{\textrm{tot}})
\end{equation}
Similarly, let us denote
\begin{equation}\label{SNa}
\mathcal S(N,N_a) = \ln(\mathcal{W}(N,N_A))
\end{equation}
that allows us to write
\begin{equation}\label{pNa2}
p(N,N_A) = e^{\mathcal S(N,N_A)-\mathcal S(N)} 
\end{equation}
This later equation shows that the most probable value for $N_A$ is the one that maximizes $\mathcal S(N,N_a)$.
In other words, the value taken by the internal variable $N_A$ at the equilibrium
is the one that maximizes the entropy $\mathcal S(N,N_a)$.

Alternatively, it can be expressed as follows. Consider a macroscopic state characterized by an internal variable $N_A$. The set of microstates compatible with this state occupies a volume $\mathcal{W}(N_A)$ in the phase-space.
Let us call the quantity $S(N_A)=\ln(\mathcal{W}(N_A))$ ($N$ has been removed because it is constant) the entropy of this state. The equilibrium is the state of maximum volume $\mathcal{W}(N_A)$ and maximum entropy $S(N_A)$.

Multiplicity $\mathcal{W}(N,N_A)$ and probability $p(N,N_A)$ of a given state only differ by the factor $\mathcal{W}_{\textrm{tot}}(N)$ that  is constant for given volume $\mathcal V$ and number of molecules $N$. 
For this reason, the multiplicity is sometimes called "thermodynamical probability"\,\cite{Tien_1979}. Actually this denomination is in the spirit of Boltzmann and Planck who first made the connection between entropy and probabilities\,\cite{Planck_1914}.

\subsection{Irreversibility and H-theorem}\label{irrev2}

Let us examine the case of the free expansion of a perfect gas (\S\ref{irrev}) from an initial volume $\mathcal V_i$ to a final volume $\mathcal V_f$ at the equilibrium.
When a hole suddenly opens, the number of discernable allowed microstates increases from $\mathcal{W}_i = \mathcal V_i^N$ to $\mathcal{W}_f = \mathcal V_f^N$. The equilibrium is the state of maximum multiplicity described in the previous section. It corresponds to molecules uniformly distributed among the two compartments.
In agreement with thermodynamics, the difference of Boltzmann entropies between the final and initial equilibria is\,:
$$\Delta S = \ln(\mathcal{W}_f) - \ln(\mathcal{W}_i) = N \ln (\mathcal V_f /\mathcal V_i)$$ 
The entropy difference is positive.
So that it is tempting to say that it is growing with time during the process.
Unfortunately, the statistical entropy is only defined at the equilibrium, not during the transient stage.
In other words, the time-variable is present neither in Eq.\ref{S} nor in Eq.\ref{Sboltzmann}.

\begin{figure}[!htbp]
\begin{center}
	\includegraphics[width=1\linewidth]{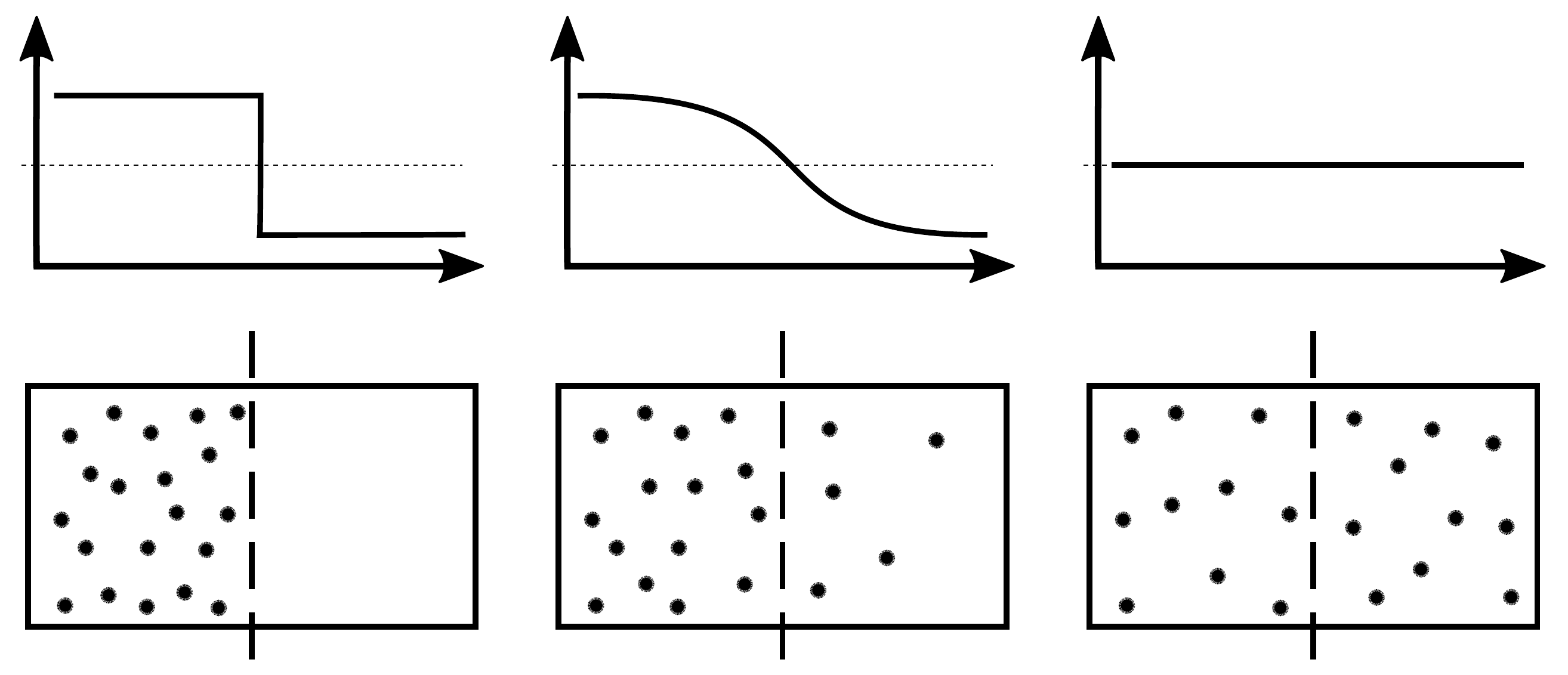} 
	\caption{Sequence of free expansion of a gas. The challenge of Boltzmann's H-theorem is to account for the time evolution of the density profile (top) and for the irreversibility of the process from the statistics of particles collisions (after \cite{Lebowitz_1993}).}
	\label{prof}
\end{center}
\end{figure}

For Boltzmann the challenge was to account for this time dependence and its irreversible character\,\cite{Boltzmann_Lectures} in the framework of the kinetics theory of gases developed by Maxwell. 
For that, let us consider the phase-space of one single particle (in $\mathbb{R}^{6}$) and divide it into elementary \textquote{volumes} of size $\mathfrak{h}^3$.
Denote $f_i(t)$  the probability of finding a particle in elementary \textquote{volume} $i$ at time~$t$.
Boltzmann introduces a quantity $H$, defined at any time, that writes
\begin{equation}\label{defH}
H = -\sum_i f_i(t)\ln f_i(t)
\end{equation}
\nomenclature{$H$}{H-function of Boltzmann}%
where the summation is done over all elementary volumes of the phase-space, to be compared to Eq.\ref{S} where it is done over microstates. It is important to emphasized that the probabilities involved in the two equations are not the same\,: statistical entropy deals with an energy distribution, whereas H-function deals with a density distribution.
Note that the original Boltzmann's H-function was defined without the minus sign. Whether or not Eq.\ref{S} and \ref{defH} should be presented with the same sign is a matter of taste. Here, this option has been preferred for reason of consistency with \S\ref{Shannon}.

By assuming a transport equation which basically conjectures the independence of random variables that govern a collision before it happens, but a post-collision fully deterministic and obeying the Newton's laws, Boltzmann gave evidence that $H$ always increases with time and is constant at the equilibrium\,:
\begin{equation}\label{H2}
\frac{\diff H}{\diff t}\ge 0
\end{equation}
He also showed that at the equilibrium, $f_i$ is uniform and that the corresponding velocity distribution of gas particles is the one of Maxwell.
Eq.\ref{H2} is known as the H-theorem.
Provided that, to a constant, the $H$-function can be identified with the Gibbs entropy,
it is potentially an extension of this notion to non-equilibrium and a demonstration of the second principle of thermodynamics from microscopic postulates. 

Beyond the mathematical proofs of Eq.\ref{H2} and of the equality $H=S$, that is an active research field\,\cite{Villani_2002}, mainly two objections have been raised under the form of two famous paradoxes. The first is known as the recurrence paradox (or Poincaré-Zermelo paradox).
In virtue of ergodicity, the probability that all molecules go back to the initial compartment is non-zero, but equal to the ratio $\mathcal{W}_i/\mathcal{W}_f=(\mathcal V_i /\mathcal V_f)^N$. This is actually very small for $N\simeq 10^{23}$, but nevertheless the thermodynamical irreversibility, that is the impossibility of returning to the original state without energy compensation, is turned into a very low probability. 
In practice, this should never happen, the interval of recurrence being so huge, but conceptually it means that $H$ should be recurrent (statistically periodical) rather than monotonically increasing. 
This problem of consistency comes down to\,:

\myparadox{Poincaré-Zermelo paradox}{}{Ergodic systems behave recurrently.}{The H-function is monotonically increasing.}{}
If we accept that finally there is no contradiction between these statements (i.e. it is a veridical paradox, see \S\ref{GBdiscuss}), then the paradox obliges us to regard probabilities $f_i$ involved in Eq.\ref{defH}, as \textit{a priori} probabilities or human expectations.

The second paradox is due to J. Loschmidt.
Open the hole allowing free expansion at time~0. After a certain time $\tau$, imagine that the direction of the velocity of each molecule is reversed, without changing its magnitude. The operation does not affect the macroscopic properties of the gas. But then, the gas goes backward through the same sequence of collisions than the previous one. So that at time $2\tau$ its original state is restored. Boltzmann gave different answers to this objection and not all of them are very clear (\cite{Uffink_2006}~\S6, for clarifications see \cite{Darrigol_2021}).
In fact, as we will see in \S\ref{demons}, Loschmidt's demon, capable of reversing velocities, is not the only one that seems to violate the second law of thermodynamics.
The exorcism will be the same.

The H-theorem intends to make the economy of the postulate of the second law of thermodynamics by deriving the irreversibility from microscopic and objective properties of matter. In fact, Boltzmann himself admits that \myquote{The one-sidedness of this process is clearly not based on the equations of motion of the molecules. For these do not change when the time changes its sign. This one-sidedness rather lies uniquely and solely in the initial conditions.} (L.~Boltzmann \cite{Boltzmann_Lectures}~\S87). 
Initial conditions about which nothing is known at the microscopic level, but which are only expected \textit{a priori} to obey certain laws, such as the principle of insufficient reason and its variations like probability independence, equiprobability or isotropy etc.
In brief, the postulate (i.e. the second law) has simply been displaced.

\subsection{Gibbs paradox \#1 and \#2}\label{GBphystat}

Let us now reexamine the mixing of two gases, which we know from thermodynamics is closely related to free expansion (\S\ref{mixing}).
First consider the difference of Boltzmann entropy between final and initial states in the case where the gas is the same in both compartments. Initially, the entropy of each compartment is $S_A=S_B=N \ln \mathcal V$, so that the entropy before the separation was removed is $S_i=S_{A\cup B}=S_A+S_B=2N \ln \mathcal V$. After the removing of the separation and at the equilibrium, one has $2N$ particles in a volume $2\mathcal V$, so that the final entropy is $S_f=2N\ln(2\mathcal V)$. The difference is
\begin{equation}\label{Smixstatphys}
\Delta S = 2N\ln 2
\end{equation}
If the two compartments are initially filled with two different gas species provided that their proportions are constant, a unit of volume as used in Eq.\ref{W1bis} can be still defined in	 average. So that the calculation would be the same and also the result. Quite simply because the equations does not permit to account whether the two gases are identical or not.
In both cases, the entropy of mixing is the same and non-zero.

The good point is that adopting Boltzmann entropy allows us to solve the Gibbs paradox \#1\,: the entropy of mixing is independent of the degree of similarity of the two gases, there is therefore no longer any discontinuity. The same or different gases are treated in exactly the same way.

The bad point is that it gives rise to a new paradox that can be stated as\,:

\myparadox{Gibbs paradox~\#2}{Mixing two identical gases}{leaves Clausius entropy unchanged.}{increases Boltzmann entropy.}{}
That is contradictory since both entropies are expected to be the same.
This paradox will find a simple solution in \S\ref{mixthi} and \S\ref{GBthi}.

\subsection{What statistical entropy is}

Statistical mechanics makes the bridge from microscopic to macroscopic levels in terms of probability distributions. It aims at accounting for equilibrium and irreversibility.
With statistical mechanics, the equilibrium is the most probable state.
Translated in terms of a thermodynamical observable, this amounts to say that\,:

\statement{Entropy is the thermodynamical quantity that is maximized at equilibrium.}

\noindent With the Boltzmann H-function, statistical mechanics clarifies the spreading metaphore introduced in \S\ref{whatis1}, that is to say specifies what should be spread out.

A mechanical analogy arises when using the free energy that embed the negative of entropy (negentropy). The free energy is a potential that is minimized at equilibrium. So that a system that is moved away from equilibrium experiences a restoring \textquote{entropic force}.
Interestingly, this force does not correspond to any microscopic potential.
Because entropy is not the property of a microstate but the property of an ensemble, the property of a distribution, the property of a macroscopic state.
Entropy is an emergent property.
A somewhat discordant point is that starting from the microscopic with the aim of explaining everything from there, statistical mechanics makes entropy appear as emergent and in return demonstrates that the attempt is in vain.
The bottom-up approach does not totally emancipate from the human and macroscopic scale.

To this last remark, we must add the arbitrariness of the premises, which ultimately all come down to the principle of insufficient reason and to the manner probabilities are regarded as \textit{a priori} expectations.
This principle has been criticized for a long time and the most disturbing argument is undoubtedly this one\,:
\myquote{It cannot be that because we are ignorant of the matter we know something about it.} (R.L. Ellis \cite{Ellis_1850}).
This is very difficult to refute, but Poincaré does it.
\myquote{You ask me to predict for you the phenomena about to happen. If, unluckily, I knew the laws of these phenomena I could make the prediction only by inextricable calculations and would have to renounce attempting to answer you; but as I have the good fortune not to know them, I will answer you at once. And what is most surprising, my answer will be right.} (H. Poincar\'e\,\cite{Poincare_1913} p.396). This is exactly the situation statistical mechanics finds itself. Starting from uncertain assumptions, it works wonderfully and gives us right answers. It allows us to retrieve all traditional thermodynamics by extending its results and its fields of application to chemistry, condensed matter physics and even more from biology to astrophysics.

The point is that statistical mechanics can do the same thing (and even more by solving certain paradoxes) but with even more economy of thought and a gain in self-consistency, by consenting to subjectivity from the start. This is what information theory does.

\section{Shannon entropy}\label{Shannon}

During the first half of the 20th century, with the increase of new communication media,  the need arose to formalize mathematically problems associated with the transmission and storage of information and more particularly to its quantification and compression in order to optimize their respective physical supports without any loss. Transmission and storage of information have a material cost that should be minimized. This was the primary goal of information theory, but the link with statistical mechanics was rapidly done.

\subsection{Information}

Pioneering work was done in 1928 by R.~Hartley\,\cite{Hartley_1928} who quantified the information in terms of the number of bits needed to encode it. Suppose a device (a source) that emits a message written by using $\mathcal{W}=32$ different symbols. As $32=2^5$, one needs $\log_2(32)=5$ bits per symbol in order to have a sufficiently large number of possible combinations. More generally
\begin{equation}\label{Hartley}
I = \log_2(\mathcal{W}) 
\end{equation}
is the number of bits needed to encode one symbol. 
However, in general Eq.\ref{Hartley} does not provide the most economical way to encode a message as frequencies of appearance of letters differ. For instance in english, the 6 vowels occupy in average 50\% of texts but would need only 3 bits to be encoded instead of 5.
By denoting $p_i$ the probability of occurrence of letter $i$,  the expected average number of bits to encode any letter would be instead\,:
\begin{equation}\label{Shannon0}
I = -\sum_{i=1}^\mathcal{W}  p_i \log_2(p_i)
\end{equation}
This is the more general expression proposed by C.~Shannon\,\cite{Shannon_1948} for the material cost of the information emitted by the source, that is to say the minimum average material cost per letter (bandwidth of the transmitter channel, storage space etc) needed to avoid any information loss.
For consistency with the previous section, let us use the natural logarithm and denote\,:
\begin{equation}\label{Shannon1}
\begin{array}{rl}
	S  &=  I \ln 2 \\
	&\displaystyle = -\sum_{i=1}^\mathcal{W}  p_i \ln(p_i),
\end{array}
\end{equation}
the Shannon entropy of the source.
It was called by Shannon \textquote{measure of information} and very often in the literature \textquote{quantity of information}. But this denomination has led some authors to point out some paradoxes. For instance,
\myquote{the entropy of the probability distribution for the location of my house keys increases when I discover that they actually are not, as I held to be very probable, in the pocket of my coat.} (J. Uffink \cite{Uffink_1995}). This is presented as an example of situation where the acquisition of information would cause an increase of entropy, which would be contradictory with the idea of a \textquote{measure of information}. Actually, in this example we try to replace \textit{a priori} probabilities by \textit{a posteriori} probabilities i.e. frequencies. But these need many occurrences to be meaningful. Doing the experiment (i.e. checking if the key is in the pocket) many times statistically solves the problem. Anyway the denomination \textquote{measure of information} is probably not the best\,\cite{Jaynes_1957b} because it concerns information that we do not have.
For this reason \textquote{information cost} or \textquote{measure of uncertainty} will be preferred.

Imagine we know nothing about the probability distribution of letters, except that there are $\mathcal W$ possibilities. In order to be sure not to lose any information emitted by the source, we are obliged to encode with an average number of bits given by Eq.\ref{Hartley}.
If later, we learn that messages are actually written in english, we can recalculate and lower the information cost, so that it remains optimum.
At the other extreme, imagine that we find that the source only emits entirely predictable characters. There is no need to encode, transmit or store the message at all. The information cost is zero.
So that the material cost corresponds in fact to a quantity of information we have not. 
The information cost is actually a measure of the uncertainty on the emission of the source. In all cases it is calculated by using \textit{a priori} probabilities, those that depend on our knowledge and are subjective.

\subsection{Uncertainty}

Information is something strange compared to physical quantities. For instance, you can received it only once, because an \textquote{information} you already know is no more an information. So that the value of an information depends on how surprising it is.

Imagine you enter in a casino in order to play a game of chance. You are not a regular. You just want to play once in a rational manner. You are accompanied by a diviner who is able to know in advance the result of games. However unfortunately he does not say anything except if you pay for that. So, what price is it reasonable to pay for this information\,? It depends on the game. It is easier to guess the outcome if you are playing a coin toss than if you are playing roulette. Imagine the outcome is written using a binary encoding and you buy to the diviner each bit of the outcome. This is a good basis to calculate the maximum price for the information. For instance, you play to a roulette wheel with 32 numbers and no number reserved to the bank, so that the odds of a bet is 32/1 (you bet 1\,{\euro} and gain 32\,{\euro} if you win, otherwise you loose 1\,{\euro}). The game is thus equilibrated 
and the expected gain is 0.
If you buy 1~bit of information to the diviner, your risk is divided by 2 but the odds remains unchanged, so that your expected gain is 1\,{\euro}. This is the maximum price that it would be reasonable to pay for 1 bit of information. The total cost of the information is given by Eq.\ref{Hartley} or \ref{Shannon0}, and 
Eq.\ref{Shannon1} expresses this cost taking $\ln 2$ as currency unit.
The more uncertain the outcome, the more expensive it is to know it. 
Dividing this uncertainty by a factor 2 costs $\ln 2$.
Shannon entropy $S=I\ln 2$ is the total cost to remove all uncertainty, or a measure of this uncertainty.

In the above example also, the manner the best price is calculated is subjective and depends on our knowledge about the game. The roulette may be rigged or the die may be loaded, but \textit{a priori} it would not be rational to assume so.
Interestingly the way of regarding Shannon entropy either as a material cost for the information we do not have or as an uncertainty, automatically places us in the position of regarding probabilities as subjective expectations that depend on our knowledge.
Unlike statistical mechanics, there is no more ambiguity.

\subsection{Maximum-entropy theorem}

Shannon\,\cite{Shannon_1948} showed that an expression of the form $-\sum p_i \ln p_i$ is the only one, up to a factor, that is\,: 
\begin{enumerate}[label=\arabic*), nosep]
\item positive and continuous in $p_i$.
\item monotonically increasing in $1/p_i$, that is to say increasing with uncertainty on event $i$.
\item additive over independent sources.
\end{enumerate}
While the expression of the H-function (Eq.\ref{defH}) seems to come from nowhere except from the genius of Boltzmann, Shannon entropy is the only measure of uncertainty which has the correct mathematical properties required by the role it is supposed to play.
Now, Boltzmann's choice makes sense.

Suppose we are dealing with a source about which we have only a partial knowledge of the true probability distribution $p(x)$ of the emitted outcomes $x$.
On the one hand, we want to pay the minimum hardware cost for the information, but on the other hand, we do not want to lose any of it.
Solving the problem consist in finding which probability distribution $p(x)$ maximizes the uncertainty while being consistent with our partial knowledge, i.e.
by searching which function $p(x)$ maximizes Shannon entropy while obeying certain constraints.
This point being demonstrated mathematically, it has the status of a theorem (and not that of an axiom) that can be stated as\,: 

\law{Maximum-entropy theorem}{the only distribution $p(x)$ that maximizes the uncertainty on $x$ while being consistent with our knowledge is the one that maximizes Shannon entropy.}

\noindent The best distribution can be found by the method of Lagrange multipliers\,\cite{Shannon_1948, Jaynes_1957}. If the only thing we know about $p(x)$ is that it has a finite support, then the best choice is the uniform distribution. If $p(x)$ is only known to have a positive support and a finite expected value, then the best choice is the exponential. It $p(x)$ is only known to have a standard deviation, then the best choice is the gaussian.
All these results follow from the maximum-entropy theorem of Shannon.

\subsection{Maximum-entropy principle}

In statistical mechanics, we investigate problem involving a source (a system) which outcomes are the microstates, the position $x$ in the phase-space, with an unknown distribution $p(x)$.
To move towards solving our problem, however, we need to propose something, then start working with it and see where it leads us with respect to experiments, and eventually come back and change the starting point.
This is a common procedure in science that should not be a problem for anyone.
So here, we are wondering what is the best choice of probability distribution to start. 
The choice must not introduce more information than we have \textit{a priori}, so that it must maximizes the uncertainty, that is to say Shannon entropy, with the constraint to fit with our knowledge\,\cite{Jaynes_1957}. 
Doing so, it is possible to recover the microcanonical distribution for isolated systems, the canonical distribution for thermalized systems or the gaussian distribution that governs the fluctuations of internal variables at equilibrium.
A good thing is that after that, everything is in agreement with the experiments (as with the usual procedure in statistical mechanics) and no return to the starting point is necessary.

If the result is the same as with the usual procedure of statistical mechanics, what is the point of using the maximum-entropy theorem\,? The advantages are exposed by Jaynes\,\cite{Jaynes_1957}.
They are basically of two kinds\,: 1)~a gain of self-consistency; 2)~an economy of thought.

The classical foundations of statistical mechanics are the definition of equilibrium, the hypothesis of ergodicity, the hypothesis of independence and the fundamental postulate (\S\ref{equilibrium2} to \S\ref{fp}).
Let us consider the refoundation from the only postulate that the equilibrium is the state which maximizes the uncertainty on the actual microstate of the system. Thus, according to the maximum-entropy theorem this postulate becomes\,:

\statement{The equilibrium is the state that maximizes Shannon entropy of the distribution of microstates.}

\noindent The only requirements for the foundations of a theory are of two orders\,: 1)~self-consistency; 2)~\textit{a posteriori} validation with confrontation against experiments. 
With respect to the latter, i.e. agreement with thermodynamics, \textit{classical} or \textit{maximum-entropy} foundations are equivalent. 
But regarding to the former the maximum-entropy postulate is much better because it deals with subjective probabilities, or \textit{a priori} expectations, those that are evaluated without any precise idea of the actual normalization factor that is needed for their calculation. With subjective probabilities this normalization factor is \textit{a priori} estimated from symmetry arguments or what seems reasonable given our knowledge of the problem. We know that it is with no consequence as long as we are concerned by ratios of probabilities or difference of entropies (the normalization factor vanishes). 
But now we have no problem of self-consistency related to the assumption of metric transitivity\,\cite{Jaynes_1957}. 

Self-consistency is not the only advantage. By using the maximum-entropy postulate, the expression of the statistical entropy (Eq.\ref{S}) is totally freed from thermodynamics.
So that it is potentially valid out-of-equilibrium and for other probability distributions than that of microstates.
But here a problem arises.
Consider for instance a random variable $x\in[0,\pi]$ with a uniform distribution, $\sin(x)$ is also a random variable which distribution is not uniform but has a maximum for $x=\pi/2$. Thus, using the maximum-entropy criterion for $x$ or $\sin(x)$ can lead to contradictory results. 
Jaynes\,\cite{Jaynes_1973} outlines a crucial point. First, in the maximum-entropy problem, the solution (the distribution) we are looking for is implicitly supposed to be unique, as is the equilibrium state. This simple assumption actually automatically brings to our knowledge others crucial informations\,: the solution is not supposed to depend on the orientation of the observer (invariance under rotation), nor on its position (invariance under translation), nor on the scale it is considered (invariance under dilatation). Among all the possible variable describing a system, considering only those whose distributions are \textquote{invariant in form} under these transformations avoids all inconsistent results. 
In other words, applying the principle of maximum-entropy to any of the distributions satisfying these properties of invariance will provide the same result. So that the maximum-entropy principle can be finally stated as follows\,:

\law{Maximum-entropy principle}{the equilibrium is the state that maximizes Shannon entropy of variable-distributions whose form is invariant under rotation, translation and dilatation.}

\noindent This statement holds for a definition of the equilibrium (zeroth principle). In addition, it exempts us from the three other postulates of the classical foundation.
Lastly, it holds for the energy distribution of microscrostates (allowing to recover thermodynamics), but also for the phase-space density of particles involved in the H-function that is equally invariant in form under rotation, translation and dilatation. Thus, with the Boltzmann's H-theorem (Eq.\ref{H2}), the maximum-entropy principle can exempt us from the second law of thermodynamics and opens the door to out-of-equilibrium theory.
These two statements form the most economical kernel for thermodynamics and statistical mechanics. 

The conceptual advantages brought by the maximum-entropy principle are far from being recognized by everyone. In particular, it is argued\,\cite{Uffink_1995} that these advantages are only truly achieved if one adopts a generalization of Shannon's entropy  as a measure of uncertainty (see e.g. \cite{Arndt_2001} for an exhaustive review of generalized entropies). This field is so vast that it goes beyond the purpose of this paper. However, let us briefly outline and discuss some arguments, especially those stated in ref.\,\cite{Uffink_1995}. In my opinion, these are of two orders. The first is technical\,: the invariance constraints required for the probability distributions, as Jaynes argued, to be enforceable must involve continuous probability densities rather than discrete probabilities. This lead to some mathematical difficulties related to the choice of a unit for the density (probability densities are not adimensional and cannot be the argument of a logarithm) and to the divergence of the resulting entropy. Treatment of this point can be found in the book of Ben-Naim and Cassadei\,\cite{Ben-Naim_2016}. Basically, it amounts to renormalize the variable (i.e. the argument of the probability density) by the resolution, that is always finite in a physical world, in a way that is finally quite similar to what was done in \S\ref{statmech}.
The second argument is conceptual and concerns the justification of the principle, in particular the justification of the uniqueness of the solution argued by Jaynes as a starting point to deduce invariance requirements. First principles of a theory do not require to be \textit{a priori} justified. They form the \myquote{not rationally deducible part of the theory}\,\cite{Einstein_1934}. The only justification is the \textit{a posteriori} agreement with experiments and the economy of thought it can provide. To seek a justification would amount to seek a primary cause to an infinite causal chain. However, regarding the \textquote{uniqueness}, this should not bother physicists concerning a principle that ultimately serves as the definition of equilibrium.

\subsection{Information and energy}\label{demons}

Economy and consistency should be sufficient justifications for adopting and including information concept at the foundations of the statistical mechanics theory.
But these arguments will be further strengthened by establishing a direct link between information and energy.
This connection is made by what is called thermodynamic demons.

The family of thermodynamical demons\,\cite{Rex_2017, Ciliberto_2018} was born with the temperature-demon of Maxwell\,\cite{Maxwell_1872}.
Imagine a gas in an insulating container separated in two parts $A$ and $B$ along the $x$-axis by a thermally insulating wall having a small door. A demon is able to measure the velocity component $v_x$ of molecules; determines the median value $v^*$ for its modulus and open the door to allow passage only to those which satisfy the condition $(v_x>v^*>0) \textrm{ OR } (-v^*<v_x<0)$: swifter molecules can only pass from $A$ to $B$ whereas slower ones from $B$ to $A$. This results in a temperature difference between the two compartments, which can eventually be used for running a thermodynamic cycle and producing work.

A simplified version of this demonic device is the pressure-demon that reduces the condition to $(v_x>0)$: molecules whatever their speed can only pass from $A$ to $B$. This results in a pressure difference between the two compartments, which can be used for producing mechanical work.
Alternatively, in this simplified version the demon can be replaced by a concrete device, either by a one-way valve as proposed by Smoluchowski\,\cite{Rex_2017}, or by a ratchet-pawl mechanism\,\cite{Feynmann_Ratchet}, or by an electric diode and the gas particles by electrons\,\cite{Brillouin1950}, then if the two compartments communicate by an additional way, the device is expected to rectify thermal fluctuations and produce a net current of particles, which here again can deliver useful energy.

All these devices decrease the entropy of a system, which without cost in energy would  violate the second principle of thermodynamics. In reality, for the last two concrete devices it has been experimentally shown that they can work (even with a poor efficiency), provided that the rectifier (pawl or diode) is cooled at a lower temperature than the rest of the system\,\cite{Bang:2018we, Gunn:1969wr} in exchange for the entropy decrease. Thus, the only manner to understand how the first two demonic-devices can work in agreement with the second principle is to admit that the quantity of information needed by the demon has a cost in energy and therefore to admit an equivalence between Clausius and Shannon entropies. 
In fact, since the formula for the Shannon entropy is the same as Gibbs and Boltzmann ones, this equivalence was already done via statistical mechanics. But here, it is direct.

However, the problem with these demonic-devices is that the quantitative correspondence between the velocity measurement and the information cost is not clear.
Clarification was done by Szilard\,\cite{Szilard_1964} who proposed an even more simplified version of the Maxwell demon using only one molecule in a total volume $2V$. This time the demon does not care about the velocity, but is just able to detect in which compartment is the molecule. 
At his convenience, he is able to install a piston that encloses the molecule within the half-volume $V$ of his choice.
Doing so, the demon divides the uncertainty on the position of the molecule by a factor 2, and the entropy is decreased by $\ln 2$. The system can return to the original state by a reversible isothermal expansion that provides to the surroundings a work equal to $\ln 2$. The overall cycle is consistent with thermodynamics.

\begin{figure}[!htbp]
\begin{center}
	\includegraphics[width=1\linewidth]{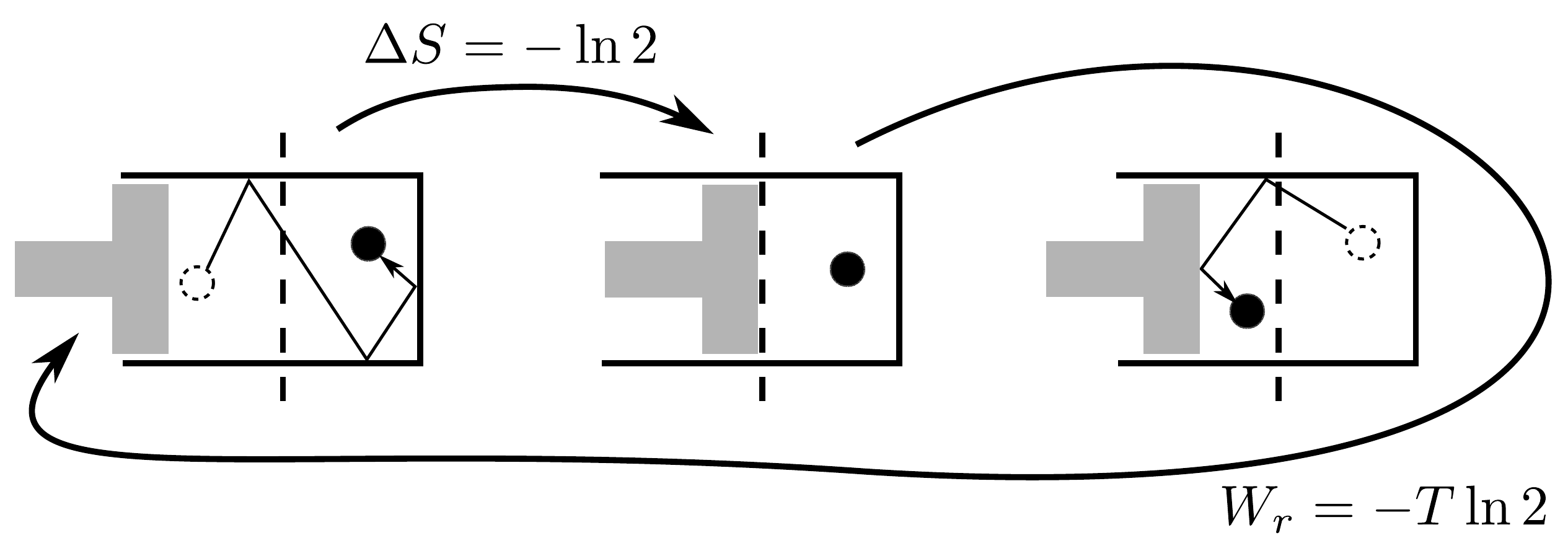}
	\caption{Szilard demon installs a piston when the molecule is in the suitable compartment, allowing the device to subsequently produce work.}
	\label{szilard}
\end{center}
\end{figure}

Let us now consider the Loschmidt paradox concerning the reversibility. 
Loschmidt imagines that after a free expansion has been occurred, the sign of the velocity of the $N$ molecules is reversed, so that their trajectories also, having for consequence that the system recovers its original state. The corresponding decrease in entropy can be now understood in term of information cost.
Consider one molecule, prior to the reversal of the velocity, its sign (with respect to an arbitrary direction) is unknown .
So that the reversal supposes that we measure it, then do the reversal.
Thus, after the operation the uncertainty on the sign is divided by 2, and the entropy by $\ln 2$. For $N$ molecules the entropy is decreased by $N\ln 2$.
As with the Szilard device, the system can provide an equivalent amount of work by isothermal expansion.

\subsection{Mixing and traceability}\label{mixthi}

As explained in \S\ref{GBthermo}, mixing two perfect gases at the same temperature and pressure, whatever their difference, never comes with any thermodynamical effect (no work and no heat exchanged). So that the mixing entropy can only be determined when the system is guided to its original state using a reversible path. A mixing-unmixing cycle must be considered. Also, the notion of cycle is meaningful only when performed repeatedly and reproducibly.
The first cycle of a series cannot be regarded as belonging to such a stationary regime, it must be at least the second for that.
Suppose the gas made of two different isotopes in same proportion. In thermodynamics, it is quite common that we do not care about that, so the gas is regarded as a simple compound. The gas fills two compartments $A$ and $B$ of same volume $V$ with a separation. Consider three cases (Figure~\ref{mixcycle})\,:

\begin{enumerate}[label={\arabic*)}]
\item A first kind of cycle consists in just moving up and down the separation. Denote $t_0$ the time just before the first cycle starts.
After the first cycle is ended, all informations about the exact contents of $A$ and $B$ at time $t_0$ is lost.
At the end of any cycle when the separation is down, the number of molecules per compartment is always $N\pm\sqrt{N}$ because of their random repartition (see \S\ref{max1}).
Also, any information about the \textquote{home-compartment} of molecules is lost after $t_0$. 
So that the uncertainty concerning these two features, exact number of molecules and traceability, is unchanged by further cycles, and so the Shannon entropy.
This is the ordinary case in thermodynamics where the Shannon entropy of mixing is zero\,:
\begin{equation}\label{Smix0}
	\Delta S_{1}= 0
\end{equation}

\item Imagine that we know for certain that  at $t_0$ the two compartments had exactly the same number of molecules $N\pm0$ and that we do not want to lose
this information.
Previous cycles are not satisfactory because the random repartition of molecules gives $N\pm \sqrt N$. To evaluate the corresponding loss of information, let us remove all molecules and put them in a separate box. Take iteratively one pair $(a,b)$, put one molecule (either $a$ or $b$) in $A$ and the other in $B$. After $N$ iterations, the two compartments have exactly the same number of molecules. 
There are four possibilities to arrange $a$ and $b$ in two boxes\,:  $\{ab| , a|b, b|a, |ab\}$, and only $a|b$ or $b|a$ are convenient. So that the number of possibilities is divided by 2 for each pair. For $N$ pairs it is divided by $2^N$ and the entropy decreases by $N \ln 2$. Finally, if we consider that the initial state of the mixing-unmixing cycle is the state where the two compartments have exactly the same number of molecules, then the mixing entropy is\,:
\begin{equation}\label{Smix1}
	\Delta S_{2} = N \ln 2
\end{equation}
Note that the pairwise procedure can be stopped at any iteration, if we are satisfied by the uncertainty on $N$ would lead a random repartition of the rest of molecules. So that depending on our wish, the entropy of mixing can take any value from 0 to $N \ln 2$ by step of $\ln 2$.

\item In the third case we know for certain that at $t_0$, compartment $A$ was filled with isotope $a$ and compartment $B$ with isotope $b$. So that, we are not satisfied by the previous unmixing and want to restore exactly the original state. In other words, we want to preserve the traceability.
To achieve this, among the two possibilities $\{a|b, b|a\}$ in the previous procedure, we must choose $a|b$. Here again the number of possibilities in divided by 2 for each pair. So that at the end, compared to the previous state the entropy has decreased by an additional amount $N\ln 2$. Finally, if we consider traceability as crucial the Shannon entropy of mixing is\,:
\begin{equation}\label{Smix2}
	\Delta S_{3} = 2N \ln 2
\end{equation}
Here again, the pairwise procedure can be stopped at any moment so that the entropy of mixing can take any value from 0 to $2N \ln 2$ by step of $\ln 2$.

\end{enumerate}
Depending on our knowledge about the original state or depending on what we consider as being important about it, the mixing-unmixing cycle differs and the Shannon entropy of mixing too.

\begin{figure}[!htbp]
\begin{center}
	\includegraphics[width=1\linewidth]{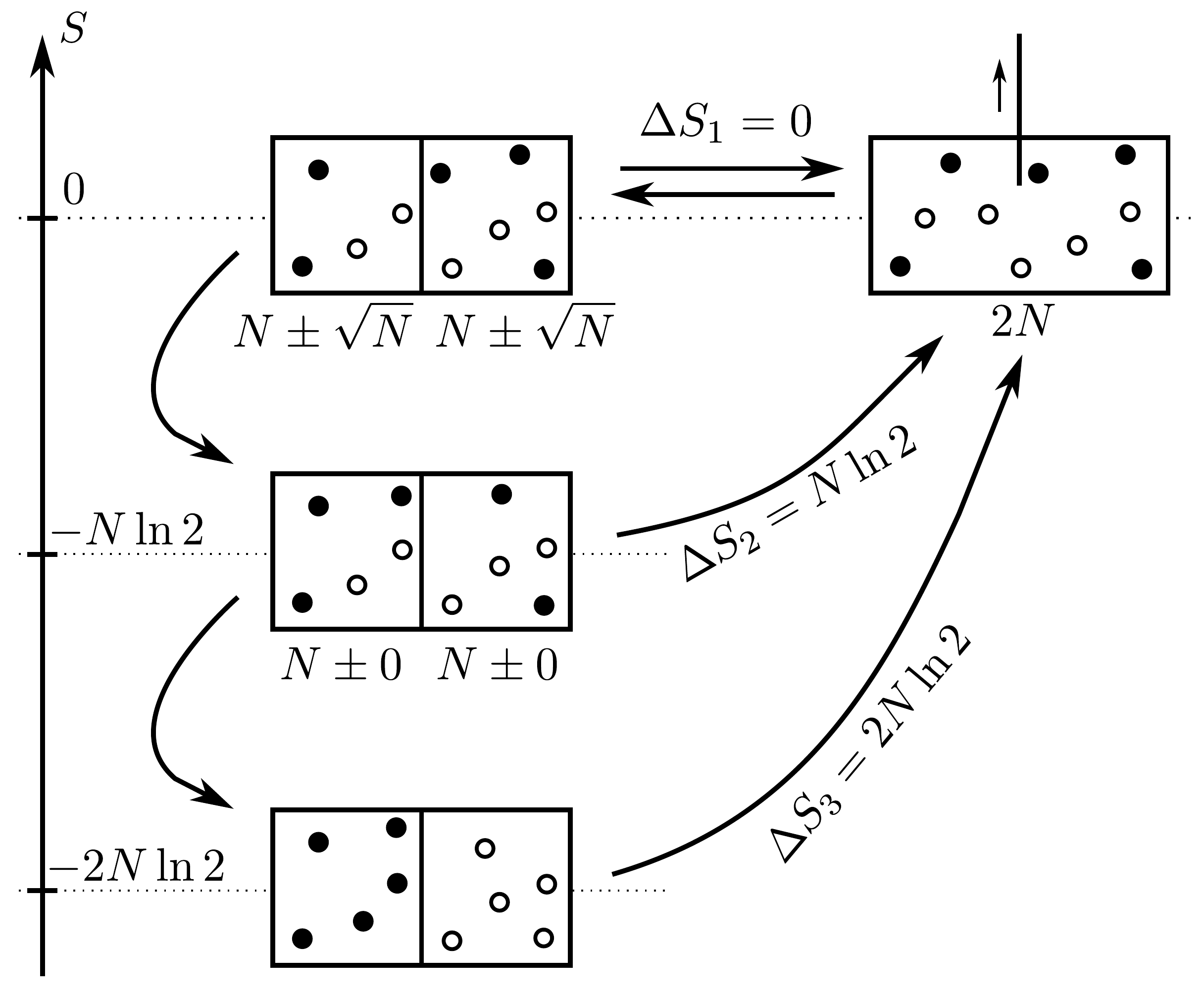}
	\caption{Mixing-unmixing cycle of a gas made of two species. The cycle, and so the mixing entropy, depends on what we consider as being the \textquote{correct} initial state, i.e. depends on the information we had and do not want to lose.}
	\label{mixcycle}
\end{center}
\end{figure}

Suppose the gas is replaced by one made of molecules, identical or not, carrying a label, let us say something like a serial number to allow their traceability.
We find all the cases above for the entropy of mixing according to our degree of requirement. So that the question is not whether the molecules constituting the gas are identical or not, but whether they are traceable or not.
For traceability, the serial number is not realistic, but filming the trajectories of particles is much more so.
In the 19th century and in the first half of the 20th it was not conceivable, in the second half of the 20th it was not yet possible, but today it is. Today, with the progresses of time-resolved transmission electron microscopy performed in liquid cell\,\cite{Pu_2020}, it becomes possible to record the trajectories of particles undergoing a random motion. Not yet for atoms, but already for nanoparticles or colloids in solution. The number of trajectories that can be recorded is simply a question of computing power. Our corresponding perception of what is mixing clearly depends on our knowledge.

\subsection{Gibbs paradox \#1 and \#2}\label{GBthi}

Let us first deal with the Gibbs paradox \#2 (\S\ref{GBphystat})\,: Clausius and Boltzmann entropies behave differently when mixing two volumes of the same gas.
In the light of information theory, it just appears to be a matter of resolution at which mixing is examined. Thermodynamics works at a macroscopic scale leading to Eq.\ref{Smix0}, whereas statistical mechanics envisions the problem at a microscopic level leading to Eq.\ref{Smix2}. The two points of view are not contradictory, they just do not examine the same thing.

The Gibbs paradox \#1 (\S\ref{GBthermo}) is concerned with a thought experiment where the degree of dissemblance of the two compartments can varies continuously, but entropy is either 0 or $2N\ln 2$.
In the example of Figure~\ref{mixcycle}, this degree of dissemblance is either the accuracy within which the number of molecules of the two compartments has been measured as being the same (standard deviation), or the initial degree of isotopic purity. This degree of dissemblance is the resolution at which the mixing is probed.
It is supposed to be improvable as we wish.
It was shown that the corresponding entropy varies accordingly,
not continuously because of the atomic structure of matter, but at least gradually between the two extreme values 0 and $2N\ln 2$.
So that, here again the paradox is a question of resolution at which the problem is examined. There is neither a unique physical entropy, nor two entropies (micro versus macro\,\cite{Dieks_2013}), but as many as we want.

Considering a mixing-unmixing cycle as depending on information we have, allows the Gibbs paradoxes \#1 and \#2 to be solved. Whereas separately, thermodynamics or statistical mechanics solves only one of the two.

\subsection{What Shannon entropy is}

Shannon entropy is the measure of uncertainty related to any given distribution of probabilities, which are this time clearly understood as subjective and depending on our knowledge.
The statistical entropy of Gibbs or Boltzmann, and consequently also the entropy of Clausius, are special cases of Shannon entropy, i.e. the Shannon entropy of the probability distribution of microstates.
The H-function of Boltzmann is also a special case, it is the Shannon entropy of the density of particles in the phase-space.

The most important point for statistical thermodynamics is that a reformulation of the theory is feasible that consists in replacing a statement like\,:

\statement{Entropy is the thermodynamical quantity that is maximized at equilibrium,}

\noindent that amounts to define entropy from the equilibrium, by the reverse statement\,:

\statement{The equilibrium is the state for which probability distributions (whose form is invariant under rotation, translation and dilatation) maximize their entropy.}

\noindent This reformulation\,: 1)~is more economical (it serves as zeroth law, it substitutes to the fundamental postulate and it makes the economy of the second law); 2)~allows a gain of self-consistency (no reference to metric transitivity); 3)~solves \textit{de facto} all paradoxes that have been confusing people for a very long time.

\section{Discussion}

In this section we look at what I believe are the two main sources of confusion about entropy\,: 1) the axiomatic thermodynamics; 2)~the usual way the Gibbs paradox is solved.

For this, I propose to evaluate them with respect to the following criteria\,:
\myquote{The basic concepts and laws which are not logically further reducible constitute the indispensable and not rationally deducible part of the theory. It can scarcely be denied that the supreme goal of all theory is to make the irreducible basic elements as simple and as few as possible without having to surrender the adequate representation of a single datum of experience.} (A.~Einstein\,\cite{Einstein_1934}). That is to say\,: 1)~Consistency; 2)~Economy; 3)~Agreement with experiments\,\cite{Mach_1911, Duhem_1906}.

\subsection{Axiomatic thermodynamics}

According to our evaluation criteria, the phenomenological approach is as respectable as the others. With the non-negligible advantage of being more intuitive because of its direct link to our experiences and senses.
Despite this, an axiomatic reformulation is often judged more elegant. 
For thermodynamics it can be done by putting laws listed in \S\ref{sndlaw} at the foundations and deriving all the rest by logical deductions.
With the same ingredients, the agreement with experiments and the economy of though will be the same. By axiomatic thermodynamics, I do not mean this one but rather the following.

In an axiomatic framework, the deductive work is much easier if the starting axioms are already written in terms of certain mathematical properties that the function $\mathcal S(U,V,N)$ is supposed to have.
So that alternatives are proposed in the literature.
Callen's book\,\cite{Callen_1985} is probably the first in this vein and has inspired many authors (e.g. \cite{Swendsen_2017}). Callen, postulates $\mathcal S(U,V,N)$ as being\,:
\begin{enumerate}[nosep]
\item continuous and differentiable.
\item monotonically increasing with $U$.
\item maximum at the equilibrium.
\item additive over its subparts.
\item homogeneous first-order.
\item zero at $T=0$.
\end{enumerate}
The last item is known as the Nernst's postulate and will come into play in \S\ref{disting}.
Item~5 is not explicitly postulated by Callen, but it is claimed to be a direct consequence of additivity (\cite{Callen_1985} bottom of p.28). As we will see, this claim is not exact\,\cite{Touchette_2002} and thus this property should be regarded as a postulate.
An homogeneous first-order function $\mathcal S(U,V,N)$ is such as\,:
\begin{equation}\label{extensivity}
\forall \alpha\in \mathbb{R}, \quad	\mathcal S(\alpha U, \alpha V, \alpha N) = \alpha \mathcal S(U, V, N) 
\end{equation}
This scaling property is called \textquote{extensivity} in thermodynamics.
In fact, there is a frequent confusion (e.g. \cite{Callen_1985} p.10) between additivity (over subparts) and extensivity (scaling). Let us first clarify this point.

\subsubsection{Additivity vs. extensivity}\label{addvsext}

Let  $A$ and $B$ be two disjoint subsets ($A\cap B=\varnothing$) of a system.
Entropy is additive if\,:
\begin{equation}\label{add1}
S_{A\cup B} = S_A + S_B
\end{equation}
Here, the comparison of the left-hand side and the right-hand side terms is done all things being equal. The union of the two subsets is not supposed to modify them. 
That is to say, if $A$ and $B$ correspond to two sub-volumes, or containers, $A\cup B$ is a mathematical operation that does not mean the fact of joining physically their contents. For instance, if $A$ and $B$ contain different gases, the union of $A$ and $B$ is not supposed to mix them.

Now, envisage a system made of two subparts $A$ and $B$, separated by a wall,
with additive state-variables $(U_A,V_A,N_A)$ and $(U_B,V_B,N_B)$, such as
$$
\begin{array}{c}
U_A+U_B=U,\quad V_A+V_B=V,\quad N_A+N_B=N\\
\textrm{and}\quad
\alpha=U_A/U=V_A/V =N_A/N
\end{array}
$$
If entropy is extensive, using Eq.\ref{extensivity} one can write
$$
\begin{array}{c}
\mathcal S(U_A,V_A,N_A)=\alpha \mathcal S(U,V,N)\\
\mathcal S(U_B,V_B,N_B)=(1-\alpha) \mathcal S(U,V,N)\\	
\end{array}
$$
So that\: 
\begin{equation}\label{ext20}
\mathcal S(U,V,N) = \mathcal S(U_A,V_A,N_A) + \mathcal S(U_B,V_B,N_B)
\end{equation}
If we remove the wall and join physically the contents of $A$ and $B$, we obtain a new system, denoted $J(A,B)$, different from $A\cup B$, with state-variables $(U,V,N)$. 
So that  from Eq.\ref{ext20}, if $\mathcal S$ is extensive then\,:
\begin{equation}\label{ext2}
\mathcal S_{J(A,B)} = \mathcal S(U_A,V_A,N_A) + \mathcal S(U_B,V_B,N_B)
\end{equation}
The similarity of Eq.\ref{add1} and \ref{ext2} is probably responsible for the confusion. Additivity and extensivity are different properties. For instance, one can easily check that the Boltzmann entropy $S=N\ln(\mathcal V)$ is additive, but not extensive. Clausius entropy is additive but nothing is said about extensivity because $N$ is not a true variable. If $S$ is additive and extensive then 
\begin{equation}\label{jcup}
S_{J(A,B)}  =  S_{A\cup B}
\end{equation}
In axiomatic thermodynamics, this equality is always hidden behind the postulate of extensivity.

\begin{figure}[!htbp]
\begin{center}
	\includegraphics[width=1\linewidth]{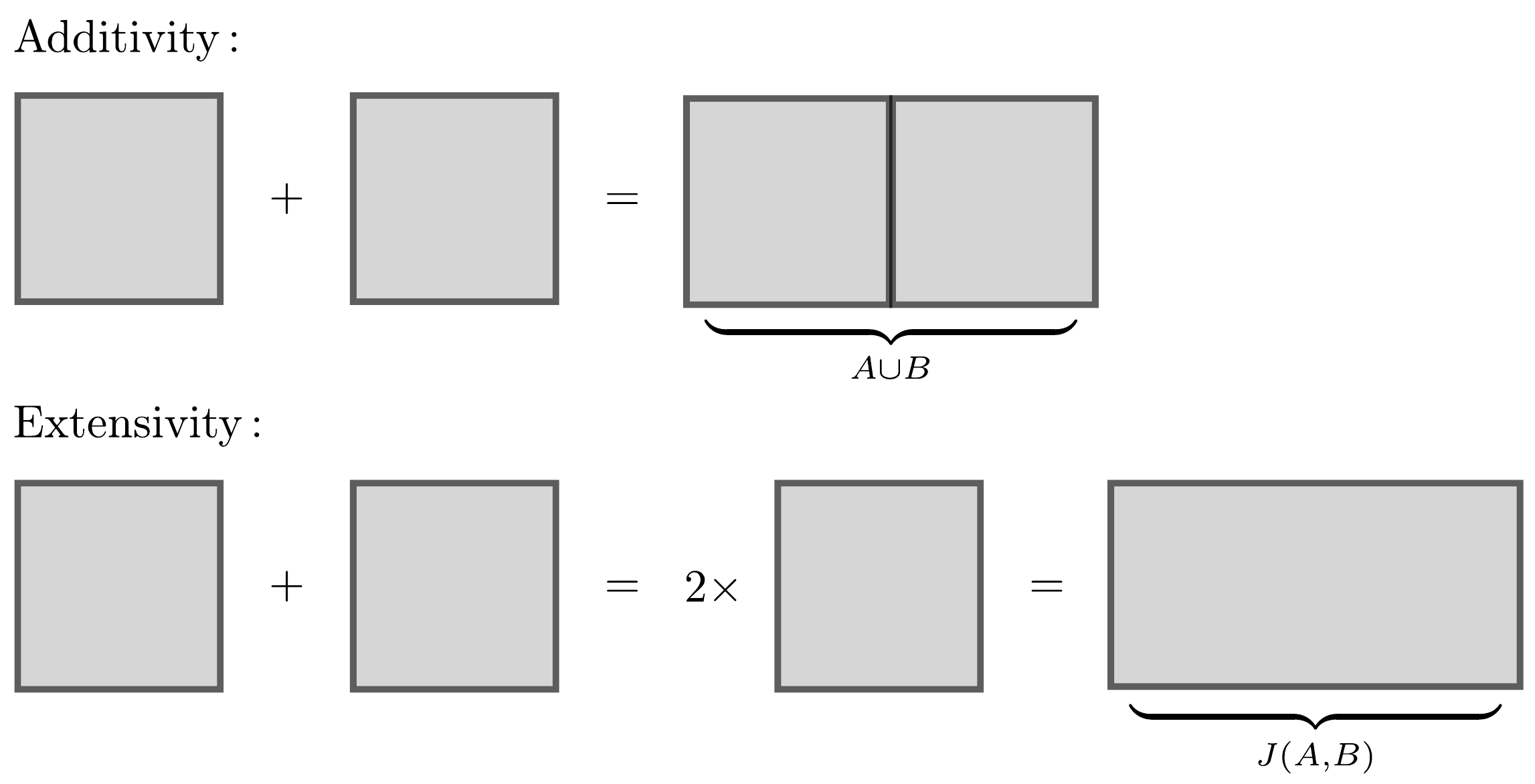}
	\caption{Additivity versus extensivity, here for two identical subsystems $A$ and $B$.}
	\label{mixcycle}
\end{center}
\end{figure}

\subsubsection{Extensivity of entropy}\label{extent}

The property of extensivity defined by Eq.\ref{extensivity}, which entropy is supposed to have, leads in particular for $\alpha=1/N$ to the equality
\begin{equation}\label{Smolar}
\mathcal S(U, V, N) =N \mathcal S(U/N, V/N, 1)
\end{equation}
By definition $V$ is extensive. Usually, if surface effects can be neglected and if interactions are only short-range, the internal energy $U$ can also be assumed extensive. So that $u=U/N$ and $v=V/N$ are internal energy and volume per particle, respectively. Allowing to write\,:
\begin{equation}\label{Smolar2}
\mathcal S(U, V, N) =Ns
\end{equation}
where $s=\mathcal S(u, v, 1)$ is the entropy of a single particle (\cite{Callen_1985} p.29). 
This is indeed a very nice property  allowing great mathematical simplifications of many problems (via Euler and Gibbs-Duhem equations). 
It is important to note that these simplifications are not conceptual but only technical. So that they cannot be regarded as bringing any economy of thought.

The physical meaning of extensivity is that it allows us to define mathematically a size for the system by only one parameter (for instance either the area or the volume but not both, except if their ratio is constant). So that, in the case of entropy, it would be proportional to this size.
Clearly, this is not the case for systems where the interplay between surface and bulk is a key feature\,: adsorption, heterogeneous chemistry, porous media, membranes, nanosciences, etc (note also that whether the surface can be neglected or not with respect to the bulk depends on the range of interactions).
So that basing the theory on the postulate of extensivity of entropy excludes them \textit{de facto} from its fields of application\,\cite{Addison_2001}.

The last point about the postulate of extensivity in thermodynamics is that it gives rise to a problem of consistency with statistical mechanics. For a perfect gas, the entropy of Eq.\ref{SN} does not fulfill the property of Eq.\ref{Smolar}, an expression of the form 
\begin{equation}\label{formext}
S=N\ln(\mathcal V/N) + \textrm{constant}
\end{equation} 
should be rather expected. This gives rise to the following problem of consistency\,:

\myparadox{Conflict of extensivity}{}{Thermodynamic entropy is postulated as being extensive.}{Boltzmann entropy is not extensive.}{}

\noindent It is usually solved by modifying the way of calculating Boltzmann entropy in order to obtain an expression like Eq.\ref{formext}.
To avoid the arbitrariness of this modification, that would otherwise amount to add a supplementary item in the foundation of statistical mechanics, Eq.\ref{formext} is justified with an argument named \textquote{correct Boltzmann counting} in close relation with the problem of mixing discussed in \S\ref{GBdiscuss}. For this reason, this conflict is often presented as being a paradox of this family. However, it originates neither from phenomenological thermodynamics, nor from the problem of mixing. It is solely introduced by the choice of axioms (so that the simplest solution is to change the axioms since alternatives exist) for this reason the term \textquote{paradox} is avoided and \textquote{conflict} preferred.

\subsubsection{Maximum of entropy}

As already mentioned in \S\ref{equilib} the idea that \textquote{the equilibrium is the state that \textquote{maximize} something with respect to something else} is tempting. 
It is reinforced by statistical mechanics for which it is the probability, and therefore the entropy, that is maximized with respect to internal variables (see \S\ref{max1}).
Phenomenological thermodynamics does not allow such a result to be directly obtained, because probabilities and fluctuations are absent and only entropy of systems at equilibrium can be calculated.

\begin{figure}[!htbp]
\begin{center}
	\includegraphics[width=1\linewidth]{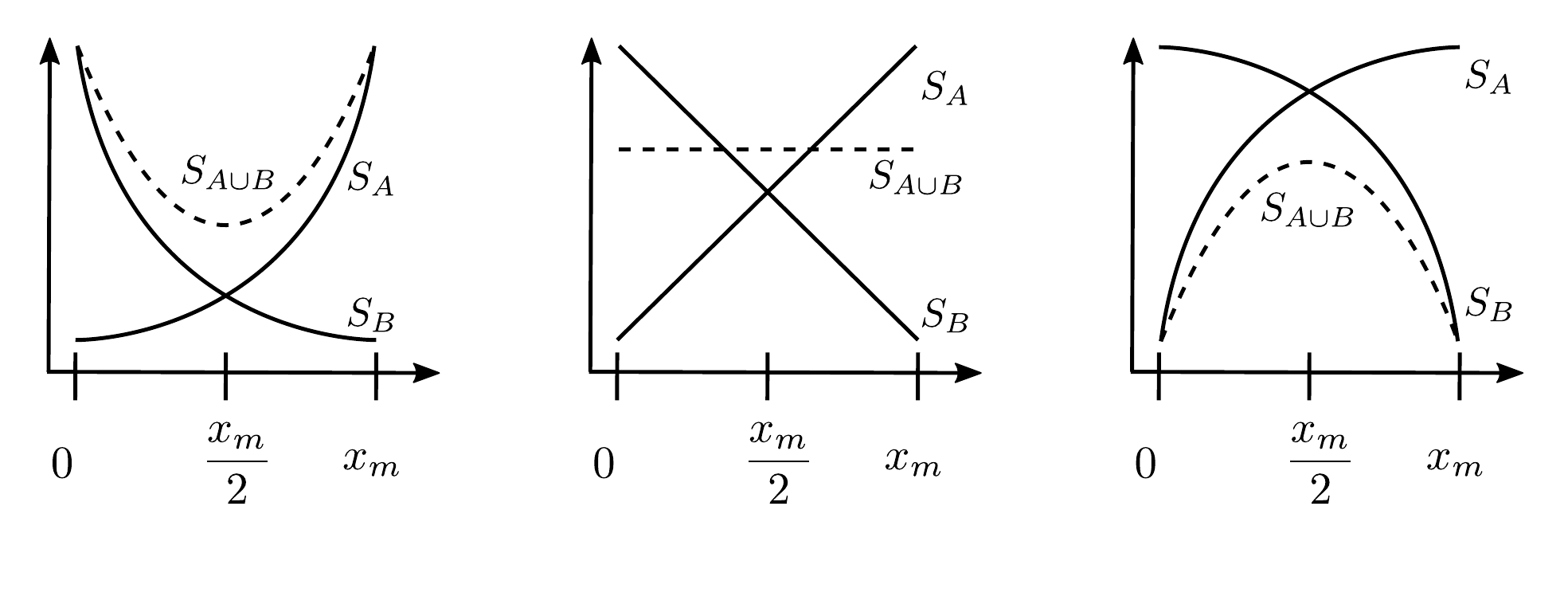}
	\caption{From left to right\,: depending on whether $\mathcal{S}_{A}$ is
		strictly-convex, linear or strictly-concave, $\mathcal{S}_{A\cup B}=\mathcal{S}_{A}+\mathcal{S}_{B}$ either goes through a minimum, is constant, or goes through a maximum.}
	\label{concave}
\end{center}
\end{figure}

To get around this obstacle, axiomatic thermodynamics usually has the following approach.
Consider two disjoint sets $A$ and $B$ at equilibrium. $A$ and $B$ are identical apart from their additive state-variables to which we apply some constraints\,:
$U_A+U_B=U$, $V_A+V_B=V$, $N_A+N_B=N$
where $(U,V,N)$ are given constants.
We wonder about the behavior of $S_{A\cup B}=S_A+S_B$ as a function of $(U_A, V_A, N_A)$.
To simplify, let us keep constant two variables, vary only the third one denoted $x_A$ and study
\begin{equation}\label{conc1}
\mathcal{S}_{A\cup B}(x_m,x_A)=\mathcal{S}(x_A) +\mathcal{S}(x_m-x_A)
\end{equation}
where $x_m$ is the maximum value allowed for $x_A$ (for instance $U_A$ and $V_A$ are kept constant, $x=N_A$ and $x_m=N$).
In Figure~\ref{concave}, $\mathcal{S}_{A\cup B}$ is plotted for strictly-convex, linear and strictly-concave function $\mathcal{S}$. 
Depending on the case, $\mathcal{S}_{A\cup B}$ goes through a minimum, is constant or goes through a maximum. Due to symmetry, stationary points are located at $x_A=(x_m-x_A)=x_m/2$.
This is general and reciprocal\,: 
a minimum or a maximum of $\mathcal{S}_{A\cup B}$ at $x_A=x_m/2$, implies the strict-convexity or strict-concavity of $\mathcal{S}$, respectively.

Suppose in addition that entropy is extensive. From Eq.\ref{jcup}, the system $J(A,B)$ resulting from the physical joining of the contents of $A$ and $B$ has an entropy
\begin{equation}\label{conc2}
\mathcal{S}_{J(A,B)}(x_m, x_A)=\mathcal{S}_{A\cup B}(x_m,x_A)
\end{equation}
For the system $J(A,B)$, $x_A$ is an internal variable that is allowed to vary (to fluctuate). If $\mathcal{S}$ is strictly-concave, $S_{J(A,B)}$ is maximized for the state for which $x_A=x_m/2$, i.e. for the most symmetrical states of the two subparts. In addition, if $\mathcal{S}$ cannot spontaneously decrease in virtue of the second law of thermodynamics, this maximum corresponds to a stable equilibrium.

In summary, if $\mathcal{S}$ is supposed to be extensive and strictly-concave, then we can prove that there exists a stable equilibrium for which intensive state-quantities are spatially uniform and allowing entropy to be maximized.
Reciprocally, if $\mathcal{S}$ is supposed to be extensive and maximum at the equilibrium, then we can prove that it is strictly-concave and that the equilibrium  is stable and corresponds to intensive state-quantities spatially uniform. This is the option of Callen's book\,\cite{Callen_1985}.

The point is that in the framework of pure thermodynamics (that is to say without invoking probabilities or information theory), if instead of postulating the stability of equilibrium (zeroth law), we prefer a \textquote{principle of maximum entropy}, we have to postulate extensivity and concavity, because there is no other way to prove the stability of equilibrium. These postulates are not required by statistical mechanics and information theory thanks to fluctuations and probabilities.

\subsubsection{Concavity of entropy}\label{concave1}

The concavity of $\mathcal S$, that has been discussed in the previous section as a function of a single variable for the sake of simplicity, must be in reality examined with respect to any direction. For the unicity and the stability of equilibrium, what is required is the strict-concavity of the hyper-surface $\mathcal S(U, V, N)$.

A multi-variable function is strictly-concave if its Hessian matrix (constructed with the partial second derivatives) is negative-definite. That is to say if\,: 
$$(-1)^k D_k > 0 \quad \textrm{for}\quad k=1\dots n$$ 
where $D_k$ is the $k$-th leading principal minor (a leading principal minor $D_k$ of a $n\times n$ matrix is the determinant of the submatrix obtained with the first $k$-th columns and rows).

Let us first examine $\mathcal S(U,V)$.
From Eq.\ref{gradient}  the gradient and Hessian are\,:
\begin{eqnarray}
&	\nabla\mathcal{S}(U,V) = \left(\begin{array}{c} c_vNU^{-1} \\  NV^{-1}\end{array}\right)\\  
&  \nabla^2 \mathcal S(U,V)= \left(\begin{array}{cc} 
	-c_vNU^{-2} & 0  \\ 
	0 &-NV^{-2}
\end{array}\right)
\end{eqnarray}
So that $D_1=-c_vNU^{-2}$ is negative and $D_2=c_vN^2U^{-2}V^{-2}$ is positive. The Hessian of $\mathcal S(U,V)$ is negative-definite thus the entropy is a strictly-concave function of $(U,V)$. 

The above result can be found in Callen's book (\cite{Callen_1985} chap.\,8). As for the global concavity of entropy with respect to all its variables, it cannot be directly examined in classical thermodynamics, because there is no way to calculate \textit{a priori} the second derivatives with respect to $N$ (it lacks for an equation of state of the form $\mu=f(N)$).
Concavity is thus either demonstrated\,\cite{Galvani_1970} from the postulate of extensivity (and a supplementary property named super-additivity) or directly postulated, which amounts to the same from a conceptual point of view.

However, it is possible to check the consistency with statistical mechanics by calculating the Hessian of $\mathcal S(U,V,N)$ given by Eq.\ref{sackur0}. With $c_v=3/2$, one obtains\,:

\begin{equation}
\nabla\mathcal{S}(U,V, N) = \left(\begin{array}{c}
	c_vNU^{-1} \\  
	NV^{-1} \\
	c_v\ln(U/N) +\ln(V)  - c_v + c
\end{array}\right) 
\end{equation}
and
\begin{equation}\label{Hessian1}
\nabla^2 \mathcal S(U,V,N)= \left(\begin{array}{ccc} 
	-c_vNU^{-2} & 0  &  c_v U^{-1}\\ 
	0 &-NV^{-2} & V^{-1}\\
	c_v U^{-1} &V^{-1} & -c_v N^{-1}
\end{array}\right)
\end{equation}
So that $D_1<0$, $D_2>0$, but $D_3= c_vU^{-2}V^{-2}N>0$. Thus the entropy given by Eq.\ref{sackur0} is not strictly-concave.

If entropy is postulated as being extensive, from Eq.\ref{formext} a term equal to $N\ln N$ must be subtracted to Eq.\ref{sackur0} leading to the alternative equality\,:
\begin{equation}\label{SackTet}
\mathcal S(U,V,N)  =\frac{3}{2}N\ln(U) + N\ln(V)  
- \frac{5}{2}N\ln(N) + N(c(m,\mathfrak{h})+1)
\end{equation}
If $\mathfrak{h}$ is taken as equal to the Planck constant $h$, this equation is known as the Sackur-Tetrode equation. For the calculation of the partial derivatives, only $\mathcal S_{NN}=-(1+c_v)N^{-1}$ is changed. One obtains for the third leading principal minor $D_3=0$. Thus in this case too, $\mathcal{S}(U,V, N)$ is not strictly-concave. Actually, it is concave (not strictly) which means that there exists a direction in the space of $(U, V, N)$ with an infinity of values for internal constraints compatible with the equilibrium.

Let us summarized as follows.
\noindent In axiomatic thermodynamics, 
\begin{itemize}
\item One postulates\,:
\end{itemize}
$$
\begin{array}{rl}
\textrm{additivity\,:}	& \mathcal{S}_{A\cup B}(N,N_A)=\mathcal{S}(N_A) +\mathcal{S}(N-N_A)\\
\textrm{extensivity\,:}	&	\mathcal{S}_{J(A,B)}(N,N_A)=\mathcal{S}_{A\cup B}(N,N_A)
\end{array}
$$
\begin{itemize}
\item Then one demonstrates\,:
\end{itemize}
$$
\begin{array}{rcl}
\left.
\begin{array}{c}
	\mathcal{S}(N_A) \\  \textrm{strictly-concave}
\end{array}\right\}
\Leftrightarrow 
\left\{{\begin{array}{c}
		\mathcal{S}_{J(A,B)}(N,N_A) \\ \textrm{has a maximum}
\end{array}}\right.
\end{array}
$$

\noindent In statistical mechanics (section \ref{max1}) one has\,:
$$
\begin{array}{c}
\mathcal{S}(N_A)  \textrm{ is not (always) strictly-concave (e.g. Eq.\ref{sackur0})}  \\ 
\mathcal{S}_{J(A,B)}(N,N_A)  \textrm{ has a maximum vs. $N_A$ (Eq.\ref{SNa})}
\end{array}
$$
Hence the following problem of consistency\,:
\myparadox{Conflict of concavity}{}{Axiomatic thermodynamics forces $\mathcal{S}$ to be strictly concave so that it is maximum at equilibrium.}{Statistical mechanics shows that it is not strictly concave, but also that it is unnecessary.}{}

\noindent Here again this conflict is solely introduced by the choice of axioms.

\subsubsection{Classical vs. axiomatic thermodynamics}

In conclusion of this section, the axiomatic thermodynamics \textit{\`a la} Callen poses the following problems\,:
\begin{enumerate}
\item Extensivity obliged us to restrict the field of application of thermodynamics.
\item  The conflict of extensivity arises.
\item The conflict of concavity arises.
\end{enumerate}
Item 1 is not a problem in itself, it is simply a pity. Item 2 and 3, also do not pose any problem, as far as we are only concerned with macroscopic thermodynamics and do not intend to make microscopic interpretations. They just pose a problem of consistency with statistical mechanics. However, in my opinion these problems do not facilitate the understanding of the concept of entropy in a global manner. For a gain that reduces to certain technical advantages.

This approach is not mandatory in thermodynamics. Alternatively phenomenological laws (see \S\ref{sndlaw}) can be raised to the status of principles, with the advantage to avoid all the above problems. But, if we are not satisfied with the phenomenological approach, and want the entropy to be maximum at equilibrium, we could not do without statistical mechanics and information theory.

\subsection{Gibbs paradoxes}\label{GBdiscuss}

Following W.V. Quine\,\cite{Quine1976} paradoxes can be classified into three categories: antinomies, falsidical paradoxes and veridical paradoxes.
Antinomies are those which contain a self-reference such as "This sentence is false". A famous one is that of Epimenides of Cnossos who was Cretan and said "Cretans, always liars...". None of the Gibbs paradoxes fit into this category. 
The last two categories can be understood in this way. Consider the three assertions\,:

\begin{enumerate}[label=\Alph*), nosep]
\item A is true.
\item B is true.
\item A and B seem to be contradictory.
\end{enumerate}

Falsidical paradoxes are those for which close examination shows the contradiction to be actual because statement A or B (or both) is not correct. Famous examples are the Zeno paradoxes.

Veridical paradoxes are those where we can finally show that A and B  are in fact not contradictory but seem to be only because of a naive preconceived idea.
A typical example is that of the Hilbert's hotel:
\begin{enumerate}[label=\Alph*),  leftmargin=*, nosep]
\item A hotel has an infinite number of rooms but all of them are occupied.
\item A traveler arrives and subject to some adjustments can finally find a room.
\end{enumerate}
There is no contradiction because by moving the person from room number $n$ to room $n+1$, by doing this for every $n$, leaves room number 1 free for a new arrival.

Solving a falsidical paradox makes it possible to correct an error. Solving a veridical paradox sheds light on a problem in a different way. The distinction is therefore very important, but unfortunately the paradox in itself gives no indication about its nature. 

In this section we come back to Gibbs paradoxes \#2 concerning the mixing entropy of two volumes of the same gas that differs in thermodynamics and statistical mechanics (\S\ref{GBphystat}). 
Literature on the subject can be classified into two categories: 
\begin{enumerate}[]
\item Papers (including this one) that consider it is veridical.
\item Papers (the majority) that consider it is falsidical. 
\end{enumerate}
In \S\ref{GBthi} we have shown how the apparent contradiction can be understood. So that in this section we examine the latter option. 

When one considers Gibbs paradox \#2 as falsidical, thermodynamics is not questioned but considered as providing the correct result towards which the statistical entropy must tend.
One is looking for a way to correct the Boltzmann entropy in order to obtain an equality of the form\,:
$S_{J(A,B)}  = S_{A\cup B}$, whereas the classical Boltzmann entropy gives instead $S_{J(A,B)}  =  S_{A\cup B}+2N\ln(2)$. One is looking for a way to make statistical entropy extensive. 
Before to examine this option, let us note that even if it allows Gibbs paradox \#2 to be resolved, it will rise up the version \#1 concerning the discontinuity\,\cite{Uffink_2006}.

\subsubsection{\textquote{Correct} Boltzmann counting}\label{CBCext}

The \textquote{correct} Boltzmann counting is supposed to rectify a miss-counting of the number $\mathcal W$ of possible allowed microstates of a set of $N$ identical classical particles.
It is believed that the classical counting overestimates $\mathcal W$ which should in fact be divided by the number $N!$ of permutations. Instead of Eq.\ref{WN} one should write\,:
\begin{equation}\label{correctPerm}
\mathcal W = \frac{\mathcal V^N}{N!}
\end{equation}
The justification of Eq.\ref{correctPerm} is discussed in \S\ref{disting}.
Here, it is not questioned and we only examine if it makes statistical entropy extensive, at least asymptotically for $N\to\infty$, named thermodynamic limit. We wonder if $ \ln\left({{\mathcal V^N}/{N!}}\right)$ admits an asymptote of the form $Ns$, that is to say if there exists $s$, independent of $N$, such as
\begin{equation}
\lim_{N\to\infty}  \left[ \ln\left({ \frac{\mathcal V^N}{N!}}\right) - Ns \right]= 0
\end{equation}
For this, the Stirling formula is used\,:
\begin{equation}
\lim_{N\to\infty}\left[\frac{N!}{\sqrt{2\pi N}\left({\frac{N}{e}}\right)^N}\right] = 1
\end{equation}
We rewrite as 
\begin{equation}\label{Stirform}
\ln N! \iseq{\infty}   N\ln(N/e)+\frac{1}{2}\ln (2\pi N)
\end{equation}
So that with Eq.\ref{correctPerm}, the entropy is
\begin{equation}
\ln\left({ \frac{\mathcal V^N}{N!}}\right) \iseq{\infty}  N\ln{\frac{\mathcal V}{N/e}} - \frac{1}{2}\ln (2\pi N)
\end{equation}
Finally one gets\,:
\begin{equation}
\lim_{N\to\infty}  \left[ \ln\left({ \frac{\mathcal V^N}{N!}}\right) - Ns \right]=  \lim_{N\to\infty}  \left[-\frac{1}{2}\ln (2\pi N)\right] = -\infty
\end{equation}
where $s=\ln(\mathcal V/N)+1$ is independent of $N$ because $\mathcal V$ is extensive.
Therefore, the \textquote{correct} Boltzmann counting (Eq.\ref{correctPerm}) does not make statistical entropy extensive.

Actually, what would makes statistical entropy extensive\,\cite{Riek_2016} is the  so-called \textquote{Stirling approximation} (not to be confused with Stirling formula Eq.\ref{Stirform})\,:
\begin{equation}\label{Stirling1}
\ln(N!)\iseq{\infty} N\ln(N/e)
\end{equation}
But even if a term $\ln(N!)$ in entropy can be justified (see below \S\ref{Swendsen} and \S\ref{disting}) there is absolutely no justification for $N\ln(N)$.

Let us suppose it is legitimate and go further by examining the different manners to obtain this correction. There are basically two approaches. 
The first, due to R. Swendsen, is technical and does not fundamentally change the way in which statistical mechanics can be considered. The second, largely in the majority, invokes quantum mechanics and the notion of indistinguishability.

\subsubsection{Swendsen approach}\label{Swendsen}

Let us come back to the equilibrium of a system made of two communicating compartments (\S\ref{max1}) and rewrite Eq.\ref{WNa} and \ref{pNa1} as:
\begin{eqnarray}
\mathcal{W}(N_A) \displaystyle = \left({\frac{\mathcal{V}_A^{N_A}}{N_A!}}\right)\left({\frac{\mathcal{V}_B^{N_B}}{N_B!}}\right)N!\\
p(N_A)=\left({\frac{\mathcal{V}_A^{N_A}}{N_A!}}\right)  \left({\frac{\mathcal{V}_B^{N_B}}{N_B!}}\right)  \left({\frac{N!}{\mathcal{V}^N}}\right) \label{pNa3}
\end{eqnarray}
Clearly, as far as $N$ and $\mathcal{V}$ are constant, defining the Boltzmann entropy as $\ln(\mathcal{W}(N_A))$ or as $\ln(p(N_A))$ gives exactly the same result except for a constant, which is already present anyway and here is hidden behind volumes expressed in unit $\lambda^3$.
Consider the latter option. Eq.\ref{pNa3} can be rewritten as:
\begin{equation}\label{pNa4}
p(N_A)   \left({\frac{\mathcal{V}^N}{N!}}\right) =\left({\frac{\mathcal{V}_A^{N_A}}{N_A!}}\right)  \left({\frac{\mathcal{V}_B^{N_B}}{N_B!}}\right)
\end{equation}
which displays an evident symmetrical treatment for the two compartments and for the whole, via the ratio $\mathcal{V}^{N}/N!$.
This observation led R. Swendsen\,\cite{Swendsen_2002,Swendsen_2006,Swendsen_2018} to go further and to propose a redefinition of the Boltzmann entropy as:
\begin{equation}
S_{\textrm{Sw}}= \ln\left({p(N_A)}\right)+c
\end{equation}
where the subscript "$\textrm{Sw}$" stands for "Swendsen". The unknown constant $c$, that has been until now interpreted as a volume-unit independent of $N$ is now arbitrarily chosen as
being an explicit function of $N$ equal to
\begin{equation}
c(N)=\ln(\mathcal{V}^{N}/N!)
\end{equation}
The motivation of this choice is that at the equilibrium and at the thermodynamic limit, $ \ln\left({p(N_A)}\right)$ scales as $\ln(N)$ and becomes negligible compared to $c(N)$ which scales as $N\ln(N)$. So that the entropy of $N$ identical particles in a volume $\mathcal{V}$ becomes:
\begin{equation}
S_{\textrm{Sw}}(N,\mathcal{V})\thlim\ln \left({\frac{\mathcal{V}^N}{N!}}\right)
\end{equation}
Technically it works (disregarding the use of Stirling approximation) but it feels a little artificial. The choice of the form that $c(N)$ must take is completely arbitrary and must be considered on the same level as a supplementary postulate. In fact, it is presented like this by Swendsen.

\subsubsection{Indistinguishability}\label{disting}

The most common argument to introduce the \textquote{correct} Boltzmann counting is that identical particles lose their individuality and traceability. When identical, particles (even very large molecules) become indistinguishable. As a consequence, microstates obtained by the $N!$ permutations of particles are all the same and must be counted as one.
This is fully contradictory to the initial approach of the kinetics theory and statistical mechanics that considers particles in the framework of classical mechanics of Newton. In this initial way of seeing things, probabilities are proportional to a volume in the phase-space and also to the time spent in this volume. With the idea of indistinguishability, it is claim that when particles are identical these volume and time spent must be divided by $N!$. As explained by K. Huang (\cite{Huang_1987} p.141), there is absolutely no way to understand this in a classical manner, the reason is fundamentally quantum, and must be accepted as is.

Beyond the fact that quantum explanations of macroscopic observations are probably not the most illuminating, from an epistemological point of view this intrusion of a quantum argument is quite arbitrary and must be considered as an additional postulate. Or alternatively, everything must be refunded starting from quantum physics.

In fact, in the classical approach what is indistinguishable is not the molecules but the way in which they manifest themselves on a macroscopic scale, which of course depends on what we are able to perceive of them. Even if the result is the same for the entropy of mixing, it is conceptually quite different.

\subsubsection{Absolute measurements of entropy}\label{SakTet}

The idea of an absolute expression of entropy and its comparison with experimental results was initiated independently by O.~Sackur and H.~Tetrode\,\cite{Grimus_2013}.  Their formula for the entropy of a perfect gas is the following\,:
\begin{equation}
S_\textrm{ST} =N\ln \frac{V}{(h{(2\pi mTe)^{-1/2}})^3}  -  \ln N!
\end{equation}
The first term is Eq.\ref{Sgaz} with the quantum of action $h$. The last term is the correction 
for indistinguishability of particles.
Entropy variations of a closed system ($N$ constant) undergoing reversible processes are equal to heat exchanges. For a transformation from $T_0$ to $T_1$, one can write\,:
\begin{equation}
S(T_1)- S(T_0)=\int_{T_0}^{T_1} \frac{1}{T}\diff Q_r 
\end{equation}
The next step is to set $T_0=0$ and use the Nernst's postulate\,: $S(0)=0$ (often named third law of thermodynamics). So that $S(T_1)$ becomes an absolute measurement of entropy that can be compared to Sackur-Tetrode formula.
The agreement is impressive\,\cite{Grimus_2013,Panos_2015} and clearly validates the choice of the Planck constant for the discretization of the phase-space.

Whether or not these measurements also validate the \textquote{correct} Boltzmann counting is a different question. Some authors claim so\,\cite{Panos_2015}, but it is incorrect. Actually, as explained by O. Stern\,\cite{Stern_1949}, the Nernst's postulates implicitly assumes the indistinguishability of particles. At $T=0$, the entropy of a solid is zero only if particles are indistinguishable. If we consider a gas of classical distinguishable particles, for the sake of consistency we must also consider a solid of distinguishable particles and write $S(0)= \ln N!$ to account for the uncertainty due to their permutations. Thus in both cases, distinguishable or not, the difference $S(T_1)- S(0)$ is the same and this question cannot be answered by such experiments.

\subsubsection{Falsidical vs.veridical}

Solving the Gibbs paradox \#2 by considering it as falsidical amounts to modify the definition of statistical entropy to make it extensive. This rises three problems\,:
\begin{enumerate}
\item There is no experimental clue for a systematic extensivity of entropy.
\item Even if the approach is successful it rises up the paradox of discontinuity (Gibbs paradox \#1) and it is unable to account for all intermediate resolutions of the mixing-unmixing cycles uncovered in \S\ref{mixthi} and \S\ref{GBthi}.
\item The approach rises a problem either of self-consistency or arbitrariness\,: all solutions of this family use in one way or another the Stirling approximation $\ln N! - N\ln N=0$, that is wrong in the thermodynamic limit of very large $N$. So that that the use of this approximation is arbitrary.
\end{enumerate}

Information theory consider Gibbs paradoxes as veridical and directly gives the correct result (\S\ref{mixthi} and \S\ref{GBthi}) for the mixing entropy without the use of Stirling approximation, with all intermediates values between 0 and $2N\ln 2$ depending on the resolution at which the problem is considered. Economy, consistency, no over interpretation of experiment, the point of view of information theory is far superior on all these aspects.

\section{Conclusion}

Confusion about the nature of entropy is maintained by a number of statements that dot the literature, which seem innocuous but in fact introduce inconsistencies, conflicts or paradoxes, and ultimately do not help in a clear understanding of the concept. 
For instance, the statement that \textquote{entropy is extensive} should be only an approximation valid in certain circumstances, but not a general principle.
The statement that \textquote{molecules are indistinguishable}, and its corollary\,: the term $\ln(N!)$, are confusing because classical particles are always distinguishable (there is no fundamental impossibility to distinguish them), but it happens that they are not distinguished, nuance is important.
The statement that \textquote{according to the second law of thermodynamics, entropy is maximum at equilibrium} should be avoided as it is only understandable once probabilities are introduced, i.e. not in the sole framework of thermodynamics, etc.

Probably one of the most common statements is the metaphoric link between entropy and \textquote{order}. How expanding a gas changes its \textquote{order} or not, is unclear. The metaphor of \textquote{spreading} and \textquote{flattening} is probably better. 
In thermodynamics, the exact nature of what is spread out, even if it remains unclear, is related to our knowledge and senses\,: it is subjective.
Its exact nature is clarified by statistical mechanics\,: it is the distribution of microstates in the overall volume of the phase-space. 
However, this result is obtained at the cost of a conflict with thermodynamics because this distribution is thought of as a fully objective property and no place is left for subjectivity.
This conflict is only solved by information theory and the maximum-entropy principle.

\myquote{If all the individual facts --- all the individual phenomena, knowledge of which we desire --- were immediately accessible to us, a science would never have arisen.
Because the mental power, the memory, of the individual is limited, the material must be arranged.} (E. Mach\,\cite{Mach_1911}).
This is the fundamental motivation of any theory, that finally arrives to an irreducible kernel of basic elements\,\cite{Einstein_1934} on which our mind can operate.
These basic elements form the fundamental postulates of the theory. The only requirement for them to serve as a basis for logical reasoning is their consistency. 
Competing sets of postulates should not be compared with respect to their supposed \textquote{physical meaning} or apparent \textquote{absurdity}. 
Actually, they are by essence unintelligible. It is fallacious to believe that some are better than others on this basis.
It is just that we are more used to certain ideas than others.

The reluctance to adopt information theory is the reluctance to adopt subjective probabilities. Which amounts to judging postulates on their intelligibility. The only judgment must be in terms of consistency and economy (assuming of course an equivalent description of phenomena). On this basis, information theory probably provides the best way to understand what is entropy.
The shortcut it allows and the removal of many inconsistencies are the best guarantee that entropy finally becomes an intuitive concept.

\vspace{6pt} 


\begin{acknowledgments}
I would like to thank Pierre Lairez for the many enriching discussions we had and the always relevant comments that followed his careful reading of the manuscript, all of which greatly contributed to its improvement.
\end{acknowledgments}


\newpage

\bibliography{weri_biblio.bib}

\begin{thebibliography}{71}%
\makeatletter
\providecommand \@ifxundefined [1]{%
 \@ifx{#1\undefined}
}%
\providecommand \@ifnum [1]{%
 \ifnum #1\expandafter \@firstoftwo
 \else \expandafter \@secondoftwo
 \fi
}%
\providecommand \@ifx [1]{%
 \ifx #1\expandafter \@firstoftwo
 \else \expandafter \@secondoftwo
 \fi
}%
\providecommand \natexlab [1]{#1}%
\providecommand \enquote  [1]{``#1''}%
\providecommand \bibnamefont  [1]{#1}%
\providecommand \bibfnamefont [1]{#1}%
\providecommand \citenamefont [1]{#1}%
\providecommand \href@noop [0]{\@secondoftwo}%
\providecommand \href [0]{\begingroup \@sanitize@url \@href}%
\providecommand \@href[1]{\@@startlink{#1}\@@href}%
\providecommand \@@href[1]{\endgroup#1\@@endlink}%
\providecommand \@sanitize@url [0]{\catcode `\\12\catcode `\$12\catcode
  `\&12\catcode `\#12\catcode `\^12\catcode `\_12\catcode `\%12\relax}%
\providecommand \@@startlink[1]{}%
\providecommand \@@endlink[0]{}%
\providecommand \url  [0]{\begingroup\@sanitize@url \@url }%
\providecommand \@url [1]{\endgroup\@href {#1}{\urlprefix }}%
\providecommand \urlprefix  [0]{URL }%
\providecommand \Eprint [0]{\href }%
\providecommand \doibase [0]{http://dx.doi.org/}%
\providecommand \selectlanguage [0]{\@gobble}%
\providecommand \bibinfo  [0]{\@secondoftwo}%
\providecommand \bibfield  [0]{\@secondoftwo}%
\providecommand \translation [1]{[#1]}%
\providecommand \BibitemOpen [0]{}%
\providecommand \bibitemStop [0]{}%
\providecommand \bibitemNoStop [0]{.\EOS\space}%
\providecommand \EOS [0]{\spacefactor3000\relax}%
\providecommand \BibitemShut  [1]{\csname bibitem#1\endcsname}%
\let\auto@bib@innerbib\@empty
\bibitem [{Note1()}]{Note1}%
  \BibitemOpen
  \bibinfo {note} {This sentence is quoted in \protect \href
  {https://www.jstor.org/stable/10.2307/24923125}{M. Tribus and E. C. McIrvine
  \protect \textit {Energy and information}, In: \protect \textit {Scientific
  American}, 225 (1971), pp. 179-190}. About the fact that it is a legend read
  \protect \href {https://ethw.org/Oral-History:Claude_E._Shannon}{Claude E.
  Shannon, an oral history conducted in 1982 by Robert Price. IEEE History
  Center, Piscataway, NJ, USA.}}\BibitemShut {Stop}%
\bibitem [{\citenamefont {Schr\"odinger}(1944)}]{Schrodinger_1944}%
  \BibitemOpen
  \bibfield  {author} {\bibinfo {author} {\bibfnamefont {E.}~\bibnamefont
  {Schr\"odinger}},\ }\href@noop {} {\emph {\bibinfo {title} {What is life
  ?}}}\ (\bibinfo  {publisher} {Cambridge University Press},\ \bibinfo {year}
  {1944})\BibitemShut {NoStop}%
\bibitem [{\citenamefont {Kondepudi}\ and\ \citenamefont
  {Prigogine}(1998)}]{Prigogine_1998}%
  \BibitemOpen
  \bibfield  {author} {\bibinfo {author} {\bibfnamefont {D.}~\bibnamefont
  {Kondepudi}}\ and\ \bibinfo {author} {\bibfnamefont {I.}~\bibnamefont
  {Prigogine}},\ }\href {\doibase 10.1002/9781118698723} {\emph {\bibinfo
  {title} {Modern thermodynamics}}}\ (\bibinfo  {publisher} {J. Wiley \&
  sons},\ \bibinfo {year} {1998})\BibitemShut {NoStop}%
\bibitem [{\citenamefont {Ben-Naim}\ and\ \citenamefont
  {Casadei}(2016)}]{Ben-Naim_2016}%
  \BibitemOpen
  \bibfield  {author} {\bibinfo {author} {\bibfnamefont {A.}~\bibnamefont
  {Ben-Naim}}\ and\ \bibinfo {author} {\bibfnamefont {D.}~\bibnamefont
  {Casadei}},\ }\href {\doibase 10.1142/10300} {\emph {\bibinfo {title} {Modern
  thermodynamics}}}\ (\bibinfo  {publisher} {World Scientific},\ \bibinfo
  {year} {2016})\BibitemShut {NoStop}%
\bibitem [{\citenamefont {Bourdieu}(1997)}]{Bourdieu_1997}%
  \BibitemOpen
  \bibfield  {author} {\bibinfo {author} {\bibfnamefont {P.}~\bibnamefont
  {Bourdieu}},\ }\href@noop {} {\emph {\bibinfo {title} {M\'editations
  pascaliennes}}}\ (\bibinfo  {publisher} {Seuil},\ \bibinfo {year}
  {1997})\BibitemShut {NoStop}%
\bibitem [{\citenamefont {Mach}(1911)}]{Mach_1911}%
  \BibitemOpen
  \bibfield  {author} {\bibinfo {author} {\bibfnamefont {E.}~\bibnamefont
  {Mach}},\ }\href {\doibase 10.1017/cbo9781107338746} {\emph {\bibinfo {title}
  {History and root of the principle of the conservation of energy}}}\
  (\bibinfo  {publisher} {The open court publishing},\ \bibinfo {year}
  {1911})\BibitemShut {NoStop}%
\bibitem [{\citenamefont {Poincar\'e}(1920)}]{Poincare_1920}%
  \BibitemOpen
  \bibfield  {author} {\bibinfo {author} {\bibfnamefont {H.}~\bibnamefont
  {Poincar\'e}},\ }\href
  {https://gallica.bnf.fr/ark:/12148/bpt6k9691658b.texteImage} {\emph {\bibinfo
  {title} {Science et m\'ethode}}}\ (\bibinfo  {publisher} {Flammarion},\
  \bibinfo {year} {1920})\BibitemShut {NoStop}%
\bibitem [{\citenamefont {Deniau}(2008)}]{Deniau2008}%
  \BibitemOpen
  \bibfield  {author} {\bibinfo {author} {\bibfnamefont {G.}~\bibnamefont
  {Deniau}},\ }\href@noop {} {\emph {\bibinfo {title} {Qu'est-ce que
  comprendre}}}\ (\bibinfo  {publisher} {Librairie philosophique J. Vrin},\
  \bibinfo {address} {Paris},\ \bibinfo {year} {2008})\BibitemShut {NoStop}%
\bibitem [{\citenamefont {Balian}(2003)}]{Balian2004}%
  \BibitemOpen
  \bibfield  {author} {\bibinfo {author} {\bibfnamefont {R.}~\bibnamefont
  {Balian}},\ }\bibfield  {title} {\enquote {\bibinfo {title} {Entropy, a
  protean concept},}\ }in\ \href {http://www.bourbaphy.fr/balian2.pdf} {\emph
  {\bibinfo {booktitle} {Poincar{\'{e}} seminar 2003}}}\ (\bibinfo  {publisher}
  {Birkh{\"a}user Basel},\ \bibinfo {year} {2003})\ pp.\ \bibinfo {pages}
  {13--27}\BibitemShut {NoStop}%
\bibitem [{\citenamefont {Maxwell}(1878)}]{Maxwell_1878}%
  \BibitemOpen
  \bibfield  {author} {\bibinfo {author} {\bibfnamefont {J.~C.}\ \bibnamefont
  {Maxwell}},\ }\bibfield  {title} {\enquote {\bibinfo {title} {Diffusion},}\
  }\href {\doibase 10.1017/CBO9780511710377.064} {\bibfield  {journal}
  {\bibinfo  {journal} {Encyclopedia Britannica, reproduced in Scientific
  papers}\ }\textbf {\bibinfo {volume} {2}},\ \bibinfo {pages} {625--646}
  (\bibinfo {year} {1878})}\BibitemShut {NoStop}%
\bibitem [{\citenamefont {Callen}(1985)}]{Callen_1985}%
  \BibitemOpen
  \bibfield  {author} {\bibinfo {author} {\bibfnamefont {H.~B.}\ \bibnamefont
  {Callen}},\ }\href@noop {} {\emph {\bibinfo {title} {Thermodynamics and an
  introduction to thermostatistics}}},\ \bibinfo {edition} {2nd}\ ed.\
  (\bibinfo  {publisher} {J. Wiley \& sons},\ \bibinfo {year}
  {1985})\BibitemShut {NoStop}%
\bibitem [{\citenamefont {Einstein}(1934)}]{Einstein_1934}%
  \BibitemOpen
  \bibfield  {author} {\bibinfo {author} {\bibfnamefont {A.}~\bibnamefont
  {Einstein}},\ }\bibfield  {title} {\enquote {\bibinfo {title} {On the method
  of theoretical physics},}\ }\href {\doibase 10.1086/286316} {\bibfield
  {journal} {\bibinfo  {journal} {Philosophy of Science}\ }\textbf {\bibinfo
  {volume} {1}},\ \bibinfo {pages} {163--169} (\bibinfo {year}
  {1934})}\BibitemShut {NoStop}%
\bibitem [{\citenamefont {Shannon}(1948)}]{Shannon_1948}%
  \BibitemOpen
  \bibfield  {author} {\bibinfo {author} {\bibfnamefont {C.~E.}\ \bibnamefont
  {Shannon}},\ }\bibfield  {title} {\enquote {\bibinfo {title} {A mathematical
  theory of communication},}\ }\href {\doibase
  10.1002/j.1538-7305.1948.tb01338.x} {\bibfield  {journal} {\bibinfo
  {journal} {The Bell System Technical Journal}\ }\textbf {\bibinfo {volume}
  {27}},\ \bibinfo {pages} {379--423} (\bibinfo {year} {1948})}\BibitemShut
  {NoStop}%
\bibitem [{\citenamefont {Jaynes}(1957{\natexlab{a}})}]{Jaynes_1957}%
  \BibitemOpen
  \bibfield  {author} {\bibinfo {author} {\bibfnamefont {E.~T.}\ \bibnamefont
  {Jaynes}},\ }\bibfield  {title} {\enquote {\bibinfo {title} {Information
  theory and statistical mechanics},}\ }\href {\doibase
  10.1103/PhysRev.106.620} {\bibfield  {journal} {\bibinfo  {journal} {Phys.
  Rev.}\ }\textbf {\bibinfo {volume} {106}},\ \bibinfo {pages} {620--630}
  (\bibinfo {year} {1957}{\natexlab{a}})}\BibitemShut {NoStop}%
\bibitem [{Gib(2018)}]{Gibbsparadox2018}%
  \BibitemOpen
  \href
  {https://www.mdpi.com/journal/entropy/special_issues/Gibbs_Paradox_2018}
  {\enquote {\bibinfo {title} {Entropy, special issue '{G}ibbs paradox
  2018'},}\ } (\bibinfo {year} {2018})\BibitemShut {NoStop}%
\bibitem [{\citenamefont {Jaynes}(1992)}]{Jaynes1992}%
  \BibitemOpen
  \bibfield  {author} {\bibinfo {author} {\bibfnamefont {E.~T.}\ \bibnamefont
  {Jaynes}},\ }\bibfield  {title} {\enquote {\bibinfo {title} {The {G}ibbs
  paradox},}\ }in\ \href {\doibase 10.1007/978-94-017-2219-3_1} {\emph
  {\bibinfo {booktitle} {Maximum Entropy and Bayesian Methods}}}\ (\bibinfo
  {publisher} {Springer Netherlands},\ \bibinfo {year} {1992})\ pp.\ \bibinfo
  {pages} {1--21}\BibitemShut {NoStop}%
\bibitem [{\citenamefont {Clausius}(1879)}]{Clausius_1879}%
  \BibitemOpen
  \bibfield  {author} {\bibinfo {author} {\bibfnamefont {R.}~\bibnamefont
  {Clausius}},\ }\href
  {https://books.google.fr/books?id=8LIEAAAAYAAJ&printsec=frontcover&hl=fr&source=gbs_ge_summary_r&cad=0#v=onepage&q&f=false}
  {\emph {\bibinfo {title} {The mechanical theory of heat}}}\ (\bibinfo
  {publisher} {Mamillan \& Co},\ \bibinfo {year} {1879})\BibitemShut {NoStop}%
\bibitem [{\citenamefont {Planck}(1903)}]{Planck_1903}%
  \BibitemOpen
  \bibfield  {author} {\bibinfo {author} {\bibfnamefont {M.}~\bibnamefont
  {Planck}},\ }\href {https://www.gutenberg.org/files/50880/50880-pdf.pdf}
  {\emph {\bibinfo {title} {Treatise of thermodynamics}}}\ (\bibinfo
  {publisher} {Longmans, Green and Co.},\ \bibinfo {year} {1903})\BibitemShut
  {NoStop}%
\bibitem [{\citenamefont {Brock}\ and\ \citenamefont
  {Knight}(1965)}]{Brock_Knight1965a}%
  \BibitemOpen
  \bibfield  {author} {\bibinfo {author} {\bibfnamefont {W.~H.}\ \bibnamefont
  {Brock}}\ and\ \bibinfo {author} {\bibfnamefont {D.~M.}\ \bibnamefont
  {Knight}},\ }\bibfield  {title} {\enquote {\bibinfo {title} {The atomic
  debates: "{M}emorable and interesting evenings in the life of the chemical
  society"},}\ }\href {\doibase 10.1086/349922} {\bibfield  {journal} {\bibinfo
   {journal} {Isis}\ }\textbf {\bibinfo {volume} {56}},\ \bibinfo {pages}
  {5--25} (\bibinfo {year} {1965})}\BibitemShut {NoStop}%
\bibitem [{\citenamefont {Maxwell}(1872)}]{Maxwell_1872}%
  \BibitemOpen
  \bibfield  {author} {\bibinfo {author} {\bibfnamefont {J.~C.}\ \bibnamefont
  {Maxwell}},\ }\href
  {https://books.google.fr/books?id=5u84AAAAMAAJ&printsec=frontcover&hl=fr#v=onepage&q&f=false}
  {\emph {\bibinfo {title} {Theory of heat}}},\ \bibinfo {edition} {3rd}\ ed.\
  (\bibinfo  {publisher} {Longmans, Green and Co.},\ \bibinfo {address}
  {London},\ \bibinfo {year} {1872})\BibitemShut {NoStop}%
\bibitem [{\citenamefont {Carnot}(1872)}]{Carnot_1872}%
  \BibitemOpen
  \bibfield  {author} {\bibinfo {author} {\bibfnamefont {S.}~\bibnamefont
  {Carnot}},\ }\bibfield  {title} {\enquote {\bibinfo {title} {R\'eflexions sur
  la puissance motrice du feu et sur les machines propres \`a d\'evelopper
  cette puissance},}\ }\href {\doibase 10.24033/asens.88} {\bibfield  {journal}
  {\bibinfo  {journal} {Annales scientifiques de l'\'Ecole Normale
  Sup\'erieure}\ }\textbf {\bibinfo {volume} {2e s{\'e}rie, 1}},\ \bibinfo
  {pages} {393--457} (\bibinfo {year} {1872})}\BibitemShut {NoStop}%
\bibitem [{Note2()}]{Note2}%
  \BibitemOpen
  \bibinfo {note} {This definition of what is \textquote {simple} and
  \textquote {complex} is a lighter version of the one given by Callen\protect
  \,\cite {Callen_1985} p.9 the later being formulated in a more rigorous
  way.}\BibitemShut {Stop}%
\bibitem [{\citenamefont {Atkins}(2010)}]{Atkins_2010}%
  \BibitemOpen
  \bibfield  {author} {\bibinfo {author} {\bibfnamefont {P.}~\bibnamefont
  {Atkins}},\ }\bibfield  {title} {\enquote {\bibinfo {title} {1. the zeroth
  law},}\ }in\ \href {\doibase 10.1093/actrade/9780199572199.003.0001} {\emph
  {\bibinfo {booktitle} {The Laws of Thermodynamics}}}\ (\bibinfo  {publisher}
  {Oxford University Press},\ \bibinfo {year} {2010})\ pp.\ \bibinfo {pages}
  {1--15}\BibitemShut {NoStop}%
\bibitem [{\citenamefont {Einstein}(1906)}]{Einstein_1906}%
  \BibitemOpen
  \bibfield  {author} {\bibinfo {author} {\bibfnamefont {A.}~\bibnamefont
  {Einstein}},\ }\bibfield  {title} {\enquote {\bibinfo {title} {On the theory
  of the brownian movement},}\ }\href@noop {} {\bibfield  {journal} {\bibinfo
  {journal} {Annalen der Physik}\ }\textbf {\bibinfo {volume} {324}},\ \bibinfo
  {pages} {371} (\bibinfo {year} {1906})}\BibitemShut {NoStop}%
\bibitem [{\citenamefont {Joule}(1850)}]{Joule_1850}%
  \BibitemOpen
  \bibfield  {author} {\bibinfo {author} {\bibfnamefont {J.P.}\ \bibnamefont
  {Joule}},\ }\bibfield  {title} {\enquote {\bibinfo {title} {On the mechanical
  equivalent of heat},}\ }\href {https://www.jstor.org/stable/108427}
  {\bibfield  {journal} {\bibinfo  {journal} {Philosophical Transactions of the
  Royal Society of London}\ }\textbf {\bibinfo {volume} {140}},\ \bibinfo
  {pages} {61--82} (\bibinfo {year} {1850})}\BibitemShut {NoStop}%
\bibitem [{Note3()}]{Note3}%
  \BibitemOpen
  \bibinfo {note} {Actually, placed here in the text, Eq.\ref {ideal} contains
  two anachronisms: 1)~in a non-atomistic world $N$ should be rather expressed
  in mole and $k$ replaced by the perfect gas constant $R$; 2)~it is precisely
  the work of Clausius and the introduction of the notion of entropy that lead
  to definitely adopt the absolute temperature scale (\cite {Maxwell_1872}
  p.155 and following). But for reasons of didactics and of coherence with the
  following, the modern expression of the ideal gas law is preferable.
  Incorporating the Boltzmann constant $k$ into temperature $T$ provides the
  double advantage of being more concise and underlying the physical meaning of
  temperature that should be more conveniently called thermal energy. This will
  permit also an equality between statistical and Shannon entropies without
  this dimensional prefactor.}\BibitemShut {Stop}%
\bibitem [{\citenamefont {Gibbs}(1874)}]{Gibbs1874}%
  \BibitemOpen
  \bibfield  {author} {\bibinfo {author} {\bibfnamefont {J.~W.}\ \bibnamefont
  {Gibbs}},\ }\href {\doibase 10.5479/sil.421748.39088007099781} {\emph
  {\bibinfo {title} {On the equilibrium of heterogeneous substances : first
  [-second] part}}}\ (\bibinfo  {publisher} {Connecticut academy of arts and
  sciences},\ \bibinfo {year} {1874})\BibitemShut {NoStop}%
\bibitem [{\citenamefont {Clausius}(1865)}]{Clausius_1865}%
  \BibitemOpen
  \bibfield  {author} {\bibinfo {author} {\bibfnamefont {R.}~\bibnamefont
  {Clausius}},\ }\bibfield  {title} {\enquote {\bibinfo {title} {Sur diverses
  formes facilement applicables qu'on peut donner aux {\'e}quations
  fondamentales de la th{\'e}orie m{\'e}canique de la chaleur.}}\ }\href
  {http://eudml.org/doc/235530} {\bibfield  {journal} {\bibinfo  {journal}
  {Journal de Math{\'e}matiques Pures et Appliqu{\'e}es}\ ,\ \bibinfo {pages}
  {361--400}} (\bibinfo {year} {1865})}\BibitemShut {NoStop}%
\bibitem [{\citenamefont {Quine}(1976)}]{Quine1976}%
  \BibitemOpen
  \bibfield  {author} {\bibinfo {author} {\bibfnamefont {W.~V.}\ \bibnamefont
  {Quine}},\ }\href {https://archive.org/details/waysofparadox00quin/mode/2up}
  {\emph {\bibinfo {title} {The ways of paradox, and other essays}}}\ (\bibinfo
   {publisher} {Harvard University Press},\ \bibinfo {address} {Cambridge,
  Massachusetts},\ \bibinfo {year} {1976})\BibitemShut {NoStop}%
\bibitem [{\citenamefont {Gibbs}(1902)}]{Gibbs_1902}%
  \BibitemOpen
  \bibfield  {author} {\bibinfo {author} {\bibfnamefont {J.W.}\ \bibnamefont
  {Gibbs}},\ }\href {https://www.gutenberg.org/files/50992/50992-pdf.pdf}
  {\emph {\bibinfo {title} {Elementary principles in statistical mechanics}}}\
  (\bibinfo  {publisher} {Charles Scribner's sons},\ \bibinfo {year}
  {1902})\BibitemShut {NoStop}%
\bibitem [{\citenamefont {Keynes}(1921)}]{Keynes_1921}%
  \BibitemOpen
  \bibfield  {author} {\bibinfo {author} {\bibfnamefont {J.M.}\ \bibnamefont
  {Keynes}},\ }\href {https://www.gutenberg.org/files/32625/32625-pdf.pdf}
  {\emph {\bibinfo {title} {A treatise on probability}}}\ (\bibinfo
  {publisher} {Macmillian},\ \bibinfo {year} {1921})\BibitemShut {NoStop}%
\bibitem [{\citenamefont {Dubs}(1942)}]{Dubs_1942}%
  \BibitemOpen
  \bibfield  {author} {\bibinfo {author} {\bibfnamefont {H.H.}\ \bibnamefont
  {Dubs}},\ }\bibfield  {title} {\enquote {\bibinfo {title} {The principle of
  insufficient reason},}\ }\href {\doibase 10.1086/286754} {\bibfield
  {journal} {\bibinfo  {journal} {Philosophy of Science}\ }\textbf {\bibinfo
  {volume} {9}},\ \bibinfo {pages} {123--131} (\bibinfo {year}
  {1942})}\BibitemShut {NoStop}%
\bibitem [{\citenamefont {Uffink}(2007)}]{Uffink_2006}%
  \BibitemOpen
  \bibfield  {author} {\bibinfo {author} {\bibfnamefont {J.}~\bibnamefont
  {Uffink}},\ }\bibfield  {title} {\enquote {\bibinfo {title} {Compendium of
  the foundations of classical statistical physics},}\ }in\ \href {\doibase
  https://doi.org/10.1016/B978-044451560-5/50012-9} {\emph {\bibinfo
  {booktitle} {Philosophy of Physics}}},\ \bibinfo {series and number}
  {Handbook of the Philosophy of Science},\ \bibinfo {editor} {edited by\
  \bibinfo {editor} {\bibfnamefont {J.}~\bibnamefont {Butterfield}}\ and\
  \bibinfo {editor} {\bibfnamefont {J.}~\bibnamefont {Earman}}}\ (\bibinfo
  {publisher} {North-Holland},\ \bibinfo {address} {Amsterdam},\ \bibinfo
  {year} {2007})\ pp.\ \bibinfo {pages} {923--1074}\BibitemShut {NoStop}%
\bibitem [{\citenamefont {Moore}(2015)}]{Moore_2015}%
  \BibitemOpen
  \bibfield  {author} {\bibinfo {author} {\bibfnamefont {C.C.}\ \bibnamefont
  {Moore}},\ }\bibfield  {title} {\enquote {\bibinfo {title} {Ergodic theorem,
  ergodic theory, and statistical mechanics},}\ }\href {\doibase
  10.1073/pnas.1421798112} {\bibfield  {journal} {\bibinfo  {journal}
  {Proceedings of the National Academy of Sciences}\ }\textbf {\bibinfo
  {volume} {112}},\ \bibinfo {pages} {1907--1911} (\bibinfo {year}
  {2015})}\BibitemShut {NoStop}%
\bibitem [{Note4()}]{Note4}%
  \BibitemOpen
  \bibinfo {note} {Eq.\protect \,\ref {volunit}, with $\protect \mathfrak h$
  equal to the Planck constant, is the Heisenberg uncertainty principle. As W.
  Heisenberg is the father of quantum mechanics, it is often viewed as coming
  from quantum physics ideas. Actually, it is between two worlds, classical and
  quantum. It comes from the wave-particle duality, from its probability
  interpretation and from the fact that the probability densities of $\protect
  \mathbf r$ and $\protect \mathbf p$ of a given particle are Fourier
  transforms of each other. Then, from the mathematical property of Fourier
  transform, increasing the resolution in the domain of one variable, degrades
  that of its conjugate.}\BibitemShut {Stop}%
\bibitem [{\citenamefont {Balian}(1991)}]{Balian_1991}%
  \BibitemOpen
  \bibfield  {author} {\bibinfo {author} {\bibfnamefont {R.}~\bibnamefont
  {Balian}},\ }\href {\doibase 10.1007/978-3-540-45475-5} {\emph {\bibinfo
  {title} {From microphysics to macrophysics}}}\ (\bibinfo  {publisher}
  {Springer Berlin Heidelberg},\ \bibinfo {year} {1991})\BibitemShut {NoStop}%
\bibitem [{\citenamefont {Boltzmann}(1964)}]{Boltzmann_Lectures}%
  \BibitemOpen
  \bibfield  {author} {\bibinfo {author} {\bibfnamefont {L.}~\bibnamefont
  {Boltzmann}},\ }\href@noop {} {\emph {\bibinfo {title} {Lectures on gas
  theory}}}\ (\bibinfo  {publisher} {Dover ed.},\ \bibinfo {year}
  {1964})\BibitemShut {NoStop}%
\bibitem [{\citenamefont {Planck}(1914)}]{Planck_1914}%
  \BibitemOpen
  \bibfield  {author} {\bibinfo {author} {\bibfnamefont {M.}~\bibnamefont
  {Planck}},\ }\href {https://www.gutenberg.org/files/40030/40030-pdf.pdf}
  {\emph {\bibinfo {title} {The theory of heat radiation}}}\ (\bibinfo
  {publisher} {P. Blakiston's son},\ \bibinfo {year} {1914})\BibitemShut
  {NoStop}%
\bibitem [{\citenamefont {Tien}\ and\ \citenamefont
  {Lienhard}(1985)}]{Tien_1979}%
  \BibitemOpen
  \bibfield  {author} {\bibinfo {author} {\bibfnamefont {C.L.}\ \bibnamefont
  {Tien}}\ and\ \bibinfo {author} {\bibfnamefont {J.H.}\ \bibnamefont
  {Lienhard}},\ }\href@noop {} {\emph {\bibinfo {title} {Statistical
  Thermodynamics}}}\ (\bibinfo  {publisher} {Hemisphere Publishing
  Corporation},\ \bibinfo {year} {1985})\BibitemShut {NoStop}%
\bibitem [{\citenamefont {Lebowitz}(1993)}]{Lebowitz_1993}%
  \BibitemOpen
  \bibfield  {author} {\bibinfo {author} {\bibfnamefont {J.L.}\ \bibnamefont
  {Lebowitz}},\ }\bibfield  {title} {\enquote {\bibinfo {title} {Macroscopic
  laws, microscopic dynamics, time's arrow and boltzmann's entropy},}\ }\href
  {\doibase https://doi.org/10.1016/0378-4371(93)90336-3} {\bibfield  {journal}
  {\bibinfo  {journal} {Physica A: Statistical Mechanics and its Applications}\
  }\textbf {\bibinfo {volume} {194}},\ \bibinfo {pages} {1--27} (\bibinfo
  {year} {1993})}\BibitemShut {NoStop}%
\bibitem [{\citenamefont {Villani}(2002)}]{Villani_2002}%
  \BibitemOpen
  \bibfield  {author} {\bibinfo {author} {\bibfnamefont {C.}~\bibnamefont
  {Villani}},\ }\bibfield  {title} {\enquote {\bibinfo {title} {Chap. 2 - a
  review of mathematical topics in collisional kinetic theory},}\ }in\ \href
  {\doibase 10.1016/S1874-5792(02)80004-0} {\emph {\bibinfo {booktitle}
  {Handbook of Mathematical Fluid Dynamics}}},\ \bibinfo {series} {Handbook of
  Mathematical Fluid Dynamics}, Vol.~\bibinfo {volume} {1},\ \bibinfo {editor}
  {edited by\ \bibinfo {editor} {\bibfnamefont {S.}~\bibnamefont
  {Friedlander}}\ and\ \bibinfo {editor} {\bibfnamefont {D.}~\bibnamefont
  {Serre}}}\ (\bibinfo  {publisher} {North-Holland},\ \bibinfo {year} {2002})\
  pp.\ \bibinfo {pages} {71--74}\BibitemShut {NoStop}%
\bibitem [{\citenamefont {Darrigol}(2021)}]{Darrigol_2021}%
  \BibitemOpen
  \bibfield  {author} {\bibinfo {author} {\bibfnamefont {O.}~\bibnamefont
  {Darrigol}},\ }\bibfield  {title} {\enquote {\bibinfo {title} {Boltzmann's
  reply to the {L}oschmidt paradox: a commented translation},}\ }\href
  {\doibase 10.1140/epjh/s13129-021-00029-2} {\bibfield  {journal} {\bibinfo
  {journal} {The European Physical Journal H}\ }\textbf {\bibinfo {volume}
  {46}},\ \bibinfo {pages} {29} (\bibinfo {year} {2021})}\BibitemShut {NoStop}%
\bibitem [{\citenamefont {Ellis}(1850)}]{Ellis_1850}%
  \BibitemOpen
  \bibfield  {author} {\bibinfo {author} {\bibfnamefont {R.L.}\ \bibnamefont
  {Ellis}},\ }\bibfield  {title} {\enquote {\bibinfo {title} {Remarks on an
  alleged proof of the {\textquotedblleft}method of least
  squares,{\textquotedblright} contained in a late number of the edinburgh
  review},}\ }\href {\doibase 10.1080/14786445008646622} {\bibfield  {journal}
  {\bibinfo  {journal} {The London, Edinburgh, and Dublin Philosophical
  Magazine and Journal of Science}\ }\textbf {\bibinfo {volume} {37}},\
  \bibinfo {pages} {321--328} (\bibinfo {year} {1850})}\BibitemShut {NoStop}%
\bibitem [{\citenamefont {Poincar\'e}(1913)}]{Poincare_1913}%
  \BibitemOpen
  \bibfield  {author} {\bibinfo {author} {\bibfnamefont {H.}~\bibnamefont
  {Poincar\'e}},\ }\href
  {https://www.gutenberg.org/files/39713/39713-h/39713-h.htm#Page_395} {\emph
  {\bibinfo {title} {The foundations of science}}}\ (\bibinfo  {publisher} {The
  science Press},\ \bibinfo {year} {1913})\BibitemShut {NoStop}%
\bibitem [{\citenamefont {Hartley}(1928)}]{Hartley_1928}%
  \BibitemOpen
  \bibfield  {author} {\bibinfo {author} {\bibfnamefont {R.~V.~L.}\
  \bibnamefont {Hartley}},\ }\bibfield  {title} {\enquote {\bibinfo {title}
  {Transmission of information},}\ }\href {\doibase
  10.1002/j.1538-7305.1928.tb01236.x} {\bibfield  {journal} {\bibinfo
  {journal} {Bell System Technical Journal}\ }\textbf {\bibinfo {volume} {7}},\
  \bibinfo {pages} {535--563} (\bibinfo {year} {1928})}\BibitemShut {NoStop}%
\bibitem [{\citenamefont {Uffink}(1995)}]{Uffink_1995}%
  \BibitemOpen
  \bibfield  {author} {\bibinfo {author} {\bibfnamefont {J.}~\bibnamefont
  {Uffink}},\ }\bibfield  {title} {\enquote {\bibinfo {title} {Can the maximum
  entropy principle be explained as a consistency requirement?}}\ }\href
  {\doibase 10.1016/1355-2198(95)00015-1} {\bibfield  {journal} {\bibinfo
  {journal} {Studies in History and Philosophy of Science Part B: Studies in
  History and Philosophy of Modern Physics}\ }\textbf {\bibinfo {volume}
  {26}},\ \bibinfo {pages} {223--261} (\bibinfo {year} {1995})}\BibitemShut
  {NoStop}%
\bibitem [{\citenamefont {Jaynes}(1957{\natexlab{b}})}]{Jaynes_1957b}%
  \BibitemOpen
  \bibfield  {author} {\bibinfo {author} {\bibfnamefont {E.~T.}\ \bibnamefont
  {Jaynes}},\ }\bibfield  {title} {\enquote {\bibinfo {title} {Information
  theory and statistical mechanics. ii},}\ }\href {\doibase
  10.1103/PhysRev.108.171} {\bibfield  {journal} {\bibinfo  {journal} {Phys.
  Rev.}\ }\textbf {\bibinfo {volume} {108}},\ \bibinfo {pages} {171--190}
  (\bibinfo {year} {1957}{\natexlab{b}})}\BibitemShut {NoStop}%
\bibitem [{\citenamefont {Jaynes}(1973)}]{Jaynes_1973}%
  \BibitemOpen
  \bibfield  {author} {\bibinfo {author} {\bibfnamefont {E.~T.}\ \bibnamefont
  {Jaynes}},\ }\bibfield  {title} {\enquote {\bibinfo {title} {The well-posed
  problem},}\ }\href {\doibase 10.1007/bf00709116} {\bibfield  {journal}
  {\bibinfo  {journal} {Foundations of Physics}\ }\textbf {\bibinfo {volume}
  {3}},\ \bibinfo {pages} {477--492} (\bibinfo {year} {1973})}\BibitemShut
  {NoStop}%
\bibitem [{\citenamefont {Arndt}(2001)}]{Arndt_2001}%
  \BibitemOpen
  \bibfield  {author} {\bibinfo {author} {\bibfnamefont {C.}~\bibnamefont
  {Arndt}},\ }\href@noop {} {\emph {\bibinfo {title} {Information measures}}}\
  (\bibinfo  {publisher} {Springer-Verlag},\ \bibinfo {year}
  {2001})\BibitemShut {NoStop}%
\bibitem [{\citenamefont {Rex}(2017)}]{Rex_2017}%
  \BibitemOpen
  \bibfield  {author} {\bibinfo {author} {\bibfnamefont {A.}~\bibnamefont
  {Rex}},\ }\bibfield  {title} {\enquote {\bibinfo {title} {Maxwell's demon - a
  historical review},}\ }\href {\doibase 10.3390/e19060240} {\bibfield
  {journal} {\bibinfo  {journal} {Entropy}\ }\textbf {\bibinfo {volume} {19}},\
  \bibinfo {pages} {240} (\bibinfo {year} {2017})}\BibitemShut {NoStop}%
\bibitem [{\citenamefont {Ciliberto}\ and\ \citenamefont
  {Lutz}(2018)}]{Ciliberto_2018}%
  \BibitemOpen
  \bibfield  {author} {\bibinfo {author} {\bibfnamefont {S.}~\bibnamefont
  {Ciliberto}}\ and\ \bibinfo {author} {\bibfnamefont {E.}~\bibnamefont
  {Lutz}},\ }\bibfield  {title} {\enquote {\bibinfo {title} {The physics of
  information: From maxwell to landauer},}\ }in\ \href {\doibase
  10.1007/978-3-319-93458-7_5} {\emph {\bibinfo {booktitle} {Energy Limits in
  Computation}}}\ (\bibinfo  {publisher} {Springer International Publishing},\
  \bibinfo {year} {2018})\ pp.\ \bibinfo {pages} {155--175}\BibitemShut
  {NoStop}%
\bibitem [{\citenamefont {Feynman}\ \emph {et~al.}(1966)\citenamefont
  {Feynman}, \citenamefont {Leighton},\ and\ \citenamefont
  {Sands}}]{Feynmann_Ratchet}%
  \BibitemOpen
  \bibfield  {author} {\bibinfo {author} {\bibfnamefont {R.~P.}\ \bibnamefont
  {Feynman}}, \bibinfo {author} {\bibfnamefont {R.~B.}\ \bibnamefont
  {Leighton}}, \ and\ \bibinfo {author} {\bibfnamefont {M.}~\bibnamefont
  {Sands}},\ }\href {https://www.feynmanlectures.caltech.edu/I_46.html} {\emph
  {\bibinfo {title} {The {F}eynman lectures on physics}}}\ (\bibinfo
  {publisher} {Addison-Wesley, Reading, MA},\ \bibinfo {year} {1966})\
  Chap.~\bibinfo {chapter} {46}, pp.\ \bibinfo {pages} {1--9}\BibitemShut
  {NoStop}%
\bibitem [{\citenamefont {Brillouin}(1950)}]{Brillouin1950}%
  \BibitemOpen
  \bibfield  {author} {\bibinfo {author} {\bibfnamefont {L.}~\bibnamefont
  {Brillouin}},\ }\bibfield  {title} {\enquote {\bibinfo {title} {Can the
  rectifier become a thermodynamical demon?}}\ }\href {\doibase
  10.1103/physrev.78.627.2} {\bibfield  {journal} {\bibinfo  {journal}
  {Physical Review}\ }\textbf {\bibinfo {volume} {78}},\ \bibinfo {pages}
  {627--628} (\bibinfo {year} {1950})}\BibitemShut {NoStop}%
\bibitem [{\citenamefont {Bang}\ \emph {et~al.}(2018)\citenamefont {Bang},
  \citenamefont {Pan}, \citenamefont {Hoang}, \citenamefont {Ahn},
  \citenamefont {Jarzynski}, \citenamefont {Quan},\ and\ \citenamefont
  {Li}}]{Bang:2018we}%
  \BibitemOpen
  \bibfield  {author} {\bibinfo {author} {\bibfnamefont {J.}~\bibnamefont
  {Bang}}, \bibinfo {author} {\bibfnamefont {R.}~\bibnamefont {Pan}}, \bibinfo
  {author} {\bibfnamefont {T.M.}\ \bibnamefont {Hoang}}, \bibinfo {author}
  {\bibfnamefont {J.}~\bibnamefont {Ahn}}, \bibinfo {author} {\bibfnamefont
  {C.}~\bibnamefont {Jarzynski}}, \bibinfo {author} {\bibfnamefont {H.T.}\
  \bibnamefont {Quan}}, \ and\ \bibinfo {author} {\bibfnamefont
  {T.}~\bibnamefont {Li}},\ }\bibfield  {title} {\enquote {\bibinfo {title}
  {Experimental realization of {F}eynman's ratchet},}\ }\href {\doibase
  10.1088/1367-2630/aae71f} {\bibfield  {journal} {\bibinfo  {journal} {New
  Journal of Physics}\ }\textbf {\bibinfo {volume} {20}},\ \bibinfo {pages}
  {103032} (\bibinfo {year} {2018})}\BibitemShut {NoStop}%
\bibitem [{\citenamefont {Gunn}\ and\ \citenamefont
  {Staples}(1969)}]{Gunn:1969wr}%
  \BibitemOpen
  \bibfield  {author} {\bibinfo {author} {\bibfnamefont {J.~B.}\ \bibnamefont
  {Gunn}}\ and\ \bibinfo {author} {\bibfnamefont {J.~L.}\ \bibnamefont
  {Staples}},\ }\bibfield  {title} {\enquote {\bibinfo {title} {Spontaneous
  reverse current due to the {B}rillouin emf in a diode.}}\ }\href {\doibase
  10.1063/1.1652709} {\bibfield  {journal} {\bibinfo  {journal} {Appl. Phys.
  Lett.}\ }\textbf {\bibinfo {volume} {14}},\ \bibinfo {pages} {54--56}
  (\bibinfo {year} {1969})}\BibitemShut {NoStop}%
\bibitem [{\citenamefont {Szilard}(1964)}]{Szilard_1964}%
  \BibitemOpen
  \bibfield  {author} {\bibinfo {author} {\bibfnamefont {L.}~\bibnamefont
  {Szilard}},\ }\bibfield  {title} {\enquote {\bibinfo {title} {On the decrease
  of entropy in a thermodynamic system by the intervention of intelligent
  beings},}\ }\href {\doibase 10.1002/bs.3830090402} {\bibfield  {journal}
  {\bibinfo  {journal} {Behavioral Science}\ }\textbf {\bibinfo {volume} {9}},\
  \bibinfo {pages} {301--310} (\bibinfo {year} {1964})}\BibitemShut {NoStop}%
\bibitem [{\citenamefont {Pu}\ \emph {et~al.}(2020)\citenamefont {Pu},
  \citenamefont {Gong},\ and\ \citenamefont {Robertson}}]{Pu_2020}%
  \BibitemOpen
  \bibfield  {author} {\bibinfo {author} {\bibfnamefont {S.}~\bibnamefont
  {Pu}}, \bibinfo {author} {\bibfnamefont {C.}~\bibnamefont {Gong}}, \ and\
  \bibinfo {author} {\bibfnamefont {A.~W.}\ \bibnamefont {Robertson}},\
  }\bibfield  {title} {\enquote {\bibinfo {title} {Liquid cell transmission
  electron microscopy and its applications},}\ }\href {\doibase
  10.1098/rsos.191204} {\bibfield  {journal} {\bibinfo  {journal} {Royal
  Society Open Science}\ }\textbf {\bibinfo {volume} {7}},\ \bibinfo {pages}
  {191204} (\bibinfo {year} {2020})}\BibitemShut {NoStop}%
\bibitem [{\citenamefont {Dieks}(2013)}]{Dieks_2013}%
  \BibitemOpen
  \bibfield  {author} {\bibinfo {author} {\bibfnamefont {D.}~\bibnamefont
  {Dieks}},\ }\bibfield  {title} {\enquote {\bibinfo {title} {{I}s there a
  unique physical entropy? micro versus macro},}\ }in\ \href {\doibase
  10.1007/978-94-007-5845-2_3} {\emph {\bibinfo {booktitle} {New Challenges to
  Philosophy of Science}}}\ (\bibinfo  {publisher} {Springer {N}etherlands},\
  \bibinfo {year} {2013})\ pp.\ \bibinfo {pages} {23--34}\BibitemShut {NoStop}%
\bibitem [{\citenamefont {Duhem}(1906)}]{Duhem_1906}%
  \BibitemOpen
  \bibfield  {author} {\bibinfo {author} {\bibfnamefont {P.}~\bibnamefont
  {Duhem}},\ }\href {https://gallica.bnf.fr/ark:/12148/bpt6k951903} {\emph
  {\bibinfo {title} {La th{\'{e}}orie physique. Son objet, sa structure}}}\
  (\bibinfo  {publisher} {Chevalier \& Rivi\`{e}re},\ \bibinfo {address}
  {Paris},\ \bibinfo {year} {1906})\BibitemShut {NoStop}%
\bibitem [{\citenamefont {Swendsen}(2017)}]{Swendsen_2017}%
  \BibitemOpen
  \bibfield  {author} {\bibinfo {author} {\bibfnamefont {R.~H.}\ \bibnamefont
  {Swendsen}},\ }\bibfield  {title} {\enquote {\bibinfo {title}
  {Thermodynamics, statistical mechanics and entropy},}\ }\href {\doibase
  10.3390/e19110603} {\bibfield  {journal} {\bibinfo  {journal} {Entropy}\
  }\textbf {\bibinfo {volume} {19}},\ \bibinfo {pages} {603} (\bibinfo {year}
  {2017})}\BibitemShut {NoStop}%
\bibitem [{\citenamefont {Touchette}(2002)}]{Touchette_2002}%
  \BibitemOpen
  \bibfield  {author} {\bibinfo {author} {\bibfnamefont {H.}~\bibnamefont
  {Touchette}},\ }\bibfield  {title} {\enquote {\bibinfo {title} {When is a
  quantity additive, and when is it extensive?}}\ }\href {\doibase
  10.1016/s0378-4371(01)00644-6} {\bibfield  {journal} {\bibinfo  {journal}
  {Physica A: Statistical Mechanics and its Applications}\ }\textbf {\bibinfo
  {volume} {305}},\ \bibinfo {pages} {84--88} (\bibinfo {year}
  {2002})}\BibitemShut {NoStop}%
\bibitem [{\citenamefont {Addison}\ and\ \citenamefont
  {Gray}(2001)}]{Addison_2001}%
  \BibitemOpen
  \bibfield  {author} {\bibinfo {author} {\bibfnamefont {S.~R.}\ \bibnamefont
  {Addison}}\ and\ \bibinfo {author} {\bibfnamefont {J.~E.}\ \bibnamefont
  {Gray}},\ }\bibfield  {title} {\enquote {\bibinfo {title} {Is extensivity a
  fundamental property of entropy?}}\ }\href {\doibase
  10.1088/0305-4470/34/38/301} {\bibfield  {journal} {\bibinfo  {journal}
  {Journal of Physics A: Mathematical and General}\ }\textbf {\bibinfo {volume}
  {34}},\ \bibinfo {pages} {7733--7737} (\bibinfo {year} {2001})}\BibitemShut
  {NoStop}%
\bibitem [{\citenamefont {Galvani}\ and\ \citenamefont
  {Scotti}(1970)}]{Galvani_1970}%
  \BibitemOpen
  \bibfield  {author} {\bibinfo {author} {\bibfnamefont {L.}~\bibnamefont
  {Galvani}}\ and\ \bibinfo {author} {\bibfnamefont {A.}~\bibnamefont
  {Scotti}},\ }\bibfield  {title} {\enquote {\bibinfo {title} {On subadditivity
  and convexity properties of thermodynamic functions},}\ }\href {\doibase
  10.1351/pac197022030229} {\bibfield  {journal} {\bibinfo  {journal} {Pure
  Appl. Chem.}\ }\textbf {\bibinfo {volume} {22}},\ \bibinfo {pages} {229--236}
  (\bibinfo {year} {1970})}\BibitemShut {NoStop}%
\bibitem [{\citenamefont {Riek}\ and\ \citenamefont {Sobol}(2016)}]{Riek_2016}%
  \BibitemOpen
  \bibfield  {author} {\bibinfo {author} {\bibfnamefont {R.}~\bibnamefont
  {Riek}}\ and\ \bibinfo {author} {\bibfnamefont {A.}~\bibnamefont {Sobol}},\
  }\bibfield  {title} {\enquote {\bibinfo {title} {Comments on the extensivity
  of the {B}oltzmann entropy},}\ }\href {\doibase 10.4172/2161-0398.1000207}
  {\bibfield  {journal} {\bibinfo  {journal} {Journal of Physical Chemistry \&
  Biophysics}\ }\textbf {\bibinfo {volume} {6}},\ \bibinfo {pages} {1--7}
  (\bibinfo {year} {2016})}\BibitemShut {NoStop}%
\bibitem [{\citenamefont {{Swendsen}}(2002)}]{Swendsen_2002}%
  \BibitemOpen
  \bibfield  {author} {\bibinfo {author} {\bibfnamefont {R.~H.}\ \bibnamefont
  {{Swendsen}}},\ }\bibfield  {title} {\enquote {\bibinfo {title} {{Statistical
  mechanics of classical systems with distinguishable particles}},}\ }\href
  {\doibase 10.1023/A:1015161825292} {\bibfield  {journal} {\bibinfo  {journal}
  {Journal of Statistical Physics}\ }\textbf {\bibinfo {volume} {107}},\
  \bibinfo {pages} {1143--1166} (\bibinfo {year} {2002})}\BibitemShut {NoStop}%
\bibitem [{\citenamefont {Swendsen}(2006)}]{Swendsen_2006}%
  \BibitemOpen
  \bibfield  {author} {\bibinfo {author} {\bibfnamefont {R.~H.}\ \bibnamefont
  {Swendsen}},\ }\bibfield  {title} {\enquote {\bibinfo {title} {Statistical
  mechanics of colloids and {B}oltzmann's definition of the entropy},}\ }\href
  {\doibase 10.1119/1.2174962} {\bibfield  {journal} {\bibinfo  {journal}
  {American Journal of Physics}\ }\textbf {\bibinfo {volume} {74}},\ \bibinfo
  {pages} {187--190} (\bibinfo {year} {2006})}\BibitemShut {NoStop}%
\bibitem [{\citenamefont {Swendsen}(2018)}]{Swendsen_2018}%
  \BibitemOpen
  \bibfield  {author} {\bibinfo {author} {\bibfnamefont {R.}~\bibnamefont
  {Swendsen}},\ }\bibfield  {title} {\enquote {\bibinfo {title} {Probability,
  entropy, and {G}ibbs' paradox(es)},}\ }\href {\doibase 10.3390/e20060450}
  {\bibfield  {journal} {\bibinfo  {journal} {Entropy}\ }\textbf {\bibinfo
  {volume} {20}},\ \bibinfo {pages} {450} (\bibinfo {year} {2018})}\BibitemShut
  {NoStop}%
\bibitem [{\citenamefont {Huang}(1987)}]{Huang_1987}%
  \BibitemOpen
  \bibfield  {author} {\bibinfo {author} {\bibfnamefont {K.}~\bibnamefont
  {Huang}},\ }\href@noop {} {\emph {\bibinfo {title} {{S}tatistical
  {M}echanics}}}\ (\bibinfo  {publisher} {J. Wiley \& sons},\ \bibinfo {year}
  {1987})\BibitemShut {NoStop}%
\bibitem [{\citenamefont {Grimus}(2013)}]{Grimus_2013}%
  \BibitemOpen
  \bibfield  {author} {\bibinfo {author} {\bibfnamefont {W.}~\bibnamefont
  {Grimus}},\ }\bibfield  {title} {\enquote {\bibinfo {title} {100th
  anniversary of the {S}ackur--{T}etrode equation},}\ }\href {\doibase
  10.1002/andp.201300720} {\bibfield  {journal} {\bibinfo  {journal} {Ann.
  Phys. (Berlin)}\ }\textbf {\bibinfo {volume} {525}},\ \bibinfo {pages}
  {A32--A35} (\bibinfo {year} {2013})}\BibitemShut {NoStop}%
\bibitem [{\citenamefont {Pa{\~{n}}os}\ and\ \citenamefont
  {P{\'{e}}rez}(2015)}]{Panos_2015}%
  \BibitemOpen
  \bibfield  {author} {\bibinfo {author} {\bibfnamefont {F.~J.}\ \bibnamefont
  {Pa{\~{n}}os}}\ and\ \bibinfo {author} {\bibfnamefont {E.}~\bibnamefont
  {P{\'{e}}rez}},\ }\bibfield  {title} {\enquote {\bibinfo {title}
  {Sackur{\textendash}tetrode equation in the lab},}\ }\href {\doibase
  10.1088/0143-0807/36/5/055033} {\bibfield  {journal} {\bibinfo  {journal}
  {European Journal of Physics}\ }\textbf {\bibinfo {volume} {36}},\ \bibinfo
  {pages} {055033} (\bibinfo {year} {2015})}\BibitemShut {NoStop}%
\bibitem [{\citenamefont {Stern}(1949)}]{Stern_1949}%
  \BibitemOpen
  \bibfield  {author} {\bibinfo {author} {\bibfnamefont {O.}~\bibnamefont
  {Stern}},\ }\bibfield  {title} {\enquote {\bibinfo {title} {On the term $k$
  $\mathrm{ln}n!$ in the entropy},}\ }\href {\doibase
  10.1103/RevModPhys.21.534} {\bibfield  {journal} {\bibinfo  {journal} {Rev.
  Mod. Phys.}\ }\textbf {\bibinfo {volume} {21}},\ \bibinfo {pages} {534--535}
  (\bibinfo {year} {1949})}\BibitemShut {NoStop}%
\end{thebibliography}%

\end{document}